\documentclass[a4paper,11pt]{article}
\pdfoutput=1 % if your are submitting a pdflatex (i.e. if you have
\usepackage{jcappub} % for details on the use of the package, please
\usepackage[T1]{fontenc} % if needed

\usepackage[latin9]{inputenc}
\usepackage{float}
\usepackage{amsmath}
\usepackage{amssymb}
\usepackage{graphicx}
\usepackage{esint}
 \usepackage{hyperref}
\usepackage{comment}
\usepackage{color}
\usepackage{microtype}
\usepackage{cleveref}
\usepackage{breakurl}
\usepackage{bbm}
\newcommand{\be}{\begin{equation}}
\newcommand{\ee}{\end{equation}}
\newcommand{\ben}{\begin{displaymath}}
\newcommand{\een}{\end{displaymath}}
\newcommand{\bea}{\begin{eqnarray}}
\newcommand{\eea}{\end{eqnarray}}
\def\K{K{\"a}hler}
   \newcommand{\rf}[1]{(\ref{#1})}
\newcommand{\vp}{\varphi}

\makeatletter

\pdfpageheight\paperheight
\pdfpagewidth\paperwidth

\@ifundefined{textcolor}{}{%
 \definecolor{BLACK}{gray}{0}
 \definecolor{WHITE}{gray}{1}
 \definecolor{RED}{rgb}{1,0,0}
 \definecolor{GREEN}{rgb}{0,1,0}
 \definecolor{BLUE}{rgb}{0,0,1}
 \definecolor{CYAN}{cmyk}{1,0,0,0}
 \definecolor{MAGENTA}{cmyk}{0,1,0,0}
 \definecolor{YELLOW}{cmyk}{0,0,1,0}
}

\newcommand{\dd}{\mathrm{d}}
\newcommand{\mn}{{\mu\nu}}

\allowdisplaybreaks

\makeatother

\makeatletter
\DeclareRobustCommand{\rcite}[1]{%
  \rcite@aux#1,\@nil{#1}%
}
\def\rcite@aux#1,#2\@nil#3{%
  \if\relax#2\relax
    Ref.~\cite{#3}%
  \else
    Refs.~\cite{#3}%
  \fi
}
\makeatother

\hypersetup{
    colorlinks = true,
    citecolor = {blue},
    linkcolor = {blue},
    urlcolor = {blue},
}

\title{ \boldmath{\rm \bf  Dark energy, $\alpha$-attractors, and large-scale structure surveys}}

\author[a]{\rm  Yashar Akrami,}
\author[b]{\rm  Renata Kallosh,}
\author[b]{\rm Andrei Linde}
\author[a, c]{\rm  and Valeri Vardanyan}

\affiliation[a]{Lorentz Institute for Theoretical Physics, Leiden University, P.O. Box 9506, 2300 RA Leiden, The Netherlands}
\affiliation[b]{Stanford Institute for Theoretical Physics and Department of Physics, Stanford University, Stanford, CA 94305, USA}
\affiliation[c]{Leiden Observatory, Leiden University, P.O. Box 9513, 2300 RA Leiden, The Netherlands}

\emailAdd{akrami@lorentz.leidenuniv.nl}
\emailAdd{kallosh@stanford.edu}
\emailAdd{alinde@stanford.edu}
\emailAdd{vardanyan@lorentz.leidenuniv.nl}

\abstract{Over the last few years, a large family of cosmological attractor models has been discovered, which can successfully match the latest inflation-related observational data. Many of these  models can also describe a small cosmological constant $\Lambda$, which provides the most natural description of the present stage of the cosmological acceleration. In this paper, we study $\alpha$-attractor models with dynamical dark energy, including the cosmological constant $\Lambda$ as a free parameter. Predominantly, the models with $\Lambda > 0$ converge to the asymptotic regime with the equation of state $w=-1$. However, there are some  models  with $w\neq -1$, which are compatible with the current observations. In the simplest models with $\Lambda = 0$, one has the tensor to scalar ratio $r=\frac{12\alpha}{N^2}$ and the asymptotic equation of state $w=-1+\frac{2}{9\alpha}$ (which in general differs from its present value). For example, in the seven disk M-theory related model with $\alpha = 7/3$ one finds $r \sim 10^{-2}$ and the asymptotic equation of state is $w \sim -0.9$. Future observations, including large-scale structure surveys as well as  B-mode detectors will test these, as well as more general models presented here. We also discuss gravitational reheating in models of quintessential inflation and argue that its investigation may be interesting from the point of view of inflationary cosmology. Such models require a much greater number of $e$-folds, and therefore predict a spectral index $n_{s}$ that can exceed the value in more conventional models by about $0.006$. This suggests a way to distinguish the conventional inflationary models  from the models of quintessential inflation, even if they predict $w = -1$. 
}


\begin{document}
\maketitle

\newpage
\section{\Large Introduction}\label{sec:intro}

The discovery of dark energy in 1998 \cite{Riess:1998cb,Perlmutter:1998np} pushed the cosmological constant problem to the forefront of research. The observers found that empty space is not entirely empty, it has a tiny energy density $\sim 10^{-29} {\rm g\cdot cm^{-3}}$. This minuscule number is 120 orders of magnitude smaller than the Planck density,  and 29 orders of magnitude smaller than the density of water.  This discovery triggered an unexpected chain of events in theoretical physics.

For many decades theorists were unsuccessfully trying to find a theory which would explain why the vacuum energy density is exactly zero. But we could not do it; it was a spectacular failure.  After the discovery of dark energy/cosmological constant, we face a much more complicated problem, consisting of two equally difficult parts: One should explain why vacuum energy/cosmological constant is not exactly zero but is extremely small, 
%about $10^{{-120}}$ in Planck units, 
and why  this constant  is of the same order of magnitude as the density of normal matter in the universe, but only at the present epoch. %Arguably, the best presently known way to address these two problems without invoking incredible fine-tuning is related to the anthropic principle, and  to the theory of the inflationary multiverse/string theory landscape. % Here we will briefly review the history of related ideas, following Ref.~\cite{Linde:2015edk}, where a more detailed discussion can be found.

Arguably, the best presently available theoretical reasoning for the smallness of dark energy is based on anthropic constraints on the energy density of a  metastable  vacuum state (cosmological constant) \cite{DaviesUnwin,Linde:1984ir,Sakharov:1984ir,Banks:1984cw,Barrow:1988yia,Linde:1986dq,Weinberg:1987dv,Martel:1997vi,Garriga:1999hu}, which may take different values in the context of inflationary multiverse (string theory landscape) \cite{Linde:1984ir,Sakharov:1984ir,Bousso:2000xa,Kachru:2003aw,Douglas:2003um,Susskind:2003kw}. For a brief review of related ideas see Ref.~\cite{Linde:2015edk}. While the underlying theory is still incomplete, perhaps it is fair to say that, for many of us, the incredible smallness of the cosmological constant/dark energy  no longer looks as surprising and problematic as it was twenty years ago, at the moment of its discovery.% in Refs.~\cite{Riess:1998cb,Perlmutter:1998np}.

%Scientific community is still split in its views on  the anthropic solution of the cosmological constant problem.%, and on the theory of inflationary multiverse in general. 
%However, we are not aware of any compelling alternative, so we will take this solution very seriously. Whereas the underlying theory is still incomplete, perhaps it is fair to say that, for many of us, the incredible smallness of the cosmological constant/dark energy  no longer looks as surprising as it was 20 years ago, at the moment of its discovery  in \cite{Riess:1998cb,Perlmutter:1998np}.

A closely related approach to the cosmological constant problem was proposed back in 1986 \cite{Linde:1986dq}. It was based on a combination of eternal chaotic inflation \cite{Linde:1986fd} and a subsequent slow roll of what was later called  `quintessence' field $\phi$. 
The model described a field $\phi$ with an extremely flat effective potential $V(\phi) =  \gamma  \phi$, with $\gamma \ll 10^{{-120}}$, an inflaton field $\sigma$ with an  inflaton  potential $V(\sigma)$ vanishing at its minimum,  and an arbitrary cosmological constant $\Lambda$:
\be\label{1}
V(\phi,\sigma) = V(\sigma) +  \gamma  \phi +\Lambda \,.
\ee
During eternal inflation supported by the field $\sigma$, the field $\phi$ experiences inflationary quantum fluctuations, which change its local values. As a result, the universe in this scenario becomes divided into  exponentially many  exponentially large parts (`universes') containing {\it all} possible values of the field $\phi$. After inflation, the  energy density $\rho$ of the scalar field inside these  universes (i.e. dark energy) is given by 
\be\label{linear}
V(\phi) =  \gamma  \phi +\Lambda \,.
\ee
Since the field $\phi$ can take any value and changes extremely slowly because of the smallness of $V'(\phi) = \gamma \ll 10^{{-120}}$,  the potential $ \gamma  \phi +\Lambda$ behaves as an effective cosmological constant taking all possible values in different parts of the universe. It was argued in Ref.~\cite{Linde:1986dq} that life as we know it can exist only in those parts of the universe where $|V(\phi)|=  |\gamma  \phi + \Lambda| \lesssim 10^{{-120}} \sim 10^{-29} {\rm g\cdot cm^{-3}}$. Thus, the absolute value of the effective cosmological constant in the observable part of the universe must be smaller than $\mathcal{O}(10^{-29})\ {\rm g\cdot cm^{-3}}$. This solves the cosmological constant/dark energy problem in this model,  independently of the value of the original `cosmological constant' $\Lambda$ \cite{Linde:1986dq}. A detailed investigation of cosmological consequences of this simple model and its generalizations was performed later in Refs.~\cite{Kallosh:2003bq,Wang:2004nm,Kallosh:2003mt,Garriga:2003hj}.

However, unlike the earlier proposed mechanisms \cite{Linde:1984ir,Sakharov:1984ir} and the string theory landscape scenario \cite{Bousso:2000xa,Kachru:2003aw,Douglas:2003um,Susskind:2003kw}, the quintessence-related mechanism of Ref.~\cite{Linde:1986dq} requires  fine-tuning of the parameter $|V' |= |\gamma| \lesssim 10^{{-120}}$,   in addition  to the standard anthropic constraint $|V(\phi)| \lesssim 10^{{-120}}$.  One may argue that the requirement $V'  < 10^{{-120}}$ in this scenario is also anthropic  \cite{Linde:1986dq,Kallosh:2003bq,Wang:2004nm,Kallosh:2003mt,Garriga:2003hj}. Indeed,  for $V'   \gg 10^{{-120}}$, the field $\phi$ in the regime with $|V(\phi)| \lesssim 10^{{-120}}$ moves fast, the potential $V(\phi)$ quickly becomes  negative and the universe collapses too early. %This argument may work in other similar models \cite{Kallosh:2003bq,Wang:2004nm,Kallosh:2003mt,Garriga:2003hj}. 
But in  the vast majority of the subsequently proposed  models of dark energy   \cite{Wetterich:1987fm,Ratra:1987rm,Brax:2017idh}  one has $V(\phi) \geq 0$,  and therefore, in addition to the problem of explaining why $V(\phi)   \lesssim 10^{{-120}}$, one should solve an equally difficult problem and explain why $V'(\phi)   \lesssim 10^{{-120}}$.  Thus models of dynamical dark energy often bring more problems than they are trying to solve. 

%Scientific community is still split in its views on  the anthropic solution of the cosmological constant problem, and on the theory of inflationary multiverse in general. However, we are unaware of any compelling alternative, so we will take this solution very seriously. Whereas the underlying theory is still incomplete, perhaps it is fair to say that, for many of us, the incredible smallness of the cosmological constant/dark energy  no longer looks as surprising as it was 20 years ago, at the moment of its discovery  in \cite{Riess:1998cb,Perlmutter:1998np}.

Some of these problems may go away if one considers dynamical dark energy/quintessence not as an alternative to string theory landscape, but as a possible addition to it. Indeed, in string theory one has many moduli fields, some of which can be extremely light. If their mass is sufficiently small, they may stay away from their minima. Thus, we may have an exponentially large multiplicity of discrete vacuum energy levels, and, in addition, a slowly varying contribution of light moduli to dark energy. %This may further increase the flexibility of the string landscape construction.

This scenario would describe quintessence with an additional provision: The potential of dark energy/quintessence may contain an arbitrary string theory contribution to the vacuum energy, i.e. to the cosmological constant. Moreover, in the context of the KKLT construction \cite{Kachru:2003aw}, vacuum energy in string theory is a result of a (generically) huge negative vacuum energy of a supersymmetric AdS vacuum state and of a huge positive contribution of uplifting. The sum of these two   contributions can equally easily undershoot or overshoot the level $\Lambda = 0$. This suggests that, after averaging over an exponentially large number of positive and negative contributions in the landscape, the probability of  a tiny negative cosmological constant $\Lambda \sim -10^{{-120}}$ should be approximately equal to the probability of a tiny positive cosmological constant $\Lambda \sim +10^{{-120}}$. This is similar to the conjecture made in a different context in Ref.~\cite{Martel:1997vi}.

In practical terms, this means that instead of limiting our attention to dark energy models with  potentials $V(\phi)$ vanishing in the large field limit, one should study predictions of a large class of models with  potentials $V(\phi) + \Lambda$, where $\Lambda$ can take a wide range of values. Admittedly, this is a very primitive model of what may actually happen in the landscape, but we will keep this model in mind when discussing what different theories may actually predict.

This simple provision immediately improves some of the previously proposed models. Consider for example the simplest dark energy potential \rf{1} proposed in Ref.~\cite{Linde:1986dq}. An important part of this model was the stage of eternal inflation driven by the scalar field $\sigma$, which pushed the scalar field $\phi$ in different directions in different parts of the universe and created parts of the universe with the post-inflationary values of the  potential $\gamma  \phi +\Lambda \lesssim 10^{{-120}}$. In order to cancel the naturally large value $\Lambda = \mathcal{O}(1)$ in this theory with $V' = \gamma < 10^{{-120}}$, one would need to trust the simple linear expression for the potential \rf{1} in the incredibly large range of variation of the field $\Delta \phi   \gtrsim 10^{{120}}$. This is a very challenging requirement.

In the new scenario,  the scalar field $\sigma$ is no longer required. Its only role was to create fluctuations of the field $\phi$ which provide the variability of the effective cosmological constant, but this variability is already present in the string theory landscape. Similarly, the huge range of variation of the field $\phi$ is no longer required. It is sufficient  to have $V'(\phi) \ll 10^{{-120}}$ in a small vicinity of some point $\phi = \phi_{0}$. In other words, once we delegate the solution of the cosmological constant problem to the string theory landscape \cite{Bousso:2000xa,Kachru:2003aw,Douglas:2003um,Susskind:2003kw}, the remaining problem of constructing a viable model of dark energy becomes much simpler.

Of course, if we assume that the cosmological constant problem is already solved, then one may wonder whether we need quintessence at all. And the answer is that we may not need it now, but we might need it later, if future cosmological data  indicate that the equation of state of dark energy  differs from $w = -1$. Also, from a purely theoretical point of view, one should not discard a possibility that we live not at the absolute minimum of a potential, but somewhere along a flat direction. This may  further enrich the spectrum of different possibilities available in the string theory landscape.

Looking at the observational trends over the last decade, it seems most likely that we will end up with an increasing observational support for the standard model of cosmology ($\Lambda$CDM). Many modified gravity models are now ruled out~\cite{Creminelli:2017sry,Sakstein:2017xjx,Ezquiaga:2017ekz,Baker:2017hug,Nojiri:2017hai,Amendola:2017orw,Crisostomi:2017lbg,Langlois:2017dyl,Gumrukcuoglu:2017ijh,Heisenberg:2017qka,Kreisch:2017uet,Dima:2017pwp,Peirone:2017ywi,Crisostomi:2017pjs,Akrami:2018yjz} by the coincident detection of gravitational waves from a neutron star merger and their electromagnetric counterpart, events GW170817~\cite{TheLIGOScientific:2017qsa} and GRB 170817A~\cite{Goldstein:2017mmi}; see also Refs.~\cite{Lombriser:2015sxa,Brax:2015dma,Lombriser:2016yzn,Pogosian:2016pwr} where the implications of such gravitational wave measurements for modified gravity were discussed before the actual observations. This discovery gives a strong  support to General Relativity.  The models of dark energy which we study here, are also likely to be ruled out in favor of the cosmological constant. %But before the large-scale structure surveys are performed, we might  prepare for what we believe is an unlikely outcome, deviation from $\Lambda$CDM, still based on  General Relativity. \comYA{Reminder to myself: add a description of future surveys, with references.}  
Nevertheless, in view of the upcoming large-scale structure (LSS) surveys, it makes sense to prepare some phenomenological models of quintessential inflation, which may deviate from $\Lambda$CDM, but  do not require a deviation from General Relativity. 
%Moreover,  the construction of realistic models may be simplified if we do not insist that all scalar fields must be strongly stabilized at the minimum of their potential. 

Thus, it would be interesting to try to construct viable dark energy models in this new context, using some novel ideas which have recently been discovered in inflationary cosmology. In particular, recent investigations have found a broad class of theories, cosmological $\alpha$-attractors, which are based on models where the kinetic term of a scalar field has a pole  \cite{Kallosh:2013hoa,Ferrara:2013rsa,Kallosh:2013yoa,Cecotti:2014ipa,Galante:2014ifa,Kallosh:2015zsa}. 
In such theories, the potential has a plateau shape, exponentially rapidly approaching a constant at large values of the inflaton field $\vp$. These models, to be described in section~\ref{basic} of this paper, are favored by the recent inflation-related cosmological observations \cite{Ade:2015lrj}.

Because of the extreme flatness of the potential in $\alpha$-attractors, these models can be suitable not only for describing inflation but also to describe dark energy,  see e.g. Refs.~\cite{Linder:2015qxa,Dimopoulos:2017zvq,Mishra:2017ehw,Bag:2017vjp,vandeBruck:2017voa,Dimopoulos:2017tud}. Moreover, it may also be possible to find $\alpha$-attractor models which can simultaneously describe inflation and dark energy  \cite{Dimopoulos:2017zvq,vandeBruck:2017voa,Dimopoulos:2017tud} in the context of the quintessential inflation \cite{Peebles:1998qn}.

In this paper, we extend the investigation of the quintessential inflation models based on $\alpha$-attractors. We study models with arbitrary $\Lambda$, % take into account the instant preheating mechanism \cite{Felder:1998vq,Felder:1999pv,Kofman:2004yc}, 
 relax some of the assumptions made in Refs.~\cite{Dimopoulos:2017zvq,vandeBruck:2017voa,Dimopoulos:2017tud}, and consider a much more general class of theories. In particular, we describe the $\alpha$-attractor version of the simplest linear dark energy model \rf{linear},  a model with exponential potential with two shoulders proposed in Ref.~\cite{Carrasco:2015rva}, and a generalized version of the model studied in Refs.~\cite{Dimopoulos:2017zvq,Dimopoulos:2017tud}. 

The asymptotic value $w_{\infty}$ of the parameter $w$ in the equation of state $p = w \rho$ for quintessential inflation depends on the limiting value of the quintessence potential. If this value is negative, the universe eventually collapses, but under certain conditions it may pass through a temporary but long stage of acceleration. Here we call $w_{\infty}$ the asymptotic value of $w$ for dark energy, to distinguish it from the time-dependent dark energy equation of state $w_{\rm DE}$ and the observable ``all-inclusive'' effective equation of state $w_{\rm eff}$.

If the potential $V$ of the quintessential inflation models asymptotically vanishes (i.e. if the cosmological constant is zero), the value of $w_{\infty}$ in the simplest models is given by 
\be
w_{\infty}=   -1+  \frac{2}{9\alpha}\, .  
\label{cute}\ee 
Interestingly, the difference between  $w_{\infty}$ and the equation of state $w = -1$ for the cosmological constant is inversely proportional to $\alpha$, whereas the tensor to scalar ratio is directly proportional to it,
\be
r =  \frac{12\alpha}{N^2} \, ,
\label{cute1}\ee
where $N$ corresponds to the remaining number of $e$-folds from the end of inflation at the moment of generation of perturbations studied by WMAP and Planck. This may help us either to rule out, or to confirm theories of that type by a combination of searches for B-modes and investigation of dark energy. %Note, however, that the equations above describe the asymptotic values of $w_{\rm DE}$, and the present value of $w_{\rm DE}$ may be quite different from  $w_{\infty}$.

Note that this result is valid only if the cosmological constant is zero, which provides us with an intriguing possibility to test this hypothesis. Meanwhile in the theories with a negative cosmological constant, the universe eventually collapses. However, in some cases one may have a prolonged state of accelerated expansion, just as in the model proposed in Ref.~\cite{Linde:1986dq}.

If the asymptotic value of the potential is positive  (i.e. if the cosmological constant is positive), and the quintessence field slowly rolls towards infinity, the universe asymptotically approaches a de Sitter regime with 
\be
w_{\infty}=-1\, . 
\ee 
This is the most general regime that is relatively easy to achieve. Of course, if these models correctly describe our world, the observations looking for deviations of quintessence from  the cosmological constant will not bring us anything exciting. But there may be a silver lining here.

Indeed, the process of reheating in the models of quintessential inflation is non-standard, and it can be very inefficient. In that case, the inflaton field after the end of inflation may enter a long stage when its energy density is  dominated by the kinetic energy with $w = +1$. This simple fact affects the  number of $e$-folds $N$ \cite{Dimopoulos:2017zvq}. Indeed, as we will show, the number of $e$-folds in the $\alpha$-attractor models of quintessential inflation with gravitational reheating can be greater than the corresponding number in the conventional (non-quintessential) versions of $\alpha$-attractors and in the Starobinsky model by $\Delta N \sim 10$. This is a significant difference, which may have important observational consequences.

In particular, the general prediction of $\alpha$ attractors for $n_{s}$ is
\be\label{nsns}
n_{s} = 1-{2\over N} \, .
\ee
One can easily check that the difference between $n_{s}$ for conventional $\alpha$-attractors with $N \sim 50$ and $\alpha$-attractor models of quintessential inflation with $N \sim 60$ is about $0.006$, which coincides with $1\sigma$ error bar in the Planck 2015 results  \cite{Ade:2015lrj}. This increase in the value of $n_s$ and $N$ is not very easy to achieve otherwise, see e.g. Refs.~\cite{Ueno:2016dim,Eshaghi:2016kne}.

This suggests that future observations may be able to differentiate between the regular versions of inflationary $\alpha$-attractors and their quintessential generalizations. More generally, we might be able to differentiate, though somewhat indirectly,  the cosmological constant and   quintessence  without relying on extreme accuracy in measuring $w$. This is a rather intriguing byproduct of the present investigation.

In this paper we will also describe the models which involve two different fields with $\alpha$-attractor potentials. The first of these two fields (or the combination of the two) will be responsible for inflation, and the second field will be responsible for quintessence. The resulting models are very flexible; they are close in spirit to the models of multi-field cascade inflation proposed in Ref.~\cite{Kallosh:2017wnt}. 

In addition to the current cosmic microwave background (CMB) experiments, such as WMAP~\cite{Bennett:2012zja}, Planck~\cite{Adam:2015rua}, ACTPol~\cite{Naess:2014wtr} and SPT-Pol~\cite{Keisler:2015hfa}, as well as the Stage III CMB experiments like AdvACT~\cite{Henderson:2015nzj} and SPT-3G~\cite{Benson:2014qhw}, and the future CMB Stage IV ground~\cite{Abitbol:2017nao} and space based experiments such as LiteBIRD~\cite{Matsumura:2016sri,Matsumura:2013aja}, aiming at more precise measurements of the CMB B-modes, arguably the next leading cosmological probes are the large-scale structure surveys, measuring baryon acoustic oscillations (BAO) and the growth of structure through redshift-space distortions (RSD), as well as weak gravitational lensing. There is a classification of the LSS surveys similar to that of the CMB experiments. This includes Stage III experiments currently taking data and continuing to do so for the next two or three years, as well as Stage IV experiments that are currently being designed and constructed to provide a large amount of high quality data in the next five to ten years. The Stage III experiments include, for example, the Canada-France Hawaii Telescope Lensing Survey (CFHTLenS)~\cite{Heymans:2012gg,Fu:2014loa}, the Kilo Degree Survey (KiDS)~\cite{Hildebrandt:2016iqg,Kohlinger:2017sxk}, the Extended Baryon Oscillation Spectroscopic Survey (eBOSS)~\cite{Dawson:2015wdb}, and the Dark Energy Survey (DES)~\cite{Abbott:2015swa,Troxel:2017xyo,Abbott:2017wau}. We however expect an exciting time to come when the Stage IV LSS surveys start to deliver data. These include several ground based experiments such as the Dark Energy Spectroscopic Instrument (DESI)~\cite{Aghamousa:2016zmz,Aghamousa:2016sne}, the Large Synoptic Survey Telescope (LSST)~\cite{Abell:2009aa,Marshall:2017wph}, and the Square Kilometre Array (SKA)~\cite{Bull:2014rha,Jarvis:2015tqa,Bacon:2015dqe,Kitching:2015fra,Yahya:2014yva,Santos:2015gra}, as well as the space based experiments Euclid~\cite{Laureijs:2011gra,Amendola:2016saw} and the Wide Field InfraRed Survey Telescope (WFIRST)~\cite{Spergel:2015sza,Hounsell:2017ejq}. A synergy of all these various probes of both early- and late-time observables will provide invaluable information about the models of inflation and dark energy.

In this paper, we perform an analysis of our $\alpha$-attractor models of dark energy in view of their implications for the current and future large-scale structure surveys. We do not intend here to perform a comprehensive comparison of our models to the current data or a detailed forecast analysis of the models for the future LSS experiments (see Ref.~\cite{Casas:2017wjh} for an example of such an exhaustive analysis for models connecting inflation and dark energy). For some models, we base our discussions solely on simple numerical computations of cosmic histories as well as dark energy and effective equations of state, without going through a detailed comparison to observations, to see whether these models can potentially provide viable cosmologies. For some others, though, we perform a statistical analysis and compare their predictions to geometrical constraints on the cosmic history at the background level using a combination of current observational data, which we believe can provide a sufficiently good understanding of our models and their viability. We leave an extensive statistical study of the models for future work where a perturbative analysis will be performed. We also discuss the implications of our findings for future cosmological surveys and in particular ask the question of whether the more precise measurements of dark energy properties will enable us to test our models against $\Lambda$CDM. Here we similarly do not perform a detailed forecast analysis of the models and are interested only in a rough estimate of the testability of the models using future data. We again leave a comprehensive forecast analysis of the models for future work.

\section{\Large\boldmath {Asymmetric cosmological $\alpha$-attractors}}\label{basic}
There are many different ways to introduce the theory of $\alpha$-attractors, see Refs.~\cite{Kallosh:2013hoa,Ferrara:2013rsa,Kallosh:2013yoa,Cecotti:2014ipa,Galante:2014ifa,Kallosh:2015zsa}. On a purely phenomenological level, the main features of  all of these models can be represented in terms of a single-field model with the Lagrangian  \cite{Galante:2014ifa,Kallosh:2015zsa}
 \be
 {1\over \sqrt{-g}} \mathcal{L} = { R\over 2}   -  {(\partial_{\mu} \phi)^2\over 2\bigl(1-{\phi^{2}\over 6\alpha}\bigr)^{2}} - V(\phi)   \,  .
\label{cosmo}\ee
Here $\phi(x)$ is the scalar field, and we use $M_\text{Pl}=1$ units.  The origin of the pole in the kinetic term can be explained in the context of hyperbolic geometry. These geometries are natural in extended supergravity, although they may also describe cosmological models unrelated to  supergravity.
The parameter  $\alpha$  can take any positive value in the minimal ${\cal N}=1$ supergravity, but recent developments based on extended supergravity, M-theory, and string theory favor 7 particular choices: $3\alpha = 1,2,3,...,7$ \cite{Ferrara:2016fwe,Kallosh:2017ced,Kallosh:2017wnt}.  

In the limit $\alpha \rightarrow \infty$ this model coincides with the standard chaotic inflation  with a canonically normalized field $\phi$ and the inflaton potential $V(\phi)$  \cite{Linde:1983gd}. 
However, for any finite values of $\alpha$, the field $\phi$ in \rf{cosmo} is not canonically normalized, and must satisfy the condition $\phi^2<6\alpha$.  

\begin{figure}[h!]
\begin{center}
\includegraphics[scale=0.55]{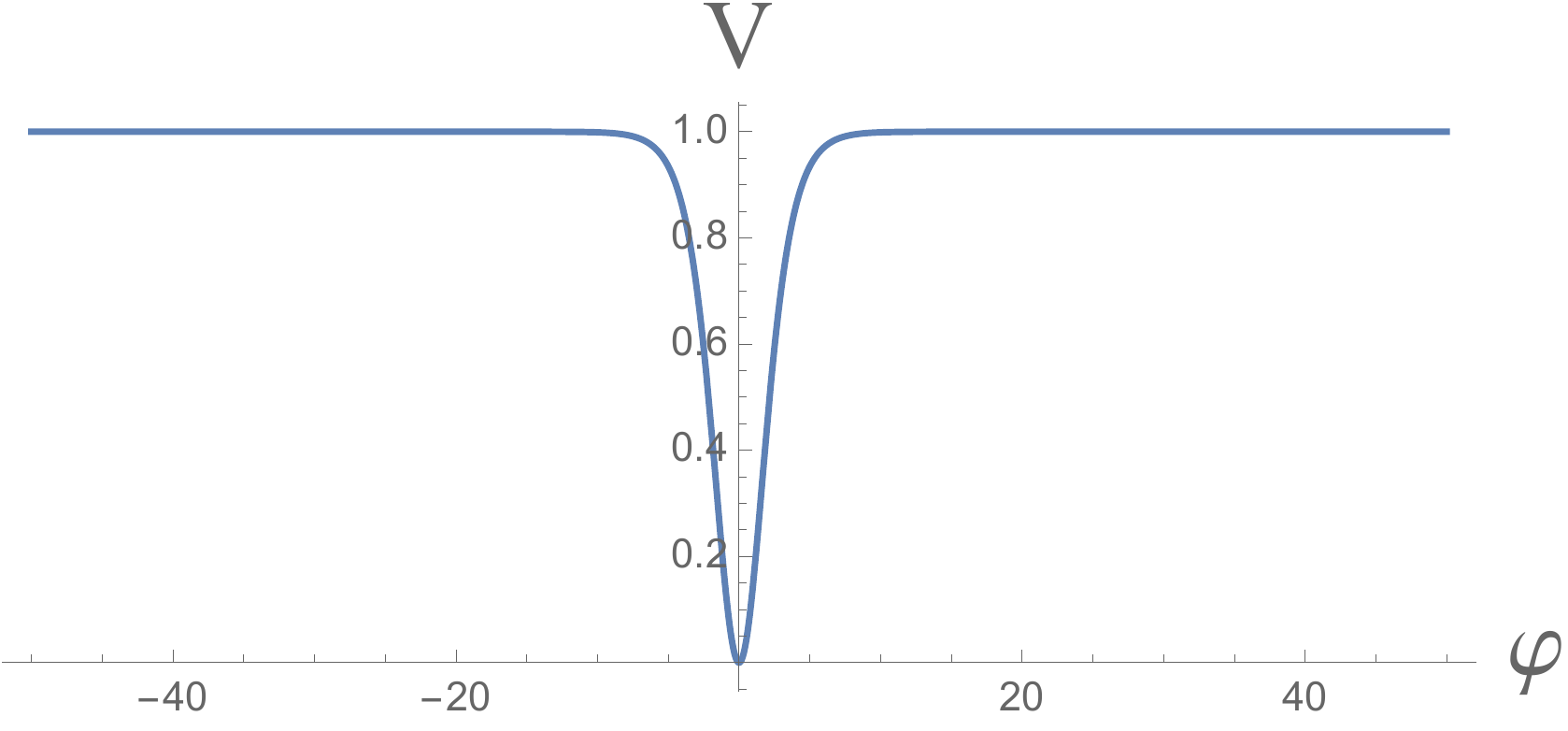}
\end{center}
\caption{\footnotesize The potential $V(\vp) =3\alpha  m^{2 }  \, \tanh^{2}{\varphi\over\sqrt {6 \alpha}}$ for $\alpha = 1$, shown in units of $3m^{2}$, with $\vp$ in Planck units.  For  $1/3< \alpha <10$   one has $n_{s}\sim 0.965$ and the tensor to scalar ratio $r$ is in the range from $3\times 10^{-2}$ to $10^{-3}$, providing a good match to the Planck data.}
\label{F1}
\end{figure}

Instead of the variable $\phi$, one can use a canonically normalized field $\vp$ by solving the equation ${\partial \phi\over 1-{\phi^{2}\over 6\alpha}} = \partial\vp$, which yields
\be\label{tanh} 
\phi = \sqrt {6 \alpha}\, \tanh{\varphi\over\sqrt {6 \alpha}} \, .
\ee
The full theory, in terms of the canonical variables, becomes
 \be
 {1\over \sqrt{-g}} \mathcal{L} = { R\over 2}   -  {(\partial_{\mu}\varphi)^{2} \over 2}  - V\big(\sqrt {6 \alpha}\, \tanh{\varphi\over\sqrt {6 \alpha}}\big)   \,  .
\label{cosmoqq}\ee
Note that in the limit $\phi \to 0$ the variables $\phi$ and $\varphi$ coincide; the main difference appears in the limit $\phi^2 \to 6 \alpha$: In terms of the new variables, a tiny vicinity of the boundary of the moduli space at $\phi^2=6\alpha$ stretches and extends to infinitely large $|\varphi|$. We will assume that the potential $V(\phi)$ and its derivatives are non-singular for $\phi^2\leq6\alpha$. 
In that case, generic potentials $V(\phi) = V(\sqrt {6 \alpha}\, \tanh{\varphi\over\sqrt {6 \alpha}})$ at large $|\vp|$ approach two infinitely long plateaus with the heights corresponding to the values of $V(\phi)$ at the two boundaries,
\be
V_{\pm} \equiv V(\phi)|_{\phi = \pm \sqrt {6 \alpha}} \, .
\ee  
The simplest example of such a theory is given by the model with $V(\phi) = m^{2}\phi^{2}/2$. In terms of the canonically normalized field $\vp$, the potential is given by
\be\label{shape}
 V(\varphi)=  3\alpha  m^{2 }  \, \tanh^{2}{\varphi\over\sqrt {6 \alpha}} \,. 
 \ee 
This is the simplest representative of the so-called T-models, with the T-shaped potential shown in  Fig. \ref{F1}.
For any values of $\alpha \lesssim 10$, the amplitude of the inflationary perturbations, the prediction for the spectral index $n_{s}$, and the tensor to scalar ratio $r$ match observational data under a single condition  \cite{Kallosh:2015lwa}
\be
{V_{\pm}\over  \alpha} \sim 3 m^{2} \sim 10^{{-10}} \, .
\ee
To understand what is going on in this class of theories for general potentials $V(\phi)$, let us consider, for definiteness, positive values of $\phi$ and study a small vicinity of the point $\phi = \sqrt {6 \alpha}$, which  becomes stretched to infinitely large values of the canonical field $\vp$  upon the change of variables $\phi \to \vp$. If the potential $V(\phi)$ is non-singular at the boundary  $\phi = \sqrt {6 \alpha}$, we can expand it in series with respect to the distance from the boundary,
\be
V(\phi) = V_{+} + (\phi-\sqrt {6 \alpha})\, V'_{+} +\mathcal{O}\left((\phi-\sqrt {6 \alpha})^{2}\right) \,,
\ee
where we denote $V'_{+} \equiv \partial_{\phi}V |_{\phi = +\sqrt {6 \alpha}}$. 

In the vicinity of the boundary $\phi=\sqrt {6 \alpha}$, the relation \rf{tanh} between the original field variable $\phi$ and the canonically normalized inflaton field $\vp$ is given by
\be\label{tanh2} 
\phi = \sqrt {6 \alpha}\, \left(1 - 2 e^{-\sqrt{2\over 3\alpha} \varphi }\right)\, ,
\ee
up to the higher order terms $\mathcal{O}\bigl(e^{-2\sqrt{2\over 3\alpha} \varphi }\bigr) $. At $\vp \gg \sqrt {6\alpha}$, these  terms are exponentially small as compared to the terms $\sim  e^{-\sqrt{2\over 3\alpha} \varphi }$, and the potential acquires the following asymptotic form
\be\label{plateau}
V(\vp) = V_{+} - 2  \sqrt{6\alpha}\, V'_{+}\ e^{-\sqrt{2\over 3\alpha} \varphi } \, .
\ee
The constant $2  \sqrt{6\alpha}\, V'_{+}$ in this expression can be absorbed into a redefinition of the field $\vp$. That is why if inflation occurs at large $ \vp \gg \sqrt{\alpha}$, all inflationary predictions are universal.

 In particular, the parameters $n_{s}$ and $r$ describing the spectrum of inflationary perturbations are given by \rf{cute1} and \rf{nsns},
 \be\label{aa}
r= {12\alpha\over N^{2}}\,, \qquad n_{s} =  1 - {2\over N}\,.
\ee 
These results depend only on $\alpha$ and the number of $e$-folds $N$ remaining to the end of inflation since the moment when quantum fluctuations were generated. Meanwhile, the amplitude of scalar perturbations for $\alpha$-attractors generated at the upper plateau of the potential \rf{plateau} is given by
\be
\mathcal{P}_{\mathcal{R}}(k)    = \frac{ N^2}{18\pi^2}\ {V_{+}\over \alpha}\,   \label{eq:COBEexp2}\,.
\ee
Thus the COBE/Planck normalization constrains the ratio ${V_{+}/\alpha}$ \cite{Kallosh:2015lwa}. Taking the value $(2.208\pm0.075)×10^{-9}$~\cite{Ade:2015xua,Aghanim:2016sns} for $\mathcal{P}_{\mathcal{R}}$ and $N\sim 60$ $e$-folds for inflation, we find the constraint on the height of the inflationary plateau,
\begin{eqnarray}
{V_{+}\over \alpha} \sim 10^{-10}  \label{eq:COBEconstplus}\,.
\end{eqnarray}

These results were explained in Refs.~\cite{Kallosh:2013hoa,Kallosh:2013yoa} and formulated in a particularly general way in Ref.~\cite{Galante:2014ifa}: The kinetic term in this class of models has a pole at the boundary of the moduli space. If inflation occurs in a vicinity of such a pole, and the potential near the pole has a finite first derivative, all other details of the potential $V(\phi)$ and of the kinetic term far away from the pole are not important for making cosmological predictions. That is why these models are called cosmological attractors.

The simplest model $V(\phi) = m^{2}\phi^{2}/2$ considered above is symmetric with respect to the change $\phi \to - \phi$. However, this is not a universal property. Consider, for example, its generalization~\cite{Carrasco:2015rva} with the potential 
\be \label{asymm}
V={m^{2}\over 2(1+c)^{2}}(\phi+c\sqrt {6 \alpha} )^{2} \, .
\ee 
In terms of the canonically normalized field $\vp$, the potential becomes  
\be 
V = {3\alpha  m^{2}\over (1+c)^{2}}\bigl(\tanh{\varphi\over\sqrt {6 \alpha}}+c\bigr)^{2} \, .
\ee
The coefficient $(1+c)^{-2}$ is introduced to preserve the height of the inflationary plateau at $\vp \to \infty$. 
\begin{figure}[h!]
\begin{center}
\includegraphics[scale=0.5]{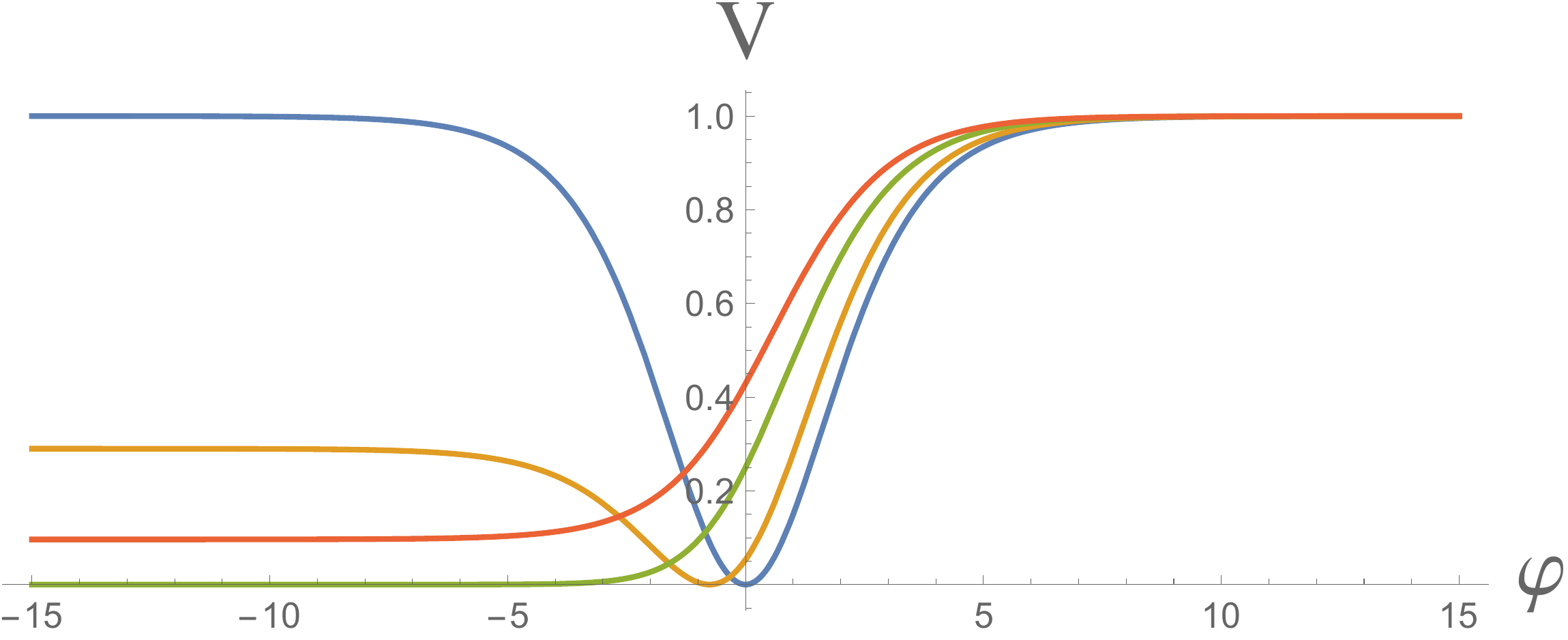}
\end{center}
\caption{\footnotesize The potential \rf{asymm}  shown in units of $\alpha m^{2}$ for $\alpha = 1$, and $c = 0$ (blue), $0.3$ (orange), $1$ (red), and $1.9$ (green). }
\label{F2}
\end{figure}

For $|c| < 1$, this potential has a minimum and two asymptotically flat shoulders of different height, as shown by the orange curve in Fig. \ref{F2}. For $c = 1$, the minimum of the potential disappears and the left shoulder describes a potential which exponentially decreases to zero at large, negative values of $\vp$.  Finally, for $c< - 1$, the potential at large, negative $\vp$ approaches a cosmological constant $V_{-} = 3\alpha m^{2}(c-1)^2/(c+1)^2$.  
One can further modify the potential by adding to it a constant of any sign, which is absolutely legitimate from the point of view of the string theory landscape.

Historically, the first versions of $\alpha$-attractor models have been developed in Refs.~\cite{Kallosh:2013hoa,Ferrara:2013rsa,Kallosh:2013yoa,Cecotti:2014ipa,Galante:2014ifa,Kallosh:2015zsa} in the supergravity context, where the potentials could be represented as $f^{2}(\phi)$, where $f(\phi)$ is a real holomorphic function. That is why we started the discussion of  $\alpha$-attractors with presenting models with a quadratic potential $V(\phi)$. However, recently a more general approach to $\alpha$-attractors in supergravity has been developed \cite{Kallosh:2017wnt,McDonough:2016der}, which allows us to describe models with arbitrary potentials $V(\phi)$, including the simplest linear dark energy potential
$V(\phi) =  \gamma  \phi + \Lambda$ proposed in Ref.~\cite{Linde:1986dq}. 

In this paper, we study $V(\vp)$ at very large, negative $\vp$. Therefore we will often identify $\Lambda$ not with  $V(0)$, but with $V_{-}$, the height of the potential in the limit of large, negative $\vp$.  This can be achieved by representing the linear potential as $V(\phi) =  \gamma  \phi + \gamma \sqrt {6 \alpha}+ \Lambda$. In terms of the canonically normalized field $\vp$, this potential is given by 
\be\label{lin}
V(\vp) =  \gamma \sqrt{6\alpha}  (\tanh {\varphi\over\sqrt {6 \alpha}}+1) + \Lambda \, , 
\ee
where $\Lambda = V_{-}$ is now the asymptotic value of the potential at $\vp \to -\infty$.

\begin{figure}[h!]
 \begin{center}
 \includegraphics[scale=0.5]{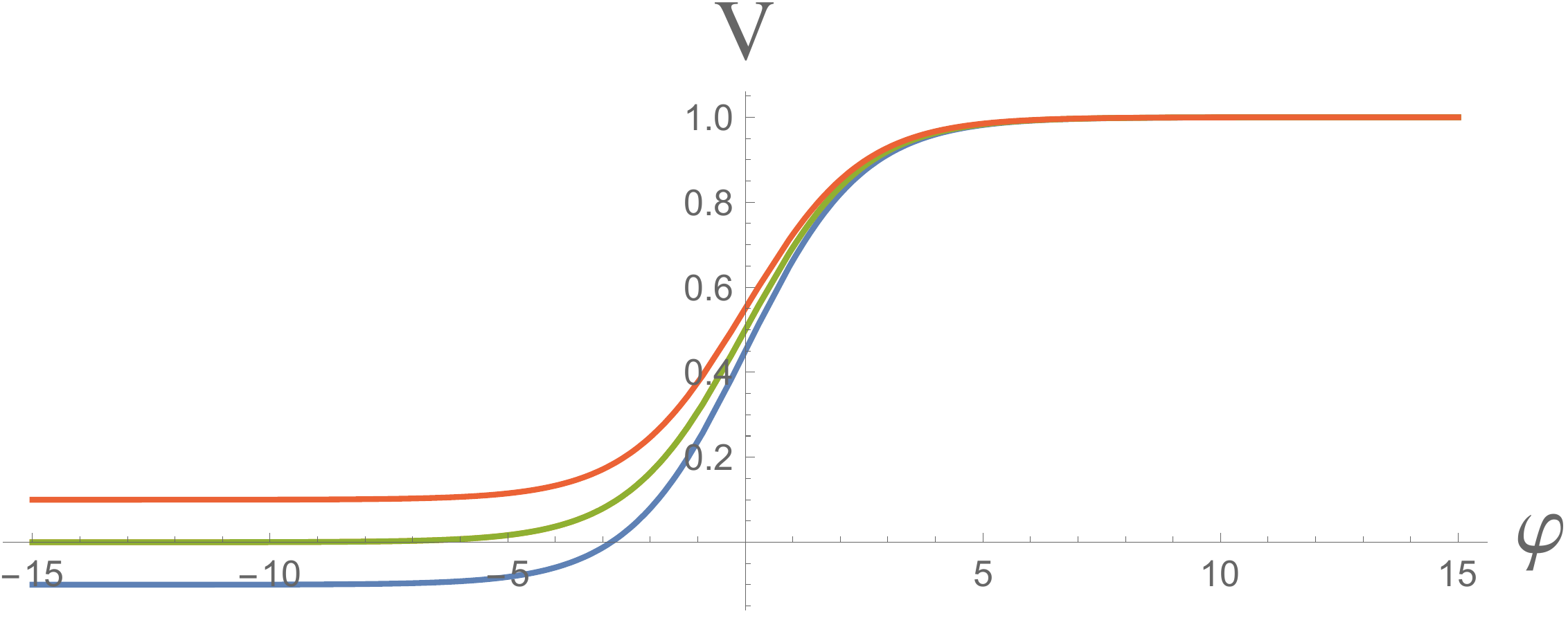}
 \end{center}
 \caption{\footnotesize The potential \rf{lin} has two plateaus, with $V = V_{\pm}$. We illustrate its values  for $V_{+}= 1$ and $V_{-}=\Lambda  = -0.1$ (blue),  $0$ (green),  and $+0.1$ (red).}
 \label{F3}
 \end{figure}

We illustrate the shape of this potential for various values of its parameters in Fig. \ref{F3}. 
At $\varphi \gg \sqrt {6\alpha}$ the potential is given by 
\be\label{pos}
V=   V_{+} - 2\gamma \sqrt{6\alpha}\  e^{-\sqrt{2\over 3\alpha} \varphi }\, ,
\ee
whereas at   $\varphi \ll - \sqrt {6\alpha}$ one has 
\be\label{mlll}
V=   V_{-} + 2\gamma \sqrt{6\alpha}\  e^{\sqrt{2\over 3\alpha} \varphi } \, .
\ee

In general, the asymptotic behavior of asymmetric potentials $V(\vp)$ at large, negative values of the field, $\vp \ll - \sqrt{6\alpha}$, is given by an expression similar to \rf{plateau},
\be\label{plateaumin}
V(\vp) = V_{-} + 2  \sqrt{6\alpha}\, V'_{-}\ e^{\sqrt{2\over 3\alpha} \varphi } \, ,
\ee
where $V'_{-} \equiv \partial_{\phi}V |_{\phi = -\sqrt {6 \alpha}}$.  Thus, as long as $V'_{-}$ is non-singular and does not vanish,\footnote{If one fine-tunes the potential $V(\phi)$ to have a minimum, or maximum, at one of the boundaries $\phi = \pm \sqrt{6\alpha}$, the first derivative $V'_{-}$ in \rf{plateaumin}, or $V'_{+}$ in \rf{plateau}, vanishes. This affects the asymptotic behavior of the potential. For example, in the theory with the quadratic potential \rf{asymm} with $c = 1$, the asymptotic behavior at $\vp \to -\infty$ is governed by the higher exponent  $e^{2\sqrt{2\over 3\alpha} \varphi }$, which is equivalent to making $\alpha$ four times smaller.} %This may make  models with  $V'_{-}=0$ less suitable for the description of dark energy.}, 
all such potentials have the same universal asymptotic behavior at large, negative $\vp$: Up to a  shift  $\vp \to  \vp - \sqrt{3\alpha\over 2} \log (2  \sqrt{6\alpha}\, V'_{-})$ and a redefinition  $ \sqrt{2\over 3\alpha}\to \lambda$, they can be represented in a more familiar way,
\be\label{plateaumin2}
V(\vp) = \Lambda + e^{\lambda \varphi } \, .
\ee
This general asymptotic expression will be very helpful in evaluation of $\alpha$-attractors as dark energy candidates. 

To explain the basic idea, let us first consider the simplest case of $\Lambda = 0$. Then we will have an exponential  potential
\footnote{The related effective models of accelerated expansion in string theory were proposed in Ref.~\cite{Dodelson:2013iba}, and they lead to $w_\text{DE} <-1/3$.}
\be 
V(\vp) =  e^{\lambda \varphi } \, ,
\ee
 where 
 \be
  \lambda =\sqrt {2\over 3 \alpha} \, .
  \ee
 This potential  vanishes in the limit $\vp \to -\infty$.  For $\lambda \ll 1$, the potential is flat, the energy density of  normal matter decreases faster than $V$, and the system eventually enters the asymptotic regime of power-law inflation with (see for example the review  \cite{Martin:2013tda})
\be\label{walpha1}
w_{\infty} =  -1 + {\lambda^{2}\over 3} = -1 + {2\over 9\alpha} \, .
\ee 
It is interesting to compare this result with the inflationary  predictions of $\alpha$-attractors \rf{aa}: $n_{s} =  1 - {2\over N}$\,, $r= {12\alpha\over N^{2}}$. Thus, in this scenario, inflationary predictions, as well as the value of $w_{\infty}$, are determined by the parameter $\alpha$.
In particular, for $\Lambda = 0$, and $\alpha = 7/3$ (i.e. $\lambda \sim 0.53$), which is one of the values advocated in Refs.~\cite{Ferrara:2016fwe,Kallosh:2017ced,Kallosh:2017wnt}, dark energy has the asymptotic equation of state
\be\label{w73}
w_{\infty} = -0.905
 \, .
\ee
Note, however, that in the derivation of \rf{walpha1} we assumed that $\Lambda = 0$. This assumption, which simplifies the investigation, is very hard to justify. For any positive $\Lambda$ one has
\be
w_{\infty} = -1\,, 
\ee
but for large $\alpha$ the transition from $w =-1 + {2\over 9\alpha}$ to $w = -1$ may take a long time. On the other hand, in the models with $\Lambda < 0$, the universe eventually collapses, but if    $\lambda \ll 1$ and $|\Lambda| \ll 10^{{-120}}$, there is a very long interval, longer than the present age of the universe, during which life as we know it can exist, and $w$ is very close to $-1$  \cite{Kallosh:2003mt}. Also, our universe may be very far from the asymptotic regime discussed above. Therefore, one should keep the estimate \rf{walpha1} in mind, but perform a more detailed analysis of different dark energy models, as we will do in this paper.

\section{\Large\boldmath{$\alpha$-attractors and supergravity}}
\subsection{General formulation, geometry, and special values of \boldmath{$\alpha$}}
One of the nice features of all  cosmological $\alpha$-attractor models which we will study here is that they can be easily embedded into the string theory motivated supergravity where the scalar fields are complex. The most advanced version of these models~\cite{Kallosh:2017wnt}  is based on anti-D3-brane induced geometric models of the following nature --- here we review these models in the simple case where a bosonic model has a single inflaton-quintessence field.

There is one complex scalar $Z$, a coordinate of the Poincar\'e disk with the following geometry
\be
\dd s^2= 3\alpha {\dd Z \dd\bar Z\over (1-Z\bar Z)^2} \, .
\label{Poin}\ee

Advanced formulations of $\alpha$-attractors in supergravity also contain a nilpotent superfield $S$ such that $S(x, \theta)^2=0$, whose \K~geometry represents the interaction between the anti-D3-brane and the background fields, including the inflaton-quintessence field $Z$.  The scalar component of it, $S(x)$, vanishes on the inflationary trajectory, since in this Volkov-Akulov multiplet the scalar is not independent but  is a bilinear of fermions. It is convenient to use the geometric \K~function formalism \cite{Kallosh:2017wnt}, where
\be
{\cal G} \equiv K + \log W +\log \overline W\, , 
\qquad 
    \mathbf{ V} = e^ {\cal   G}  ({\cal G}^{\alpha \bar \beta }  {\cal   G} _\alpha  {\cal   G} _{\bar \beta}  - 3) \, ,
\ee

\be
{\cal G}=\ln W_0^2 -{3\alpha \over 2} \log   {(1-Z\bar Z)^2\over (1-Z^2) (1-\bar Z^2)}  +S + \bar S + {W_0^2 \over |F_S|^2+f( Z,\bar Z)} S \bar S\, ,
\ee
and $f( Z,\bar Z)$ is an arbitrary, real function of $Z$ and $\bar Z$. This employs the \K~frame that has a manifest inflaton shift symmetry  \cite{Carrasco:2015uma}.
The potential has a stable minimum at $Z=\bar Z$. Its value along the inflaton direction $Z=\bar Z= \tanh {\varphi\over \sqrt{6\alpha}}$  is given by
\be\label{newpot}
{\bf V}  |_{Z=\bar Z}= f(Z,\bar Z)|_{Z=\bar Z} + \Lambda =  f(\tanh {\varphi\over \sqrt{6\alpha}}) + \Lambda \, .
\ee
Here, the cosmological constant $\Lambda$ can take arbitrary values determined by the choice of $F_S$ and $W_{0}$:   
\be
\Lambda =F_S^2- 3W_0^2 \, .
\ee
The choice of the \K~potential for $Z$ was made in Ref.~\cite{Kallosh:2017wnt} such that 
\be
K (Z, \bar Z)  |_{Z=\bar Z}= -{3\alpha \over 2} \log   {(1-Z\bar Z)^2\over (1-Z^2) (1-\bar Z^2)}  |_{Z=\bar Z}=
0\, ,  \qquad K_Z  (Z, \bar Z)  |_{Z=\bar Z}= 0 \, .
\label{flatK}\ee
This \K~frame leads to a simple relation between the inflaton potential  \rf{newpot} and the $S$-field geometry $g_{S\bar S}= {W_0^2 \over |F_S|^2+f( Z,\bar Z)}$. It also provides stabilization of the sinflaton field $Z-\bar Z$ at $Z-\bar Z = 0$. 

In the disk geometry \rf{Poin} 
$
 3\alpha=\mathcal{R}^2
$
 is a geometric parameter defining the radius square of the Poincar\'e disk of the hyperbolic geometry of the $\alpha$-attractor models, since by change of variables $Z' = Z\sqrt{3\alpha}$ one can represent the metric in the form 
\be
\dd s^2=  {\dd Z' \dd\bar Z'\over  \big(1-{Z'\bar Z'\over 3\alpha}\big)^2} \ ,  \qquad  |Z'|^2<3\alpha \, .
\ee
The parameter $\alpha$ also  defines  a curvature of the corresponding \K~manifold, $\mathbf{ R}_{K}= -{2\over 3 \alpha}$. Finally, one can return to the variables used in the previous section by representing the real part of $Z'$ as ${\phi\over \sqrt 2} = \sqrt{3\alpha}\, \tanh{\varphi\over\sqrt {6 \alpha}}$.
 
The asymptotic freedom of the interactions of the field $\vp$ with all other fields protects the asymptotic flatness of the potential for any $\alpha$. Thus, in general quantum field theory models, as well as in ${\cal N}=1$ supergravity, there are no constraints on $\alpha$, it can take any value $\alpha > 0$.\footnote{One should distinguish the general theoretical constraints on $\alpha$ and the model-dependent cosmological constraints. In Ref.~\cite{Dimopoulos:2017zvq}, the authors assumed $0.03< \alpha < 1/3$. In a subsequent paper \cite{Dimopoulos:2017tud}, they noted that these conditions did not lead to a satisfactory dark energy model in their scenario, and instead picked the range $1.5< \alpha < 4.2$. However, they admitted that the constraint $\alpha < 4.2$ is not firmly motivated because of the asymptotic freedom of the field $\vp$ in $\alpha$-attractors \cite{Kallosh:2016gqp}. Meanwhile, we find that the condition $\alpha> 1.5$  is excessive, and it completely disappears in the models with a positive cosmological constant, see  section~\ref{sec:compdataexp}. In particular, in section~\ref{linsec} we will present  a model with a positive cosmological constant where one can have quintessential inflation  for $\alpha \lesssim 10^{{-2}}$.}
   
From the point of view of maximal supergravity, string theory, and M-theory, the most interesting values of $\alpha$   are  \cite{Ferrara:2016fwe,Kallosh:2017ced,Kallosh:2017wnt}
 \be
3\alpha= 1,2, 3, 4, 5, 6, 7 \,.
 \ee
An interpretation of this family of models is rather interesting. These models describe 7 unit size Poincar\'e disks with $3\alpha = 1$ for seven different  fields $Z_{i}$.  The basic choice of $\alpha = 1/3$ corresponds to a single unit size disk model with $Z_{1}\bar Z_{1} < 1$. If all other fields are stabilized and  cannot move, one has a single attractor with $\alpha = 1/3$, where the corresponding field $\phi_{1}$ can change from $-\sqrt{2}$ to $+\sqrt{2}$. %, without  violating the somewhat speculative weak gravity conjecture
If all seven of them interact and are forced dynamically to move together \cite{Kallosh:2017ced,Kallosh:2017wnt}, then each of them also moves from $-\sqrt{2}$ to $+\sqrt{2}$, but the combination of these fields changes from $-\sqrt{14}$ to $+\sqrt{14}$, along the diagonal of a 7-dimensional cube.
 
The choice of $\alpha = 1$ describes $\alpha$-attractor formulations of the Starobinsky model and Higgs inflation. The fibre inflation model, which is based on the large volume compactification in string theory, corresponds to $\alpha = 2$ \cite{Cicoli:2008gp,Kallosh:2017wku}. The choice of $\alpha = 7/3$, which we will sometimes use in various examples, corresponds to the maximally symmetric realization of the 7-disk M-theory model  \cite{Ferrara:2016fwe,Kallosh:2017ced,Kallosh:2017wnt}.

\subsection{Suppressing the fifth force}\label{sec:fifth}

There is a well known issue with quintessence regarding the fifth force problem. This problem appears if the masses of particles in the standard model depend on the quintessence field $\phi$. 

Consider first an unrealistic example and assume that the electron mass $m_\text{e}$ receives a contribution $\Delta m_\text{e}  = g\, \phi$. Then (in addition to electromagnetic interactions) electrons would attract each other through the gravitational force $\sim {(m_\text{e}+g \phi)^{2} \over   r^{2}}$, as well as through an additional fifth force $F_{5}\sim {g^{2}\over r^{2}}$ due to the interactions via the nearly massless quintessence  field $\phi$. This force will have the same dependence on  $r$ as the gravitational attraction, but it will not be proportional to $m_\text{e}^{2}$, which would violate the equivalence principle.

An obvious way to avoid this problem is to suppress the interaction of the standard model fields with quintessence. For example, as was already observed  in Ref.~\cite{Dimopoulos:2017tud}, the asymptotic freedom of the field $\vp$ in $\alpha$-attractors \cite{Kallosh:2016gqp} allows to exponentially suppress this coupling even if it were present. However, the suppression of the fifth force should be extremely strong, which may require very large values of $\vp$.  In the  $\alpha$-attractor models to be discussed in this paper, this may not be a problem since we do not introduce any direct coupling between $\phi$ and electrons or quarks, which would lead to the force $F_{5}\sim {g^{2}\over r^{2}}$ discussed above.  

However, one may wonder whether this coupling may appear in supergravity even if the field $\vp$ belongs to the hidden sector, without a direct coupling to the standard model fields. Fortunately, there is a specific feature of our underlying supergravity models which helps to avoid the fifth force issues. The coupling of the inflationary sector to matter in these models has been studied in Ref.~\cite{Kallosh:2016ndd}. The inflaton-quintessence field is $Z$, and there is also a nilpotent superfield $S$, as explained above. It has been found how to construct the interaction between matter and the inflationary sector so that the presence of the matter fields does not affect a successful inflationary evolution and that  there are no tachyons in the matter sector during and after inflation.

One of the most important features of this class of models is the requirement of the flatness of the \K~potential for the inflaton-quintessence field  $Z$, shown in Eq.~\rf{flatK}. In particular, since the field $Z-\bar Z$ orthogonal to the inflaton direction is heavy and is stabilized at the inflaton trajectory $Z = \bar Z$, one finds that
\be
e^{K(Z=\bar Z)} =1 \ ,
\ee
and there is no dependence of the mass of the matter fields on the inflaton field via the \K~potential since 
\be
K_Z(Z=\bar Z)=0 \ .
\ee
These features of the \K~potential have been discussed in Ref.~\cite{Brax:2009kd} as the reason for the fifth force problem to be alleviated in supergravity. Our models, which were constructed with the purpose of stabilization of the sinflaton field $Z-\bar Z$ during the cosmological evolution, just satisfy the properties required from the \K~potentials in Ref.~\cite{Brax:2009kd}.

Moreover, according to Ref.~\cite{Kallosh:2016ndd} one can construct satisfactory cosmological models where the mass of the matter field $U$ does not depend on the inflaton-quintessence field $Z$. Examples of such models in Ref.~\cite{Kallosh:2016ndd} include the following \K~potential and superpotential:
\be
K (Z, \bar Z) = -{3\alpha \over 2} \log   {(1-Z\bar Z)^2\over (1-Z^2) (1-\bar Z^2)} +S\bar S + U\bar U \label{K} \ ,
\ee
\be
W= g(Z) + S f(Z) + {m\over 2} U^2 \ .
\ee
For our purposes, we need to assume that 
 $g(Z) $ has a negligible dependence on $Z$ or is $Z$-independent, and the same for the parameter $m$ in the superpotential.
 The mass eigenvalues of the scalar field $U$ are
\be
\mu^2= V+ |g|^2 \pm |g| m + m^2 \ .
\ee
The value of the potential $V$ during the quintessence stage is negligible, $V \sim 10^{{-120}}$. The rest of the mass formula is $Z$-independent by the choice of the parameters in the superpotential. The situation with fermions is similar, their masses are $Z$-independent. This means that with a proper embedding of the standard model in our theory, matter fields decouple from quintessence. Such models do not suffer from the fifth force problem.

\section{\Large{Single-field quintessential inflation models}}\label{sec:1field-quint-inf}

\subsection{Inflationary dynamics, late-time evolution, and cosmic acceleration}\label{sec:evolution}

In this section, we focus on some models where a single scalar field $\phi$ is responsible for both inflation and dark energy.

The action for these single-field, $\alpha$-attractor, quintessential inflation models has the general structure
\begin{equation}\label{eq:lagrangian1}
S = \frac{1}{2}\int \dd^{4}x\sqrt{-g}R - \int \dd^{4}x\sqrt{-g}\left(\frac{\partial_{\mu} \phi\partial^{\mu} \phi}{2\bigl(1-\frac{\phi^{2}}{6\alpha}\bigr)^2} + V(\phi)\right) + S_{\text{matter}}[g_{\mu\nu},\Psi]    \,
\end{equation}
where the scalar field $\phi$ has a potential $V(\phi)$. Here $S_{\text{matter}}$ is the matter action where matter fields are denoted collectively by $\Psi$. Note that we have absorbed any cosmological constant term $\Lambda$ into the potential.

The same action can be written as
\begin{equation}\label{eq:lagrangian2}
 S = \frac{1}{2}\int \dd^{4}x\sqrt{-g}R - \int \dd^{4}x\sqrt{-g}\left(\frac{1}{2}\partial_{\mu} \varphi\partial^{\mu} \varphi + V(\varphi)\right) + S_{\text{matter}}[g_{\mu\nu},\Psi]    \,,
\end{equation}
where the field $\varphi$ has a canonical kinetic term, and is related to the non-canonical field $\phi$ through (\ref{tanh}).

Before we discuss specific models, defined by assuming specific forms for the potential $V(\phi)$, we briefly review the general dynamical equations and some important quantities for the studies of cosmic histories, during inflation and after that.

During inflation, matter and radiation are both negligible, and we can therefore determine the dynamics of the system by varying the action (\ref{eq:lagrangian2}) with respect to the metric and the scalar field $\varphi$. Let us assume that the universe is described by a Friedmann-Lema\^{i}tre-Robertson-Walker (FLRW) metric. Specializing to a spatially flat universe and working in cosmic time $t$, we have
\begin{align}
g_\mn \dd x^\mu \dd x^\nu &= -\dd t^2 + a^2(t)\delta_{ij}\dd x^i \dd x^j \,. \label{eq:FLRW}
\end{align}
Here, $a(t)$ is the scale factor, which is a function of time only. The Friedmann equation and the equation of motion for $\varphi$ take the forms
\begin{align}
&3H^2 =  \frac{1}{2}\dot\varphi^2+V(\varphi) \,,\label{eq:FriedmannInf1}\\
&\ddot\varphi+3H\dot\varphi+\frac{\dd}{\dd\varphi}V(\varphi)= 0 \,,\label{eq:phiEoMInf1}
\end{align}
where $H\equiv\frac{\dot a}{a}$ is the Hubble parameter, and a dot denotes derivatives with respect to cosmic time.

It is convenient and instructive to work with the number of $e$-folds $N\equiv \ln a$ as time coordinate. Denoting a derivative with respect to $N$ by a prime, we have
\begin{equation}
\frac{\dd\varphi}{\dd t}=\frac{\dd\varphi}{\dd N}\frac{\dd N}{\dd t}=\varphi^\prime H\,,
\end{equation}
and Eqs. (\ref{eq:FriedmannInf1}) and (\ref{eq:phiEoMInf1}) now become
\begin{align}
&3H^2 =  \frac{1}{2}{\varphi^\prime}^2H^2+V(\varphi) \,,\label{eq:FriedmannInf2}\\
&\varphi^{\prime\prime}H^2+\varphi^{\prime}H^\prime H+3H^2\varphi^{\prime}+\frac{\dd}{\dd\varphi}V(\varphi)= 0 \,.\label{eq:phiEoMInf2}
\end{align}
We can further simplify the equation of motion (\ref{eq:phiEoMInf2}) for $\varphi$ using the so-called slow-roll parameter $\epsilon$ with the {\it exact} expressions
\begin{equation}\label{eq:epsilon1}
\epsilon\equiv-\frac{\dot H}{H^2} = -\frac{H^\prime}{H}\,,
\end{equation}
in terms of both $t$ and $N$, and obtain the final system of inflationary equations,
\begin{align}
&H^2 = \frac{V(\varphi)}{3 -\frac{1}{2}{\varphi^\prime}^2} \,,\label{eq:FriedmannInf3}\\
&\varphi^{\prime\prime}+(3-\epsilon)\varphi^{\prime}+\frac{1}{H^2}\frac{\dd}{\dd\varphi}V(\varphi)= 0 \,,\label{eq:phiEoMInf3}\\
&\epsilon=\frac{1}{2}{\varphi^\prime}^2 \,,\label{eq:epsilon2}
\end{align}
where Eq. (\ref{eq:epsilon2}) for $\epsilon$ has been derived by taking the derivative of the Friedmann equation and using Eqs. (\ref{eq:epsilon1}) and (\ref{eq:phiEoMInf3}). Note that here we have not made any slow-roll approximation for $\epsilon$, and all the expressions are exact. The second slow-roll parameter $\eta$ also has an exact form,\footnote{Note that here we have adopted the definition of $\eta$ from e.g. Ref~\cite{Baumann:Lectures}. There exists another definition for this second slow-roll parameter, namely~\cite{Baumann:2009ds}
\begin{equation}\label{eq:eta1}
\tilde\eta\equiv-\frac{\ddot \varphi}{H\dot\varphi}=-\frac{\dd \ln |H_{,\varphi}|}{\dd N}=2\frac{H_{,\varphi\varphi}}{H}=\frac{\dd \ln |\dot\varphi|}{\dd N}\,,
\end{equation}
where $H_{,\varphi}\equiv\frac{\dd}{\dd\varphi}H$ and $H_{,\varphi\varphi}\equiv\frac{\dd}{\dd\varphi}H_{,\varphi}$. $\tilde\eta$ is related to our $\eta$ by
\begin{equation}
\tilde\eta=\epsilon-\frac{1}{2}\frac{\epsilon^\prime}{\epsilon}=\epsilon-\frac{1}{2}\eta \,.
\end{equation}
The spectral index $n_s$ now has the following expression in terms of $\epsilon$ and $\tilde\eta$:
\begin{align}
n_s &\approx 1+2\tilde\eta-4\epsilon \,,
\end{align}
and since $\epsilon\approx\epsilon_\text{v}$ and $\tilde\eta\approx\tilde\eta_\text{v}-\epsilon_\text{v}$, where $\epsilon_\text{v}$ and $\tilde\eta_\text{v}$ are the slow-roll approximations to $\epsilon$ and $\tilde\eta$, respectively, we have
\begin{align}
n_s &\approx 1+2\tilde\eta_\text{v}-6\epsilon_\text{v} \,.
\end{align}
}
\begin{equation}\label{eq:eta2}
\eta\equiv\frac{\dot\epsilon}{H\epsilon}= \frac{\epsilon^\prime}{\epsilon} \,,
\end{equation}
and can therefore be computed through $\epsilon$ and its first derivative. One can solve Eqs. (\ref{eq:FriedmannInf3})-(\ref{eq:epsilon2}) numerically to obtain the evolution of $\varphi$, $H$, $\epsilon$, and $\eta$ during inflation, as we will do for our quintessential inflation models in this paper. In addition, given $\epsilon$ and $\eta$, we can compute two other important inflationary quantities, namely the {\it spectral index} for scalar perturbations $n_s$ and the {\it tensor-to-scalar ratio} $r$ --- assuming the approximate relations between these quantities we have
\begin{align}
n_s &\approx 1-2\epsilon-\eta \,,\label{eq:nsslowrol}\\
r &\approx 16\epsilon \,.\label{eq:rslowrol}
\end{align}

Later in this paper, we will discuss several observational constraints on the parameters of the quintessential inflation models that we consider in this work, and for that we will scan over the parameters of the models and compare their theoretical predictions to the data. It is therefore important to have an idea for theoretical priors on the values of the parameters in the potential, for a given model, which can provide viable inflation. This can be achieved by applying the approximate constraint placed on the inflationary potentials from the requirement that the power spectrum of curvature fluctuations after inflation should match the COBE/Planck normalization, as discussed in section~\ref{basic}. Assuming a slow-roll regime for inflation, i.e. neglecting the terms including $\varphi^{\prime}$ and $\varphi^{\prime\prime}$ in Eqs. (\ref{eq:FriedmannInf3}) and (\ref{eq:phiEoMInf3}), respectively, the equations simplify to
\begin{align}
&H^2 = \frac{1}{3}V(\varphi) \,,\label{eq:FriedmannInfSlowRoll}\\
&3\varphi^{\prime}+\frac{1}{H^2}\frac{\dd}{\dd\varphi}V(\varphi)= 0 \,,\label{eq:phiEoMInf3SlowRoll}
\end{align}
which give
\begin{eqnarray}
\frac{\dd\varphi}{\dd N} = - \frac{1}{V(\varphi)}\frac{\dd}{\dd\varphi}V(\varphi) \label{eq:phievolvSR}\,.
\end{eqnarray}
In this slow-roll regime, the potential is related to the power spectrum of primordial curvature perturbations $\mathcal{P}_{\mathcal{R}}(k)$ through the COBE/Planck normalization equation,
\begin{eqnarray}
\frac{V(\vp)^3}{(\dd V(\vp)/\dd\varphi)^{2}} = 12\pi^2  \mathcal{P}_{\mathcal{R}}(k) \label{eq:COBEgen}\,,
\end{eqnarray}
see e.g. Ref.~\cite{Lyth:1998xn}. By solving these equations in the slow-roll approximation, one finds that in the large-$N$ approximation the results for $n_{s}$, $r$, and the amplitude of perturbations for $\alpha$ attractors are given by Eqs. \rf{aa}, \rf{eq:COBEexp2} and \rf{eq:COBEconstplus}.

In order to see whether a model of quintessential inflation is able to describe the dynamics of the universe after inflation, we need to add matter and radiation to the system of equations (\ref{eq:FriedmannInf3})-(\ref{eq:epsilon2}). In this case, the equations are modified as
\begin{align}
&H^2 = \frac{V(\varphi)+\rho_\text{M}+\rho_\text{R}}{3-\frac{1}{2}{\varphi^\prime}^2} \,,\label{eq:FriedmannLate1}\\
&\varphi^{\prime\prime}+(3-\epsilon)\varphi^{\prime}+\frac{1}{H^2}\frac{\dd}{\dd\varphi}V(\varphi)= 0 \,,\label{eq:phiEoMLate1}\\
&\epsilon=\frac{1}{2}\bigl({\varphi^\prime}^2-\frac{\rho_\text{M}^\prime+\rho_\text{R}^\prime}{6H^2}\bigr) \,,\label{eq:epsilon3}
\end{align}
where $\rho_\text{M}$ and $\rho_\text{R}$ are the energy densities of matter and radiation, respectively. They can be written as
\begin{align}
& \rho_\text{M} = 3H_0^2  \Omega_\text{M} e^{-3N}\,,\label{eq:rhoM}\\
& \rho_\text{R} = 3H_0^2  \Omega_\text{R} e^{-4N}\,,\label{eq:rhoR}
\end{align}
with $\Omega_\text{M}$ and $\Omega_\text{R}$ being the present values of density parameters for matter and radiation, respectively, and $H_0$ is the present value of the Hubble parameter. We can solve the set of Eqs. (\ref{eq:FriedmannLate1})-(\ref{eq:rhoR}) numerically and obtain the cosmic evolution in terms of $H$ for a specific model and for a set of parameters. This can then be compared to the cosmological measurements of $H$ and therefore constrain the model. We should however note that one important ingredient in solving the evolution equations is the initial conditions for the field $\varphi$. This is set by the reheating mechanism after inflation, as we will discuss in section~\ref{grpr} below.

Let us also introduce two important quantities, the evolution of which can give us deeper understanding of the dynamics of a model under investigation, the implications of the model for cosmic evolution, its observational viability, and its differences from the standard $\Lambda$CDM model.

The first quantity is the equation of state $w_\text{DE}$ for dark energy, in our case the scalar field $\varphi$. It is defined as
\begin{equation}\label{eq:wDEdef}
w_\text{DE}\equiv\frac{p_\text{DE}}{\rho_\text{DE}}=\frac{\frac{1}{2}\dot\varphi^2-V(\varphi)}{\frac{1}{2}\dot\varphi^2+V(\varphi)}=\frac{\frac{1}{2}{\varphi^\prime}^2H^2-V(\varphi)}{\frac{1}{2}{\varphi^\prime}^2H^2+V(\varphi)} \,,
\end{equation}
where $\rho_\text{DE}$ and $p_\text{DE}$ are the dark energy density and pressure, respectively, and $V(\varphi)$ is again the dark energy potential (which, as we discussed, can in principle contain a piece from the cosmological constant $\Lambda$). Note that $w_\text{DE}$ for a pure $\Lambda$ is $-1$.

Similarly to the slow-roll quantity $\epsilon$ for inflation, a useful quantity for late-time evolution of the universe is the so-called {\it effective equation of state} $w_\text{eff}$, defined as
\begin{equation}\label{eq:weffdef}
w_\text{eff}\equiv-1-\frac{2}{3}\frac{\dot H}{H^2} = -1-\frac{2}{3}\frac{H^\prime}{H}=-1+\frac{2}{3}\epsilon \,.
\end{equation} 
During radiation and matter domination epochs, $w_\text{eff}$ becomes $1/3$ and $0$, corresponding to $\epsilon=2$ and $3/2$, respectively. In $\Lambda$CDM, the dark energy domination epoch corresponds to $w_\text{eff}=-1$ ($\epsilon=0$).

We can study in more detail the behavior of dark energy in a given model by parameterizing the dark energy equation of state $w_\text{DE}$ in terms of the two so-called Chevallier-Polarski-Linder (CPL)~\cite{2001IJMPD..10..213C,2003PhRvL..90i1301L} parameters $w_{0}$ and $w_{a}$ through
\begin{equation}
w_\text{DE}(z)= w_0 + w_az/(1+z),\label{eq:CPL}
\end{equation}
where $z$ is the redshift. This parameterization is valid only near the present time (i.e. in the range $-1\lesssim N\lesssim0$, with $N=0$ corresponding to today). However, even though Eq. (\ref{eq:CPL}) cannot be used to fit the equation of state at early times or in the future, it gives a rough idea of how much the models deviate from $\Lambda$CDM at present time. $w_0$ and $w_{a}$ are also the parameters used in the definition of the {\it figure of merit} for the upcoming Stage IV large-scale structure surveys to quantify how well they can distinguish dark energy and modified gravity models from $\Lambda$CDM. We will therefore compute also $w_0$ and $w_{a}$ for our models below.

It is important to note that it is $w_\text{eff}$ (and not $w_\text{DE}$) which is used in direct comparison of the dynamics of the universe in a given model to the cosmological data, and one cannot directly constrain $w_\text{DE}$ without parametrizing it. Even though parametrizations of $w_\text{DE}$ are helpful in comparison of a model to the data, a detailed statistical analysis is always required in order to test and constrain the model; this is the approach we follow in this paper.

\subsection{Gravitational reheating versus instant preheating}\label{grpr}

The conventional mechanism of reheating after inflation is associated with a period of oscillations of the inflaton field at the minimum of its potential. In quintessential inflation, where the inflaton field does not oscillate, this mechanism does not work, and is replaced by gravitational reheating \cite{Ford:1986sy,Peebles:1998qn,Chun:2009yu}, which occurs due to particle production in changing gravitational background \cite{,ZS72,ZS77,Starobinsky:1980te}, and instant preheating~\cite{Felder:1998vq,Felder:1999pv,Kofman:2004yc}. Out of these two mechanisms, the gravitational reheating is the least efficient but the most general one, so we start with describing it here, limiting ourselves to simple estimates.

Inflationary quantum fluctuations of a light scalar field produced during inflation have the energy density of $\rho \sim {3H^{4}\over 8\pi^{2}}$ \cite{Linde:2005ht}. When inflation stops, some of this energy converts to the energy of scalar particles. This is an oversimplified way to describe the effect of particle production during inflation, but it shows a special role of the light scalar particles in this process. For example, massless vector particles are not produced, massless fermions are not produced, massive particles with masses much greater than $H$ are not produced. Following Refs.~\cite{Ford:1986sy,Peebles:1998qn}, and ignoring factors of $\mathcal{O}(1)$, one can estimate the energy of the produced particles at the end of inflation as
\be
\rho_{\rm gr} \sim 10^{{-2}} H_{\rm end}^{4} \sim 10^{-3} \rho_{\rm end}^{2} \sim 10^{-2} V_{\rm end}^{2} \ .
\ee
Here $H_{\rm end}^{4}$ and $\rho_{\rm end} \sim 2 V_{\rm end}$ are, respectively, the Hubble constant and the inflaton energy at the end of inflation, which happens at some field $\vp_{\rm end}$ when the kinetic energy of the field approaches $V_{\rm end}$ and the universe stops accelerating. 
%If we define the end of inflation by the moment when the universe stops accelerating, then this happens at the moment when $w = -1/3$, or, equivalently, when the kinetic energy density becomes equal to the potential energy density. 
The energy density $\rho_{\rm gr}$ subsequently decreases as $a^{-4}$ due to the expansion of the universe, as long as the produced particles have masses much smaller than $H$, which is the case for the flat quintessence potentials.

If the potential after inflation is very steep, as is the case in the single-field models to be considered below, soon after inflation the scalar field falls down and almost all of its energy proportional to $V$ becomes converted to its kinetic energy $\rho_{\rm kin} = {1\over 2} \dot\vp^{2}$. Thus in the first approximation  $\rho_{\rm kin} \sim V$. This kinetic energy corresponds to the equation of state $w = +1$,  and decreases as $a^{{-6}}$.

Thus, shortly after inflation the universe enters the regime of kinetic energy domination, which is sometimes called {\it kination}, but this regime ends when $\rho_{\rm kin} \sim \rho_{\rm end}^{2} a^{{-6}}$ becomes smaller than $\rho_{\rm gr} \sim 10^{-3} \rho_{\rm end}^{2} a^{{-4}}$. This happens at $a^{2} \sim 10^{3}$, when the energy density of radiation produced by reheating was $\rho_{\rm reh}  \sim 10^{-9}\,  \rho_{\rm end}^{4}$. The energy density scale $\rho_{\rm end}$ at the end of inflation in $\alpha$-attractors is typically in the range close to $\rho_{\rm end} \sim 10^{-10}$ in the Planck density units. In that case one finds $\rho_{\rm reh}  \sim 10^{-49}$ in Planck density units, or, equivalently $\rho_{\rm reh}  \sim (10^{6} {\rm GeV})^{4}$.

%Since this change of regime occurs at the onset of the stage when the energy density becomes dominated by the radiation produced during the process of gravitational reheating, one may call 

After that, the field $\vp$ continues rolling towards its large negative values until it freezes at some value $\vp_{{\rm F}}$ due to the famous Hubble friction term $3H\dot \vp$ in its equation of motion. Eventually, after the densities of radiation and cold dark matter become sufficiently small, the field $\vp$ starts rolling down again. The final results of the investigation of the equation of state of all matter in the universe depend on the value of $\vp_{{\rm F}}$. This value has been estimated in Ref.~\cite{Dimopoulos:2017zvq}, with the final result that in realistic models with gravitational preheating one may expect
\be \label{grreh}
|\Delta \vp| = | \vp_{{\rm F}}-\vp_{\rm end}| \sim 43  \ .
\ee
Note that this does not necessarily mean $|\vp_{{\rm F}}| \sim 43$ as stated in Ref.~\cite{Dimopoulos:2017zvq}, where the authors have considered the case with $\alpha  \ll 1$ rendering $\vp_{\rm end}$ negligible. Meanwhile for $\alpha = 7/3$ the end of inflation in the model studied in Ref.~\cite{Dimopoulos:2017zvq}  occurs not at $\vp_{\rm end} \sim 0$, but at $\vp_{\rm end} \sim 8$, which implies $\vp_{{\rm F}} \sim -35$.

The value of $|\vp_{{\rm F}}|$ may become  much smaller if one takes into account the possibility of instant preheating \cite{Felder:1998vq,Felder:1999pv,Kofman:2004yc}. This effect occurs if we consider interactions of the field $\vp$ with some other fields. 

For example, one may add to the original theory \rf{cosmo} a massless field $\sigma$ interacting with $\phi$ as ${g^{2}\over 2}\phi^{2}\sigma^{2}$. 
When the field $\phi$ moves through the point $\phi = 0$ with velocity $\dot\phi_{0}$, it creates particles $\sigma$ in the small vicinity of the point $\phi =0$, with the width $|\Delta \phi| \sim \sqrt{\dot\phi_{0} /g}$. The value of $\dot\phi_{0}$ in our problem is always smaller than $\sqrt {\rho_{\rm end}} \lesssim 10^{-5}$. Therefore, for sufficiently large $g$ one has $\sqrt{\dot\phi_{0} /g} < \sqrt{6\alpha}$. In that case, particle production occurs in a small region where $\phi \approx \vp$, and the old results of Refs.~\cite{Felder:1998vq,Felder:1999pv,Kofman:2004yc} derived for the canonical field $\varphi$ apply. These results show that the density of massless particles $\sigma$, created when the field $\vp$ passes through the point $\vp = 0$ is given by 
\begin{equation}\label{suppr}
 n_{\sigma}   =   {({g\dot\phi_0})^{3/2} \over
8\pi^3} \, .
\end{equation}
Then the field $\phi$ continues rolling further, giving each particle $\sigma$ a mass $g|\phi|$. This creates a gas of particles $\sigma$ with the energy density
\begin{equation}\label{supprrr}
 \rho_{\sigma}   =   {({g\dot\phi_0})^{3/2} \over
8\pi^3} \, g  |\phi| \, .
\end{equation}
This potential grows in both directions away from $\phi = 0$. For sufficiently large $g$, this may lead to a temporary trapping of the field $\phi$ near $\phi = 0$ \cite{Kofman:2004yc}. The field continues oscillating near this point until it loses some energy, particle production becomes inefficient, and the previously produced particles become diluted either by cosmic expansion or through their decay. Then the field $\phi$ resumes its rolling downhill. If instead of a single interaction term considered above  one considers a more general interaction $\sum  {g_{i}^{2}\over 2}(\phi-\phi_{i})^{2}\sigma^{2}$ with $|\phi_{i}| \ll \sqrt {6\alpha}$, one may have a chain of particle production events at each point $\phi = \phi_{i}$ \cite{Kofman:2004yc,Green:2009ds}.

It is not our goal here to study all the regimes that are possible due to instant preheating; see Refs.~\cite{Felder:1998vq,Felder:1999pv,Kofman:2004yc,Green:2009ds,Dimopoulos:2017tud} for a discussion of other possibilities. The efficiency of this process is controlled not only by the values of the couplings $g_{i}$, but also by the possibility of the decay of particles $\sigma$. This suggests that by a proper tuning of this scenario one may achieve freezing of the field $\varphi$ much earlier than in the gravitational reheating scenario. Therefore, in our subsequent analysis we will examine a broad range of possible values of $\vp_{{\rm F}}$. 

\subsection{Spectral index: Comparison with the non-quintessence scenario}\label{sp}

 The calculation of the inflationary parameters $n_{s}$ and $r$  in quintessential inflation have some distinguishing features. 
 %can be found in \cite{Dimopoulos:2017zvq}. \comYA{We have also calculated them here numerically. Also, as we checked, the same arguments are valid for the linear model aa well, so perhaps we should not restrict the discussion here to the exponential models?} 
As we will show shortly, extending the results of Refs.~\cite{Dimopoulos:2017zvq,Dimopoulos:2017tud,Maharana:2017fui}, predictions for $n_{s}$ and $r$ in quintessential inflation may differ rather significantly from the ones in the more traditional versions of $\alpha$-attractors, which do not have a stage of kination where the energy density of the universe is for a long time dominated by the kinetic energy of the inflaton field. This may give us a novel possibility to test quintessential inflation with gravitational reheating and a long stage of kination.
 
Let us remember that the values of $n_{s}$ and $r$ for $\alpha$-attractors are given by
\be
n_{s}= 1-{2\over N}\,, \qquad r = {12\alpha\over N^{2}} \, ,
\ee
where $N$ is the number of $e$-folds corresponding to the moment of production of the perturbations with momentum $k_*$ generated when the potential was equal to $V_{*} = V(\vp_{*})$.
 
We use the standard equation for the required number of $e$-folds, see Eq. (47) and a description of the notations in Ref.~\cite{Ade:2015lrj}: 
\be
\begin{aligned}
N \approx &  \; 67 - \ln \left(\frac{k_*}{a_0 H_0}\right)
%+  \frac{1}{4}\ln{\left( \frac{V_*}{\Mpl^4}\right) }
+  \frac{1}{4}\ln{\left( \frac{V_*^2}{\rho_{\rm end}}\right) } \\
&+ \frac{1-3w_\mathrm{int}}{12(1+w_\mathrm{int})}
\ln{\left(\frac{\rho_{\rm reh}}{\rho_{\rm end}} \right)} - \frac{1}{12} \ln (g_\mathrm{th} ) \, .
\label{eq:nefolds}
\end{aligned}
\ee
Using this equation, one can calculate the required number of $e$-folds $N$ for any model based on $\alpha$-attractors. Unless one studies models with extremely large or extremely small $\alpha$, one has $\rho_{\rm end} \sim V_* = \mathcal{O}(10^{-10}) $, with some variations which typically do not affect too much the value of the term $ \frac{1}{4}\ln{\left( \frac{V_*^2}{\rho_{\rm end}}\right)} $. The main difference between $N$ for different $\alpha$-attractors can be attributed to the  term $\Delta N = \frac{1-3w_\mathrm{int}}{12(1+w_\mathrm{int})}
\ln{\left(\frac{\rho_{\rm reh}}{\rho_{\rm end}} \right)}$.

In the simplest $\alpha$-attractor models, as well as in the Starobinsky model, which can be represented as an $\alpha$-attractor with $\alpha = 1$, after inflation one typically has $w_\mathrm{int} = 0$, i.e. $\Delta N = \frac{1}{12}
\ln{\left(\frac{\rho_{\rm reh}}{\rho_{\rm end}} \right)}$. In SUGRA-based $\alpha$-attractors and in the simplest versions of the Starobinsky model one often encounters an inefficient reheating with the reheating temperature $T_\text{r} \sim 10^{9} - 10^{11}$ GeV. For $T_\text{r} \sim 10^{10}$ GeV and assuming $\mathcal{O}(100)$ different types of particles in thermal equilibrium after reheating, one finds $\Delta N \sim -4$.

Meanwhile, in the quintessential $\alpha$-attractors with gravitational reheating and a long stage of kinetic energy dominance, one has $\Delta N = -\frac{1}{12}
\ln{\left(\frac{\rho_{\rm reh}}{\rho_{\rm end}} \right)}$. Notice the important sign change. Using the numerical estimates made in section~\ref{grpr}, one finds  $\Delta N = +7.5$. This particular number is rather sensitive to various assumptions on the energy scale of gravitational reheating, but let us take it at its face value. It shows that the required number of $e$-folds $N$ in the quintessential $\alpha$-attractor models can be greater than the one in the more conventional $\alpha$-attractors or in the Starobinsky model by $\Delta N \sim 10$.

As a result, the value of $n_{s}$ in quintessential $\alpha$-attractors with gravitational reheating is typically greater than in more traditional models by about 0.006 or so. This number coincides with one standard deviation in the Planck results \cite{Ade:2015lrj}. Thus, by a more precise determination of $n_{s}$, which can be achieved in the future, we may be able to distinguish quintessential $\alpha$-attractors with gravitational reheating from other models with more efficient reheating and without a long stage of kinetic energy domination. This result may become quite interesting  for development of inflationary models if more precise observations shift $n_{s}$ towards greater values as compared to the Planck 2015 results~\cite{Ade:2015lrj}. Moreover, further improvement of the accuracy of the measurement of $n_{s}$  may help us to distinguish the conventional inflationary models with the cosmological constant from the models of quintessential inflation, even if the equation of state of dark energy almost exactly coincides with $w = -1$.
 
\section{\Large {Examples of single-field models of quintessential inflation}}

\subsection{\boldmath{Linear potential}}\label{linsec}

We begin with the $\alpha$-attractor version of the simplest linear dark energy potential \cite{Linde:1986dq}
\be\label{llll}
V(\phi) =  \gamma  \phi + \Lambda\, .
\ee
In terms of the canonically normalized field $\vp$, this potential is given by
\be\label{linlin1}
V(\vp) = \gamma \sqrt{6\alpha}  (\tanh {\varphi\over\sqrt {6 \alpha}}+1)  + \Lambda\, .
\ee
At  $\varphi \gg + \sqrt {6\alpha}$ and $\Lambda \ll \gamma \sqrt{6\alpha}$ the potential is given by
\be\label{pos2}
V(\vp)=   2\gamma \sqrt{6\alpha}(1 - e^{-\sqrt{2\over 3\alpha} \varphi })\, ,
\ee
whereas at   $\varphi \ll - \sqrt {6\alpha}$ one has 
\be\label{linlin}
V(\vp)=   \Lambda + 2\gamma \sqrt{6\alpha}\  e^{\sqrt{2\over 3\alpha} \varphi } \, .
\ee
From the COBE/Planck normalization \rf{eq:COBEconstplus}, we find a constraint
\be\label{COPLA}
{\gamma\over \sqrt \alpha} \sim 2\times 10^{{-11}}\, .
\ee

\begin{figure}[h!]
\begin{center}
\includegraphics[scale=0.5]{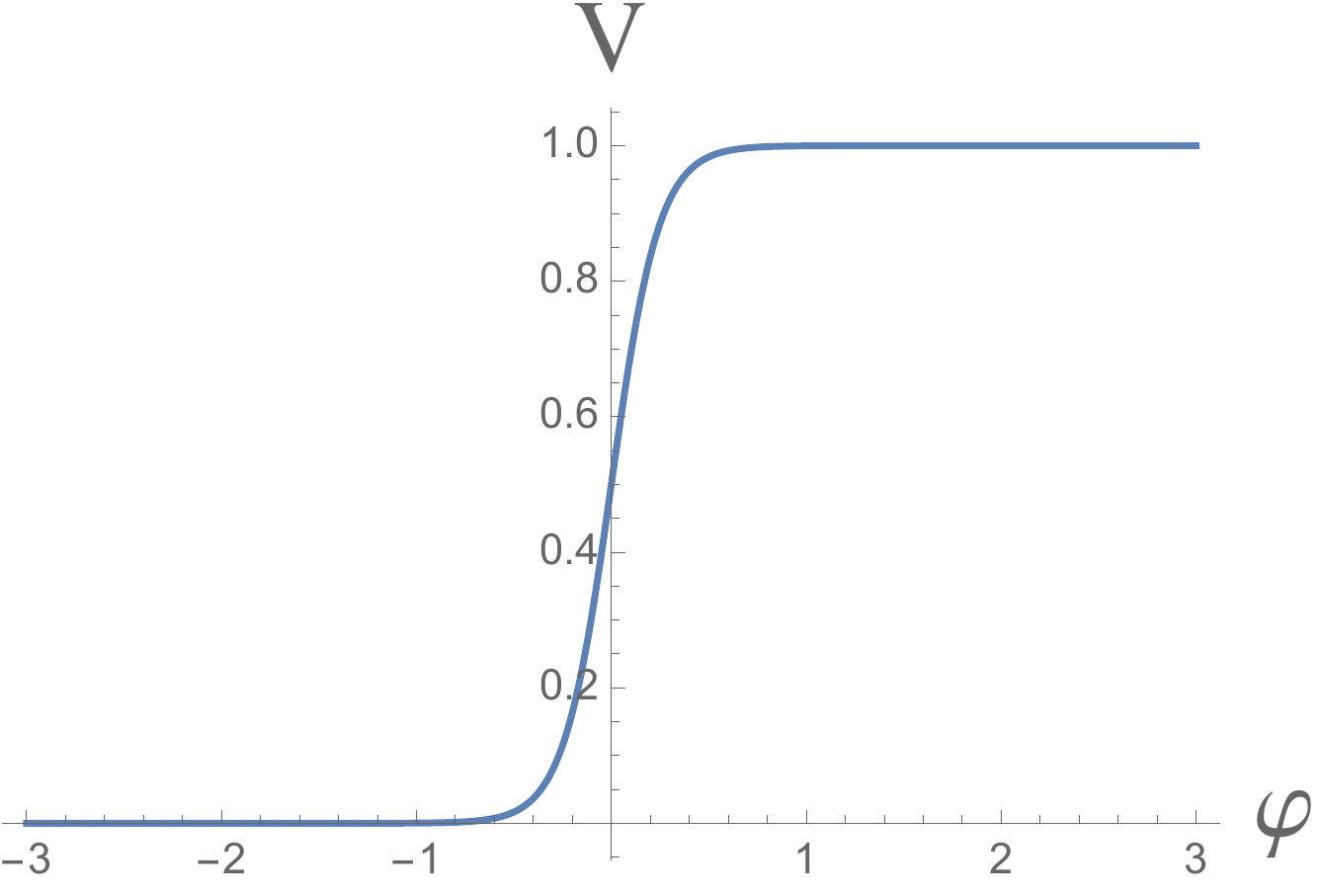}
\end{center}
\caption{\footnotesize Linear potential $V = {1\over 2\sqrt{6\alpha}}(\sqrt{6\alpha}+\phi) +\Lambda = {1\over 2} (1 + \tanh{\vp\over \sqrt{6\alpha}})+\Lambda$ for $\alpha = 10^{-2}$ and $\Lambda\sim 10^{{-120}}$.  The tiny cosmological constant $\Lambda$ is crucial for the validity of our scenario, but  $\Lambda$ is so small that it is invisible in this figure.}
\label{lin1}
\end{figure}

One could expect that the simplest linear model \rf{llll} with  $\Lambda = 0$ can be used as a model of quintessential inflation if one takes $\alpha \gtrsim 1$; see e.g. \rf{walpha1} and \rf{w73} for  $\alpha = 7/3$. However, one can easily check that in this model with $\alpha > 1/3$ the inflationary slow-roll parameter $\epsilon$ always remains smaller than 1 and inflation never ends. 

This problem can be solved by using $\alpha \ll 1$, for example $\alpha=\mathcal{O}(10^{{-2}})$, and adding a small cosmological constant $\Lambda \sim 10^{{-120}}$, see Fig. \ref{lin1}. In that case, inflation does end in a vicinity of $\vp = 0$, at $\vp_{\text{end}} \approx  \sqrt{3\alpha\over 8} \ln {1\over 3\alpha} \sim 0.2$. Then the field $\vp$ rolls down until  it freezes at some value $\vp = \varphi_\text{F}$ depending on the efficiency of reheating, see section~\ref{grpr}.  If $|\varphi_\text{F}| > \sqrt{3\alpha\over 2} \ln{\Lambda\over 2 \gamma \sqrt{6\alpha} } $, 
then the potential \rf{linlin1} is dominated by the positive cosmological constant $\Lambda$.
In that case,  at the moment when the field starts moving again, the universe gradually  enters the stage of expansion  dominated  by the cosmological constant $\Lambda$ with the equation of state $w_\text{DE} = -1$.

\begin{figure} [h!]
\center
  \includegraphics[height=3.3cm]{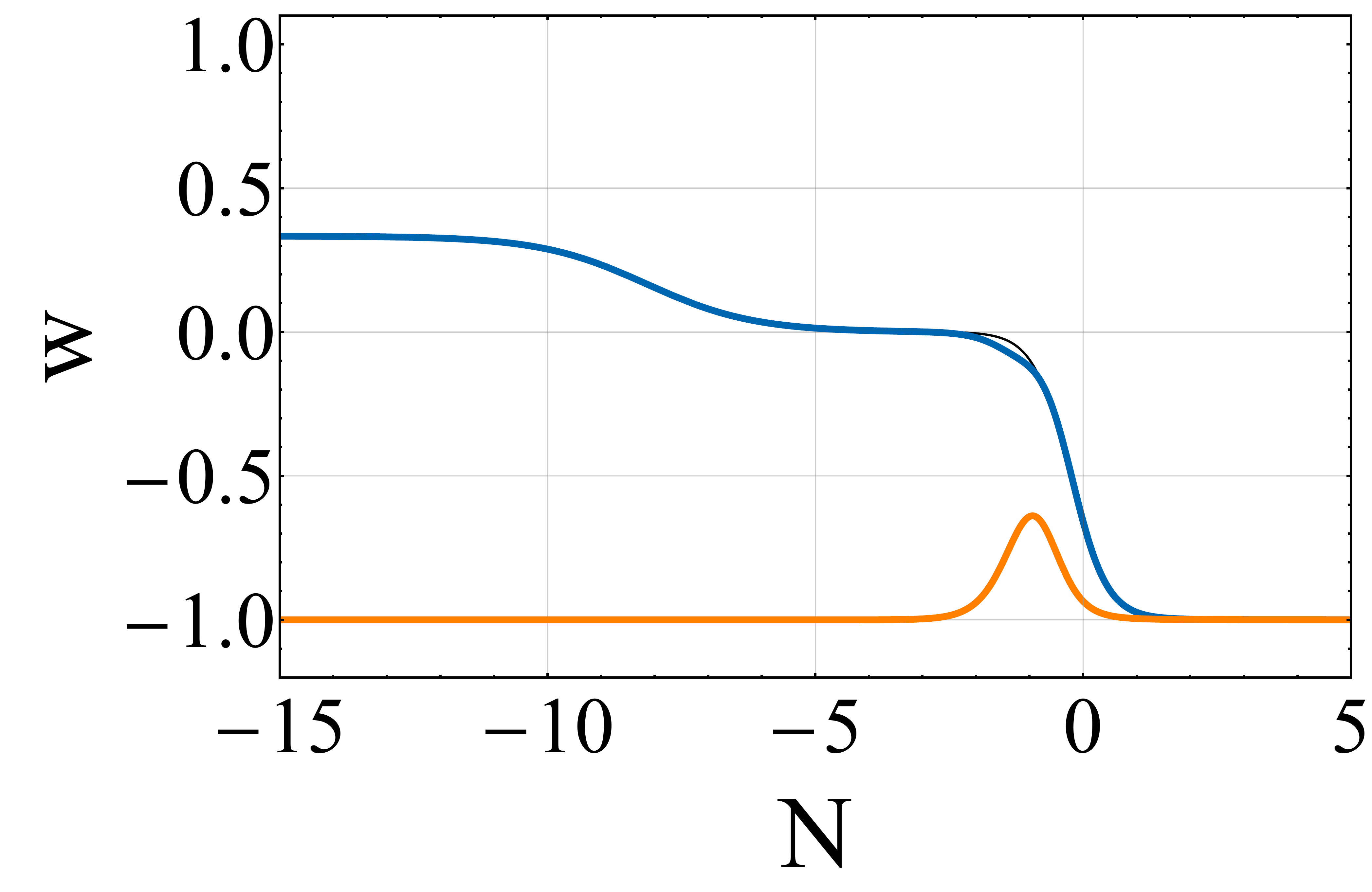}
  \includegraphics[height=3.3cm]{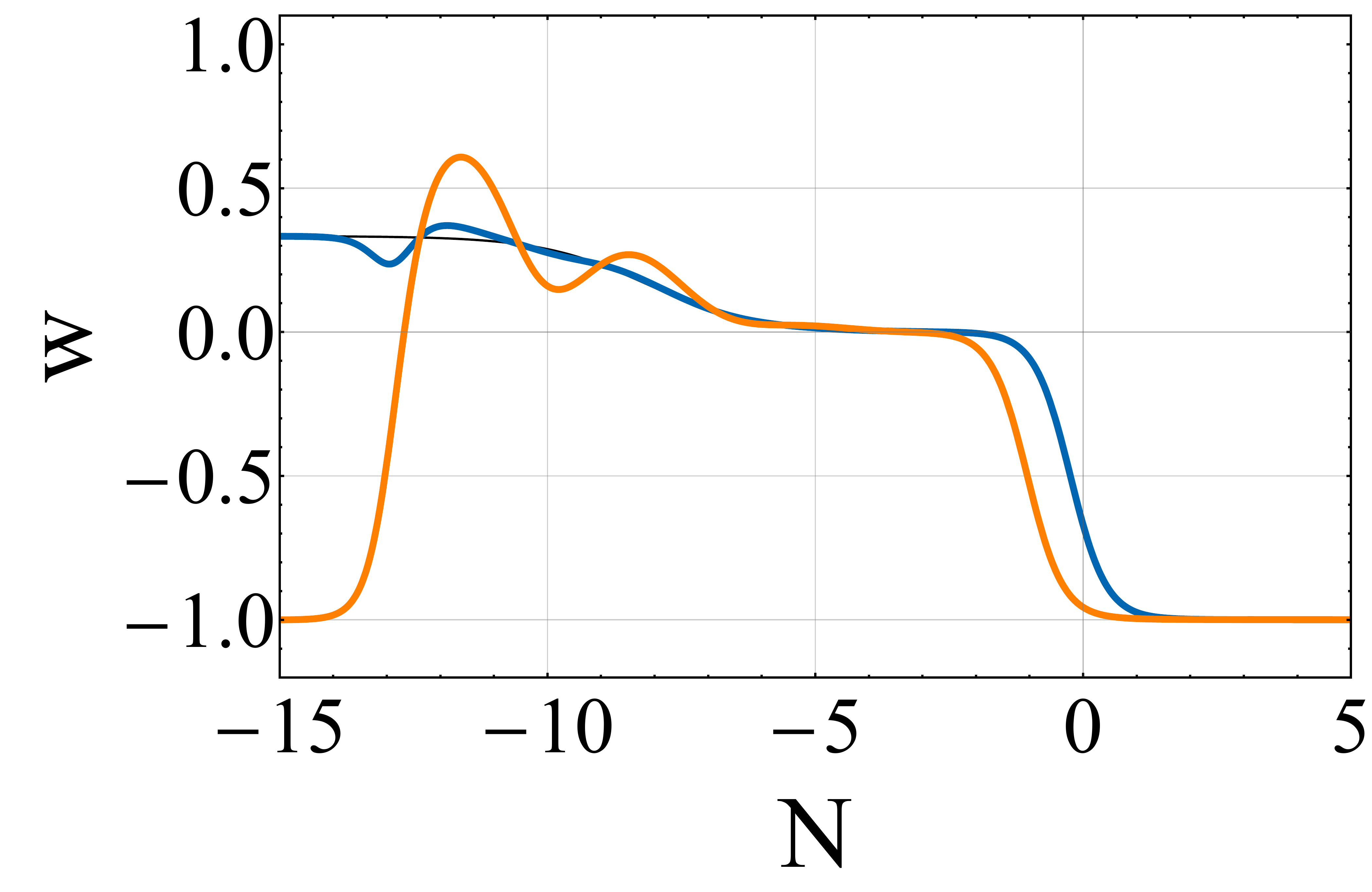}
  \includegraphics[height=3.3cm]{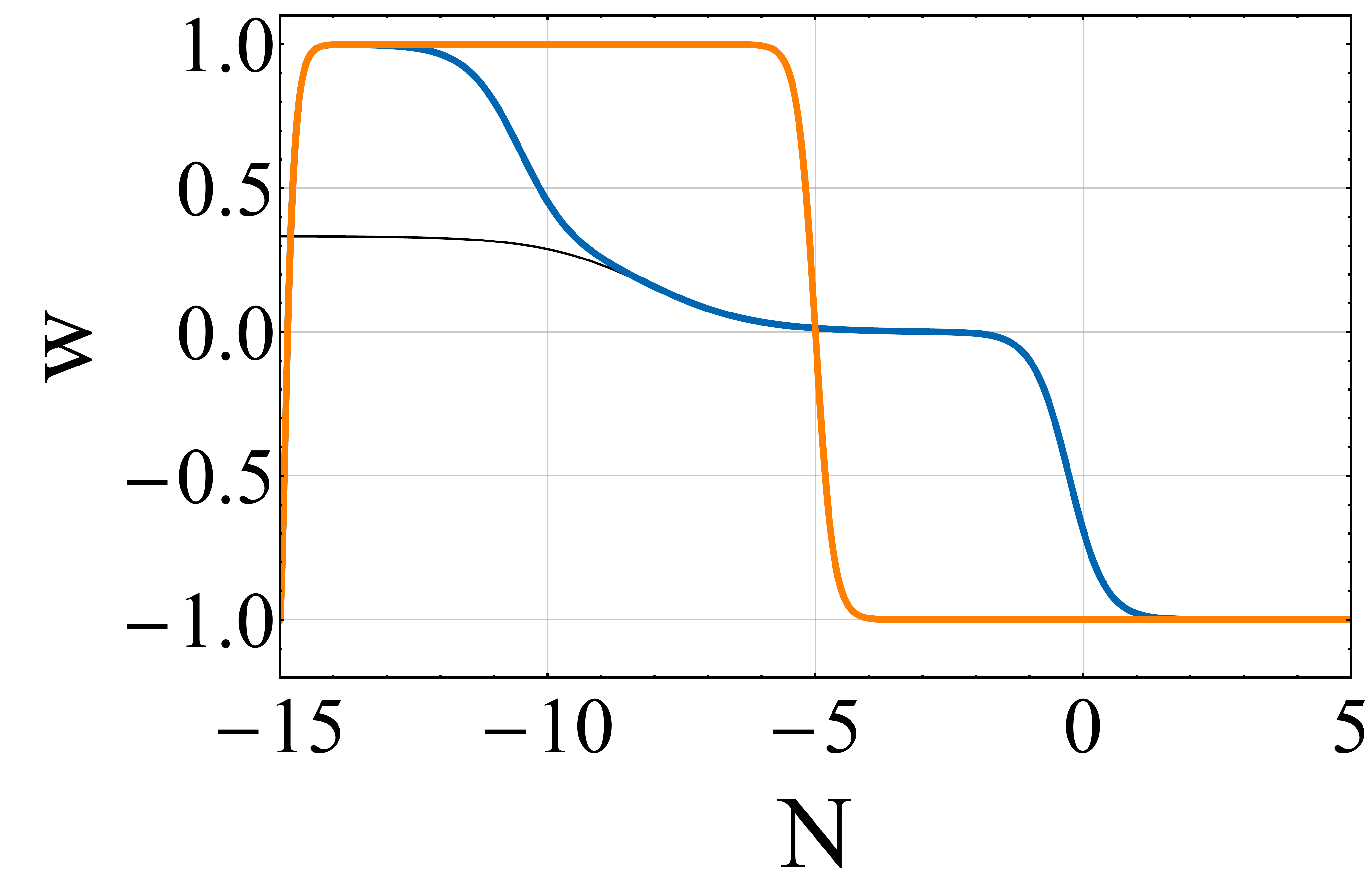}  
\caption{\footnotesize\label{fig:modelLinInf-wN-alpha002} Evolution of the equation of state as a function of the number of $e$-folds $N$ after reheating for the linear potential $\gamma\sqrt{6\alpha}(\tanh{\frac{\varphi}{\sqrt{6\alpha}}}+1) + \Lambda$ with $\Lambda = 0.7\rho_\text{c}$ and $\alpha$ set to $0.02$. The panels from left to right correspond to $\varphi_\text{F}=-43$, $\varphi_\text{F}=-36$, and $\varphi_\text{F}=-33$, respectively. The thick, blue and orange curves in each case correspond to $w_\text{eff}$ and $w_\text{DE}$, respectively, and we have also shown $w_\text{eff}$ for $\Lambda$CDM with a thin, black curve for comparison. $N=0$ corresponds to the present time.}
\end{figure}

\begin{figure} [h!]
\center
  \includegraphics[height=3.3cm]{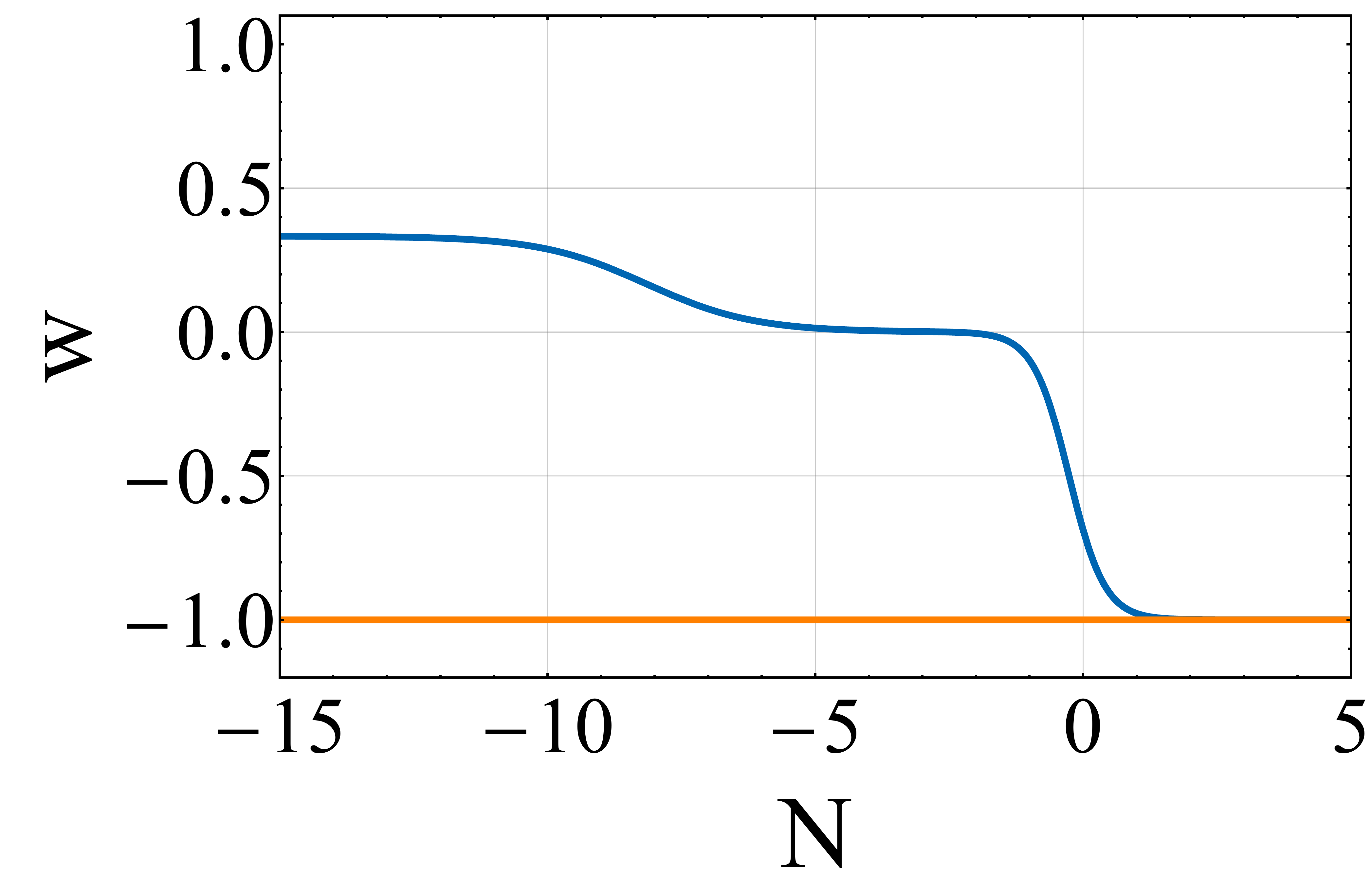}
  \includegraphics[height=3.3cm]{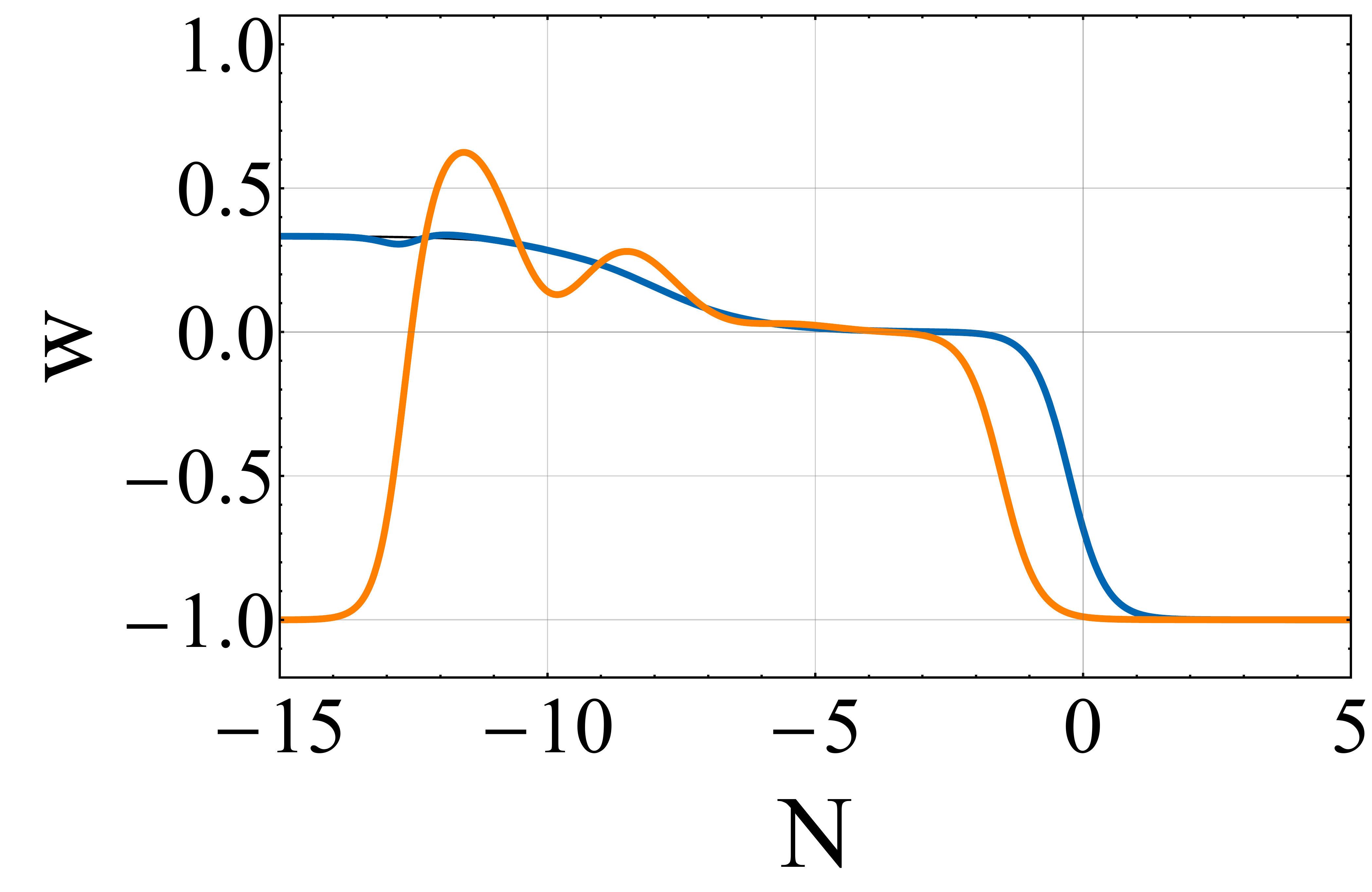}
  \includegraphics[height=3.3cm]{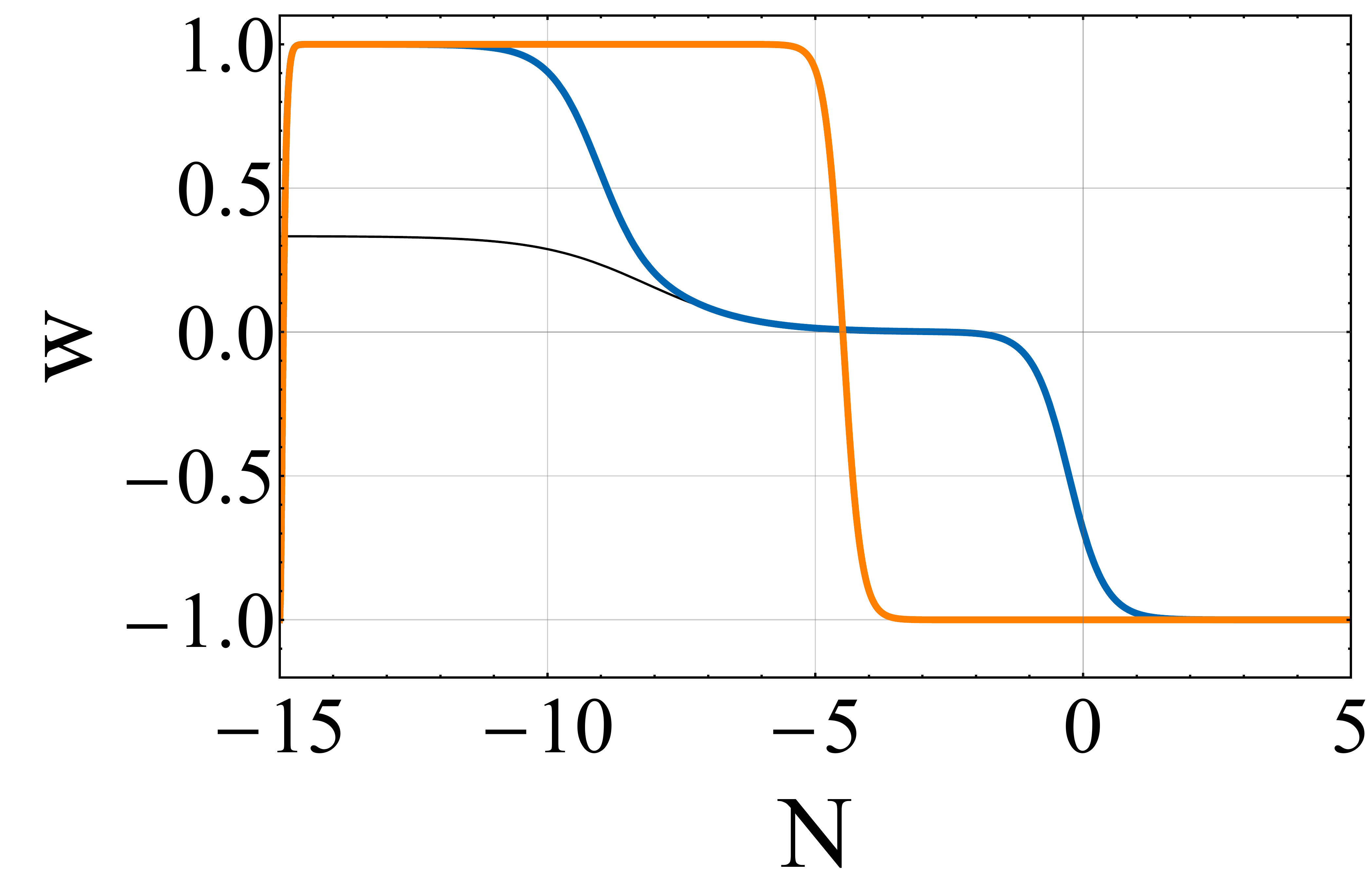}  
\caption{\footnotesize\label{fig:modelLinInf-wN-alpha0005} The same as in Fig.~\ref{fig:modelLinInf-wN-alpha002} but for $\alpha=0.005$. The panels from left to right now correspond to $\varphi_\text{F}=-22.5$, $\varphi_\text{F}=-18$, and $\varphi_\text{F}=-16$, respectively.}
\end{figure}

To go beyond the simple estimates given above and in order to determine the range of possible values of $\varphi_\text{F}$ required in this scenario, we performed a detailed numerical analysis for two different values of $\alpha = O(10^{{-2}})$.  Figs.~\ref{fig:modelLinInf-wN-alpha002} and~\ref{fig:modelLinInf-wN-alpha0005} show the effective equation of state $w_\text{eff}$ (thick, blue curves), as well as the equation of state of dark energy $w_\text{DE}$ (thick, orange curves) for this linear potential and for two illustrative choices of $\alpha=0.02$ and $\alpha=0.005$, and for different choices of $\varphi_\text{F}$. In both cases, $\Lambda$ has been set to $0.7\rho_\text{c}$, with $\rho_\text{c}\equiv 3H_{0}^2$ being the present value of the critical density, providing a total dark energy density today in agreement with observational data. The value of $\gamma \sqrt{6\alpha}$ has been set to $2.57\times10^{-12}$ and $6.4\times10^{-13}$ for $\alpha=0.02$ and $\alpha=0.005$, respectively, in order to obtain a correct inflationary scale; see (\ref{eq:COBEexp2}) and (\ref{eq:COBEconst}) which are valid also for the linear potential. In addition, we have presented $w_\text{eff}$ for $\Lambda$CDM in each case (thin, black curves) for comparison.

For $\alpha=0.02$, we have plotted three cases with $\varphi_\text{F}=-43$ (left panel), $\varphi_\text{F}=-36$ (middle panel), and $\varphi_\text{F}=-33$ (right panel). Looking first at $w_\text{eff}$ for $\varphi_\text{F}=-43$ we see that the desired cosmic history has been recovered although the evolution of $w_\text{eff}$ shows a small difference from the $\Lambda$CDM model at around $N=-2$. $w_\text{DE}$ in this scenario, however, shows a significant difference compared to the standard model --- $w_\text{DE}$ is not $-1$ always, contrary to a pure $\Lambda$, and has a pump at late times. For $\varphi_\text{F}=-36$, we see that although the late-time behavior of $w_\text{eff}$ is almost identical to that of $\Lambda$CDM, it shows a difference at early times ($N\lesssim -10$), and $w_\text{DE}$ is drastically different from a pure $\Lambda$ dark energy. By increasing $\varphi_\text{F}$ to $-33$, we now see that the times earlier than $N\sim -8$ (corresponding to the matter-radiation equality in $\Lambda$CDM) are strongly affected by the dynamics of the scalar field. We no longer recover a radiation domination epoch as in $\Lambda$CDM, and $w_\text{eff}$ goes all the way to $+1$ back in time rather than $1/3$ for radiation. This can be understood by looking at how $w_\text{DE}$ behaves at early times. The inflaton is in a kination phase at $N\lesssim -5$, and is dominant over matter and radiation at $N\lesssim -8$, hence the effective equation of state follows mainly the contribution from the inflaton and takes the value of $\sim +1$ at early times. Note that in this case the model does not give an early dark energy as $w_\text{DE}$ is $\sim +1$ and not $\sim -1$. 

Having this observation, let us systematically study different scenarios depending on the value of $\varphi_\text{F}$. Our numerical investigation of the model with $\alpha=0.02$ reveals three different possibilities:
\begin{itemize}
\item $-43 \leqslant \varphi_\text{F} \lesssim -34$: $\varphi_\text{F}\approx -43$ is the lowest value that $\varphi_\text{F}$ is allowed to take due to the reheating constraints, see section~\ref{grpr}. For the entire range of $[-43,-34]$ we obtain a dark energy which, while provides viable cosmologies over the entire history, it predicts deviations from a pure $\Lambda$ that are detectable by future observations. For example, for the two ends of the range, $\varphi_\text{F}=-43$ and $\varphi_\text{F}=-34$, we obtain $w_{0}\sim -0.936$ and $w_{a}\sim 0.192$, and $w_{0}=-0.956$ and $w_{a}=0.119$, respectively, which both should be detectable by the future Stage IV large-scale structure surveys, see section~\ref{sec:compdataexp}. In addition, for this range we recover radiation and matter domination epochs which are very similar to those of $\Lambda$CDM, with some small distortions due to the fact that the scalar field is not completely subdominant at early times; the larger the value of $\varphi_\text{F}$, the larger the distortions.  $w_\text{eff}$ and $w_\text{DE}$ for another example of $\varphi_\text{F}$ in this range are presented in Fig.~\ref{fig:modelLinInf-wN-alpha002} (middle panel) for $\varphi_\text{F}=-36$ with $w_{0}\sim -0.956$ and $w_{a}\sim 0.119$.

\item $-34 \lesssim \varphi_\text{F} \lesssim -32$: In this case, the model is viable from the point of view of late-time cosmology, with a $\Lambda$-like dark energy at late times ($w_{0}\sim-1$ and $w_{a}\sim0$), the reason being that the $\Lambda$ term is dominant over the scalar field during this period. The very early times ($N\lesssim -8$) in this range are however strongly affected by the scalar field, and behave significantly differently from that of $\Lambda$CDM, i.e. we do not get radiation domination at early times, but a domination by the inflaton in a kination phase. The model therefore gives viable cosmologies from the point of view of late-time observations, but we obtain no radiation domination epoch at early times. An example of this case has been presented in Fig.~\ref{fig:modelLinInf-wN-alpha002} (right panel) for $\varphi_\text{F}=-33$.

\item $-32 \lesssim \varphi_\text{F}$: By increasing $\varphi_\text{F}$ to values larger than $\sim -32$ the scalar field stays in the kination phase for a longer period of time, and is also dominant over matter and radiation for a longer period, resulting in an extended epoch of $w_\text{eff}=+1$ at early times. Increasing $\varphi_\text{F}$ to $-30.5$ already extends the domination of the scalar field with $w_\text{DE}=+1$ all the way to $N\approx -5$, which is the beginning of matter domination. The more we increase $\varphi_\text{F}$, the longer the period of dark energy domination (with $w_\text{DE}=+1$), so that the model will give predictions that are in clear contradiction with observations. Of course, for any values of $\varphi_\text{F}$ the energy density of dark energy will eventually be dominated by the cosmological constant with $w=-1$, but our numerical studies show that this happens later and later in time when $\varphi_\text{F}$ increases, and the $\Lambda$ domination eventually happens only in the future.

\end{itemize}

In summary, our analysis shows that the linear model with $\alpha=0.02$ provides viable cosmologies as long as $\varphi_\text{F}$ remains in the relatively broad range of $\sim [-43,-34]$, while predicting detectable deviations from $\Lambda$CDM that are sufficiently large for the model to be tested against $\Lambda$CDM. One should note that larger values of $\varphi_\text{F}$ all the way to about $-32$ can also provide viable late-time cosmologies and only affect the epoch of radiation domination in the early universe.

Let us now decrease $\alpha$ to $0.005$. Fig.~\ref{fig:modelLinInf-wN-alpha0005} shows the evolution of $w_\text{DE}$ and $w_\text{eff}$ for this scenario, but for three choices of $\varphi_\text{F}=-22.5$ (left panel), $\varphi_\text{F}=-18$ (middle panel), and $\varphi_\text{F}=-16$ (right panel). We see that for $\varphi_\text{F}=-22.5$, the model already behaves almost identically to $\Lambda$CDM, with $w_\text{DE}$ being $-1$ for the entire history. Clearly, for $ \varphi_\text{F}< -22.5$ all the way to our lower bound of $-43$, the model will remain like $\Lambda$CDM. Let us now increase $\varphi_\text{F}$ from $-22.5$ to $-21.5$ (not shown in Fig.~\ref{fig:modelLinInf-wN-alpha0005}). Our numerical analysis gives $w_0\sim-0.983$ and $w_{a}\sim 0.050$ in this case. This shows that the deviations from a pure $\Lambda$ increases by increasing $\varphi_\text{F}$. Increasing $\varphi_\text{F}$ further to $\sim -16$ still gives viable cosmologies, while the values larger than $\sim-16$ will make the early times ($N\lesssim -8$) completely affected by the kination domination of the inflaton over radiation, and radiation domination will be lost; the model, however, behaves like a pure cosmological constant at late times, i.e. with $w_{0}\sim -1$ and $w_{a}\sim 0$. An example of how $w_\text{eff}$ and $w_\text{DE}$ behaves for the range $[-21.5,-16]$ is presented for $\varphi_\text{F}=-18$ (with $w_{0}\sim -0.989$ and $w_{a}\sim 0.030$) in Fig.~\ref{fig:modelLinInf-wN-alpha0005} (middle panel), while the behavior of $w_\text{eff}$ and $w_\text{DE}$ for $\varphi_\text{F}=-16$ is given in the right panel of the figure. We see that dark energy for $\varphi_\text{F}=-18$ shows an evolution similar to the previous case of $\alpha=0.02$ with $\varphi_\text{F}=-36$. For values of $\varphi_\text{F}$ larger than $-16$ we see a behavior similar to the case of $-32 \lesssim \varphi_\text{F}$ for $\alpha=0.02$, i.e the epoch of dark energy domination in the kination phase gets extended to later times, making the model more and more unviable by increasing $\varphi_\text{F}$. We therefore conclude that the linear model with $\alpha=0.005$ provides viable cosmologies for $\varphi_\text{F}\in [\sim-21.5,\sim-16]$ with $w_{0}$ and $w_{a}$ showing deviations from $\Lambda$CDM, and for $\varphi_\text{F}\lesssim -21$ with dark energy behaving almost identically to a pure $\Lambda$. The deviations for the range $[-21.5,-16]$ are not as large as the ones we obtained for $\alpha=0.02$, but {\it might} still be detectable by the Stage IV LSS surveys.

In conclusion, we have found a realistic model of quintessential inflation based on the $\alpha$-attractor model with a linear potential. This model requires ${\gamma\over \sqrt \alpha} \sim 2\times 10^{{-11}} $,  $\alpha \lesssim 0.02$, and  a cosmological constant in the anthropically allowed range of $\Lambda \sim 10^{{-120}}$. The smaller the value of $\alpha$, the larger the range of $\varphi_\text{F}$ for which viable cosmic histories exist, although deviations from $\Lambda$CDM are expected to become less and less likely in the limit $\alpha \ll 0.01$.

This is the simplest model of quintessential inflation based on $\alpha$-attractors, so let us pause here a little, before turning to other, more complicated models.
The linear potential $V(\phi) =  \gamma  \phi + \Lambda$  is the simplest potential ever, and yet it was never used in inflationary theory until now, for a good reason: This potential is unbounded from below, so unless $\gamma$ is extraordinarily small, it leads to a rapid instability and a collapse of the universe.  A linear potential was used  in  Ref.~\cite{Linde:1986dq} for describing dark energy and solving the cosmological constant problem, but it required an extremely small constant $\gamma \lesssim 10^{-120}$ to avoid the collapse of the universe within 14 billion years. 

In our new model described in this section, we have ${\gamma\over \sqrt \alpha} \sim 2\times 10^{{-11}}$ \rf{COPLA}, which is the standard inflationary requirement for the COBE/Planck spectrum normalization. Thus $\gamma$ can be 110 orders of magnitude greater than in the quintessence model of Ref.~\cite{Linde:1986dq}. And nevertheless, we do not have any vacuum instability, because in the context of $\alpha$-attractors the potential is defined only in the finite range $|\phi| < \sqrt{6\alpha}$. The lower part of the potential in this range becomes an infinite, exponentially flat plateau in canonical variables. This gives us greater flexibility in the choice of inflationary models. 

By modifying the value of $\alpha$ and the strength of interaction of the field $\vp$ with matter, one can control the parameter $w$. One may also increase the value of the inflationary spectral index $n_{s}$ by about one standard deviation of the Planck 2015 results for $n_{s}$. The only additional fine-tuning required in this model, as compared to the more conventional models of inflationary $\alpha$-attractors, is the condition  $\alpha \lesssim 0.02$. It would be nice to find  consistent  versions of such models with $\alpha = O(1)$, and especially with $\alpha = 1/3,..., 7/3$, which are better motivated in  extended supergravity, M-theory, and string theory  \cite{Ferrara:2016fwe,Kallosh:2017ced,Kallosh:2017wnt}.  However, $N=1$ supergravity does not impose any constraints on $\alpha$.  From a purely phenomenological point of view, the requirement $\alpha \lesssim 0.02$ is not an unreasonable price to pay for a simple, unified description of inflation and dark energy.

\subsection{\boldmath{Two-shoulder model with exponential potential}}\label{should}
\begin{figure}[h!]
\begin{center}
\includegraphics[scale=0.5]{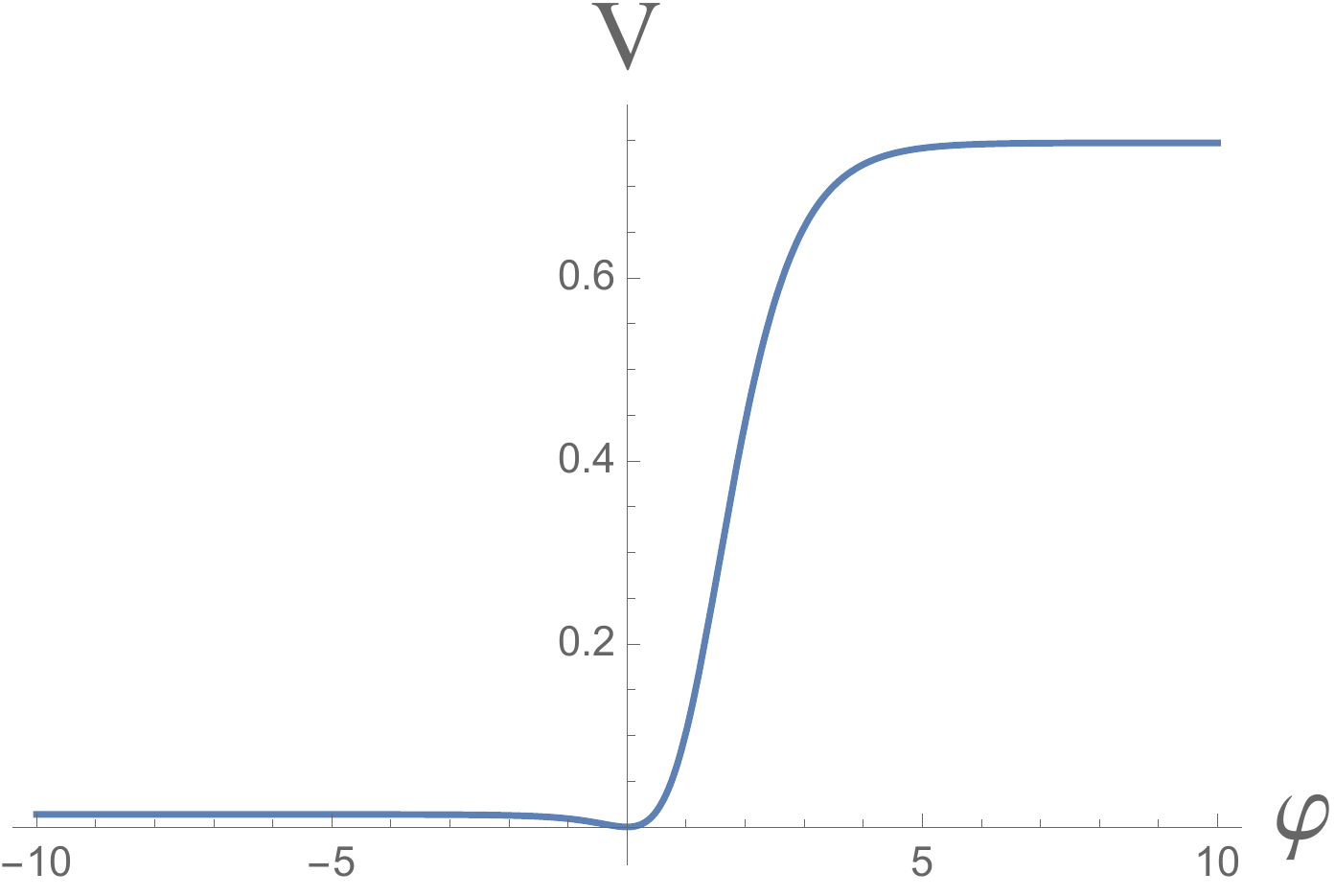}
\end{center}
\caption{\footnotesize The potential \rf{sh} shown for a toy model with $M = 1$, $\gamma = 4$, and $\alpha = 1/3$. It illustrates the main feature of the models of this class: two shoulders with an exponentially large difference in their heights. }
\label{F2shouldersBeta2}
\end{figure}

The next example to consider is the exponential two-shoulder potential introduced in Ref.~\cite{Carrasco:2015rva},
\be\label{orig}
V(\phi) = M^{2} e^{{-2\gamma}}\bigl(e^{{\gamma\phi\over \sqrt{6\alpha}}} - 1\bigr)^{2} \ .
\ee
In the canonical variables, one finds
\be\label{sh}
V(\varphi) = M^{2} e^{{-2\gamma}}\bigl(e^{ {\gamma }\tanh{\vp\over\sqrt{6\alpha}}} - 1\bigr)^{2} \ .
\ee
The potential has a minimum at $\vp = 0$. The general shape of such potentials is illustrated by Fig. \ref{F2shouldersBeta2} for a toy model with $M = 1$, $\alpha = 1/3$, and $\gamma = 2$. In realistic models, we need to take  $\gamma \gg 1$. In this limit, the right shoulder has the height $V_{+} = M^{2}$, and the left shoulder has the height $V_{-} = M^{2} e^{{-\gamma}}$.  

\begin{figure}[h!]
\begin{center}
\includegraphics[scale=0.55]{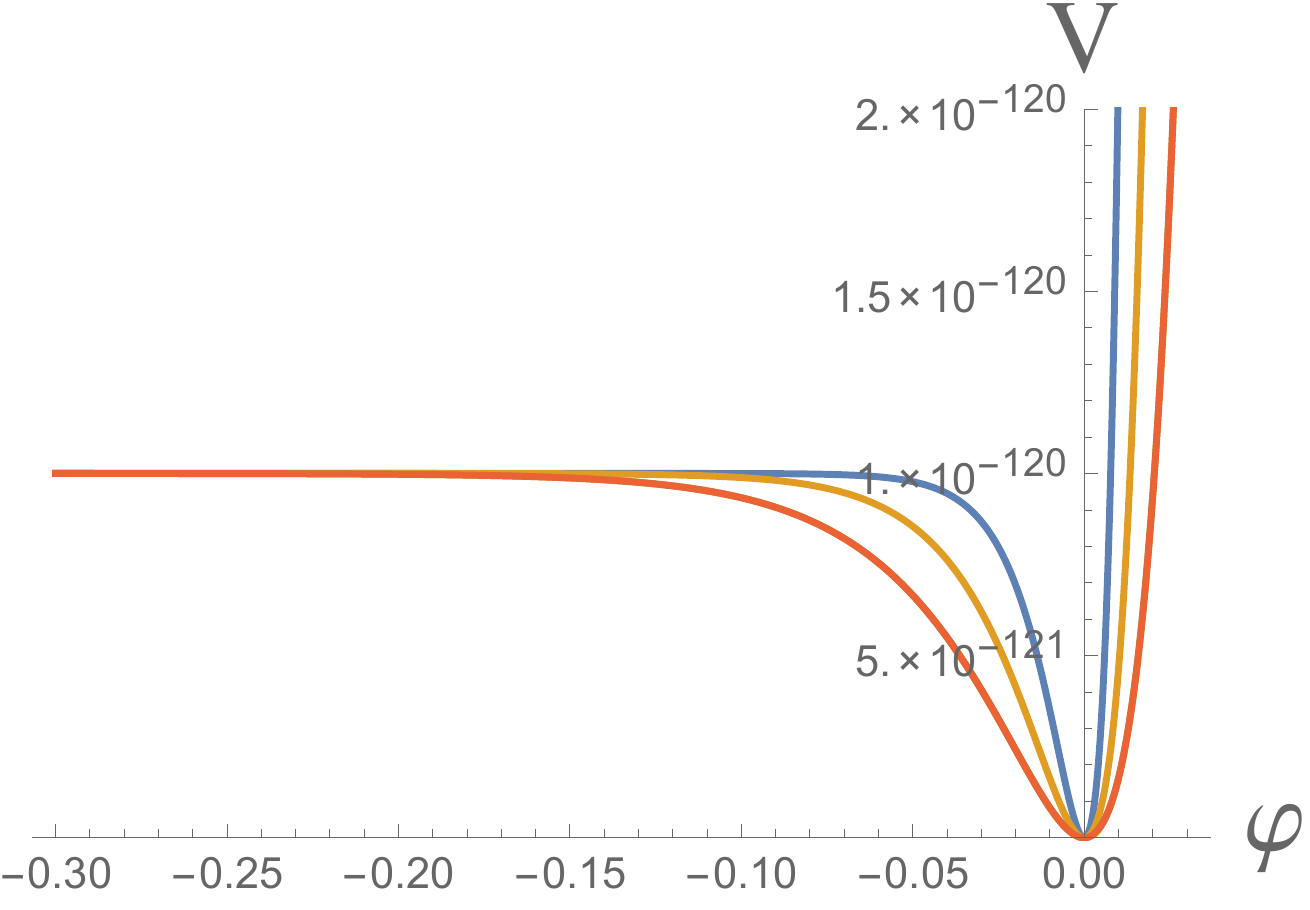}
\end{center}
\caption{\footnotesize The potential \rf{sh}  shown in Planck energy density units for $M \sim 10^{{-5}}$, $\gamma \sim 126$,  $\alpha = 1/3$ (blue curve),  $1$ (yellow curve), and $7/3$ (red curve). Inflation begins at the right shoulder of this potential, which is not shown here because it is $110$ orders of magnitude higher. After that, the field rolls to the left plateau, which almost immediately becomes flat, with an accuracy $10^{-175}$. That is why it is practically indistinguishable from the cosmological constant.}
\label{F2shoulders}
\end{figure}

An advantage of this model is that it can easily incorporate the exponentially large hierarchy $e^{{2\gamma}}$ between the inflationary energy scale $V_{+} = M^{2} \sim 10^{-10}$ and the dark energy scale  $V_{-} = M^{2} e^{{-2\gamma}} \sim 10^{-120}$.  
For $\alpha = \mathcal{O}(1)$, $M \sim 10^{{-5}}$, and $\gamma \sim 126$, this model fits all inflationary data, and describes the present stage of acceleration driven by the effective cosmological constant $V_{-} \sim 10^{{-120}}$. It is difficult to show the right and the left plateaus in one figure, because the height of the right shoulder is  110 orders of magnitude greater than the height of the left one. Therefore, we show only the left shoulder of the potential and a small vicinity of its minimum in Fig. \ref{F2shoulders}.

The shape of the left plateau shown in Fig. \ref{F2shoulders} is determined by the following asymptotic expression for $V(\vp)$ at large negative $\vp$:
\be
V=M^{2} e^{-2\gamma}  \left(1-4\gamma e^{-\gamma } e^{\sqrt{2\over 3\alpha}\vp}\right) \ .
\ee
The potential approaches $V_{-} = M^{2} e^{-2\gamma}  \sim 10^{{-120}}$, and the asymptotic deviation from this value at large, negative $\vp$ is suppressed not only by the factor $e^{\sqrt{2\over 3\alpha}\vp}$, but also by an extra factor $e^{-\gamma} \sim 10^{-55}$. This means that the potential is extremely flat everywhere outside a small vicinity  near $\vp=0$. One can check, for example, that the slow-roll parameter $\epsilon $ in this model is smaller than $10^{{-25}}$ for $\vp < 1$. The simplest way to understand it is to note that even the potential \rf{orig} in terms of the original variable $\phi$ is exponentially flat at the boundary of the moduli space $\phi = \sqrt{6\alpha}$ for $\gamma \gg 1$, and the transition to the canonical variables leads to an additional flattening. As a result, a generic prediction for dark energy in this model is $w = -1$, which is clearly consistent with all current cosmological observations. 

\begin{figure}[h!]
\begin{center}
\includegraphics[scale=0.55]{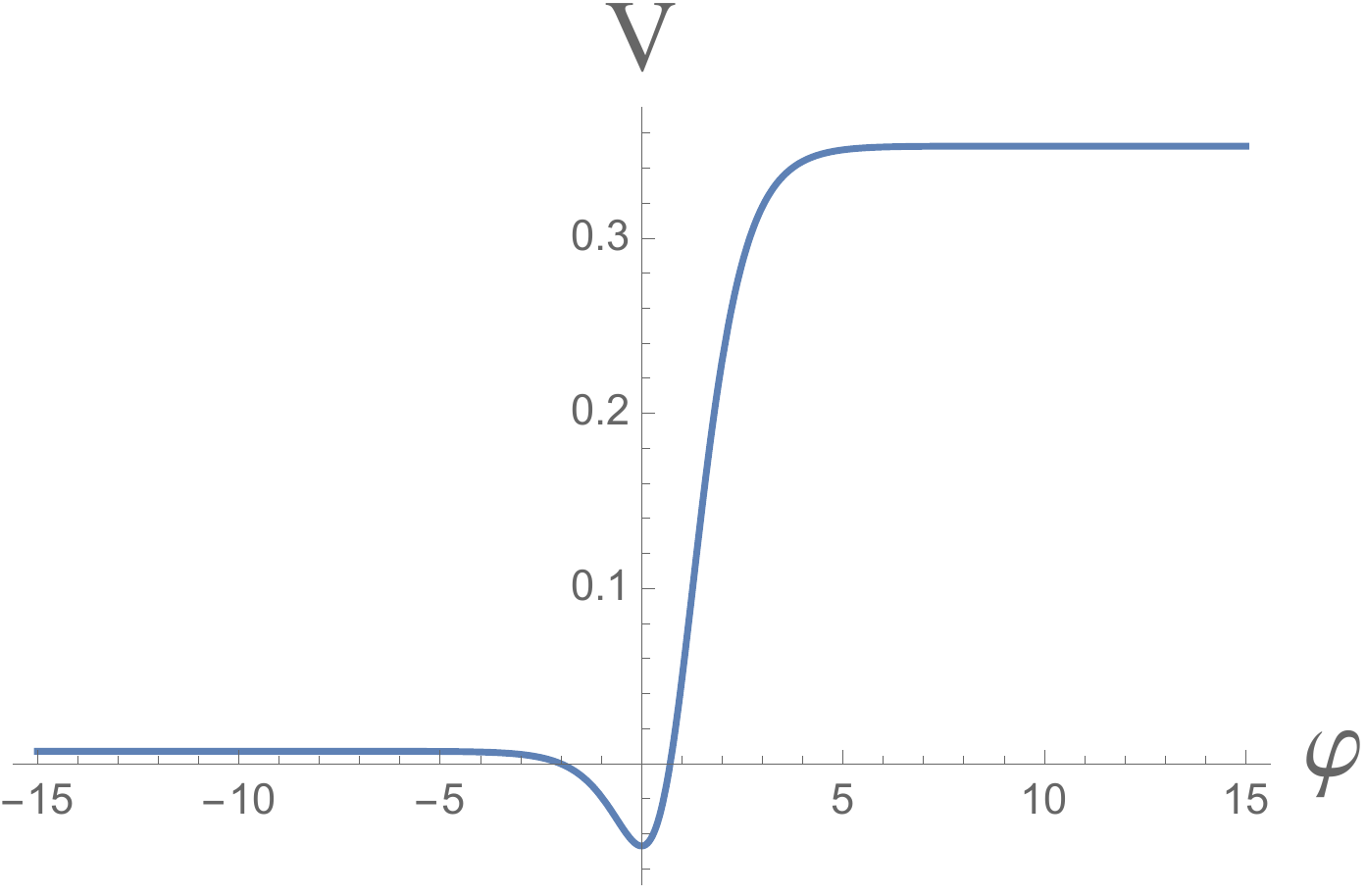}
\end{center}
\caption{\footnotesize In the asymmetric potential with a minimum at $V < 0$ one can achieve exponential hierarchy of the heights $V_{+} $ and $V_{-}$ with smaller values of $\gamma$. For illustration, in this figure we used $M = 1$, $\gamma = 1$, $\alpha = 1/3$, and added a   constant $V_{0} = -0.047$. By taking a slightly smaller value of $V_{0}$, one can easily make the asymptotic value of the potential $\Lambda = V_{-} \sim 10^{-120}$, as required by anthropic considerations.}
\label{F2shouldersMinus2}
\end{figure}

In general, one may add an arbitrary  constant $\Lambda$ to the potential \rf{sh}. By adding a negative  constant one may decrease the required value of the parameter $\gamma$. As one can see from Fig.  \ref{F2shouldersMinus2}, one can easily tune the asymptotic value of the potential to be $\Lambda = V_{-} \sim 10^{-120}$ in accordance with anthropic considerations. %We hope to return to the investigation of such models in a separate publication.

Since we generically obtain $w=-1$ in this model, one may wonder whether it has any merit over simple $\Lambda$CDM. Indeed, as we discussed in section~\ref{sec:intro}, the cosmological constant provides a much simpler interpretation for the origin of cosmic acceleration. However, the model presented here demonstrates that one can easily construct a family of inflationary models in which inflation ends without any need to stabilize the inflaton field at the minimum of its potential. Even in the models where the potential has an anti-de Sitter  minimum with a negative cosmological constant at $\vp = 0$, as in Fig. \ref{F2shouldersMinus2}, one can safely live in a de Sitter-like state on an exponentially flat low plateau. The flatness of the potential in this model, just as in all other models considered in this paper, is protected by the geometric origin of $\alpha$-attractors. 
 
As we already mentioned, further improvement of the accuracy of the measurement of $n_{s}$  may help  to distinguish this model and other models  of quintessential inflation from the more conventional $\alpha$-attractors, even if the equation of state of dark energy in quintessential inflation almost exactly coincides with $w = -1$, see section~\ref{sp}. The possibility of having a somewhat larger value of $n_{s}$ due to the long stage of kination in this scenario may become very welcome in the future, depending on the observational data.  
 
\subsection{Exponential potential}\label{sec:exp-pot}

Let us now assume a simple exponential form for the non-canonical potential $V(\phi)$ where a free cosmological constant term $\Lambda$ is also (implicitly) included. We will later fix $\Lambda$ to specific values in order to construct two specific working models with this potential.

The total potentials of our single-field, quintessential inflation models have the structure
\be
V(\phi) =M^2 e^{\gamma (\frac{\phi}{\sqrt{6\alpha}}-1)} + V_{0}\,,
\ee
which, again with $\phi= \sqrt {6 \alpha}\, \tanh{\varphi\over\sqrt {6 \alpha}}$, gives
\be
V(\vp)=  M^2 e^{\gamma\,( \tanh{\varphi\over\sqrt {6 \alpha}}-1)}+ V_{0}\,.
\ee
At large, positive $\vp$ this potential tends to the inflationary plateau with $V_{+} = M^2 +V_{0}$, and at large, negative $\vp$ it tends to the cosmological constant $\Lambda =V_{-}=M^2 e^{-2\gamma}+V_{0}$. Instead of making a general investigation for arbitrary $V_{0}$ (or $\Lambda$), we concentrate here on two particular cases, which we call Exp-model I and Exp-model II:

\begin{itemize}
	\item {\bf Exp-model I}: The constant $V_{0} $ is set to zero. In this case the potential for dark energy is solely the exponential one,
\be V = M^{2}e^{\gamma\big(\tanh{\vp\over\sqrt{6\alpha}}-1\big)} . \ee
At large, positive $\vp$ this potential tends to  $V_{+} = M^2$. Its asymptotic value at large, negative $\vp$ is given by the cosmological constant $\Lambda =V_{-} =M^2 e^{-2\gamma}$. 	
	
	\item {\bf Exp-model II}: The constant $V_{0} $  is set to $-M^{2}e^{-2\gamma}$ \cite{Dimopoulos:2017zvq}. In this case the potential for dark energy is 
\be V = M^{2} e^{{-2\gamma}} \Big(e^{\gamma\big(\tanh{\vp\over\sqrt{6\alpha}}+1\big)}- 1\Big) . \ee	
At large, positive $\vp$ in the large $\gamma$ limit it reaches $M^{2}$, as before, up to an exponentially small correction $-M^{2}e^{-2\gamma}$. It vanishes asymptotically for large, negative $\varphi$, i.e. $\Lambda =V_{-} =0$. 
		\end{itemize}

The ratio of  $V_{-}$ to $V_{+}$ in Exp-model I is given by
\be
{V_{-} \over V_{+} } = e^{-2\gamma}\approx   10^{-110} \approx e^{-252} \, .
\ee
An analogous relation should be valid for Exp-model II, but instead of $V_{-}$ one should have the present value of dark energy $V_\text{today} \sim 10^{-120}$. One can view this property of our quintessential inflation models as a drawback, since our potentials have a huge number built in. This is however the price to pay for having one plateau of the model for the early universe at about $10^{-10}$ in Planck density units, and another one for the current and future acceleration at about $10^{-120}$. In the context of a phenomenological model, however, we may view this as a parameter which is determined observationally,
\be
\gamma\approx \ln {H_{\rm infl} \over H_{\rm DE} } \, .
\ee
In such a case, we still have to find the  working models which show a consistent  deviation from the cosmological constant dynamically.

\begin{figure}[h!]
\center
  \includegraphics[height=5.5cm]{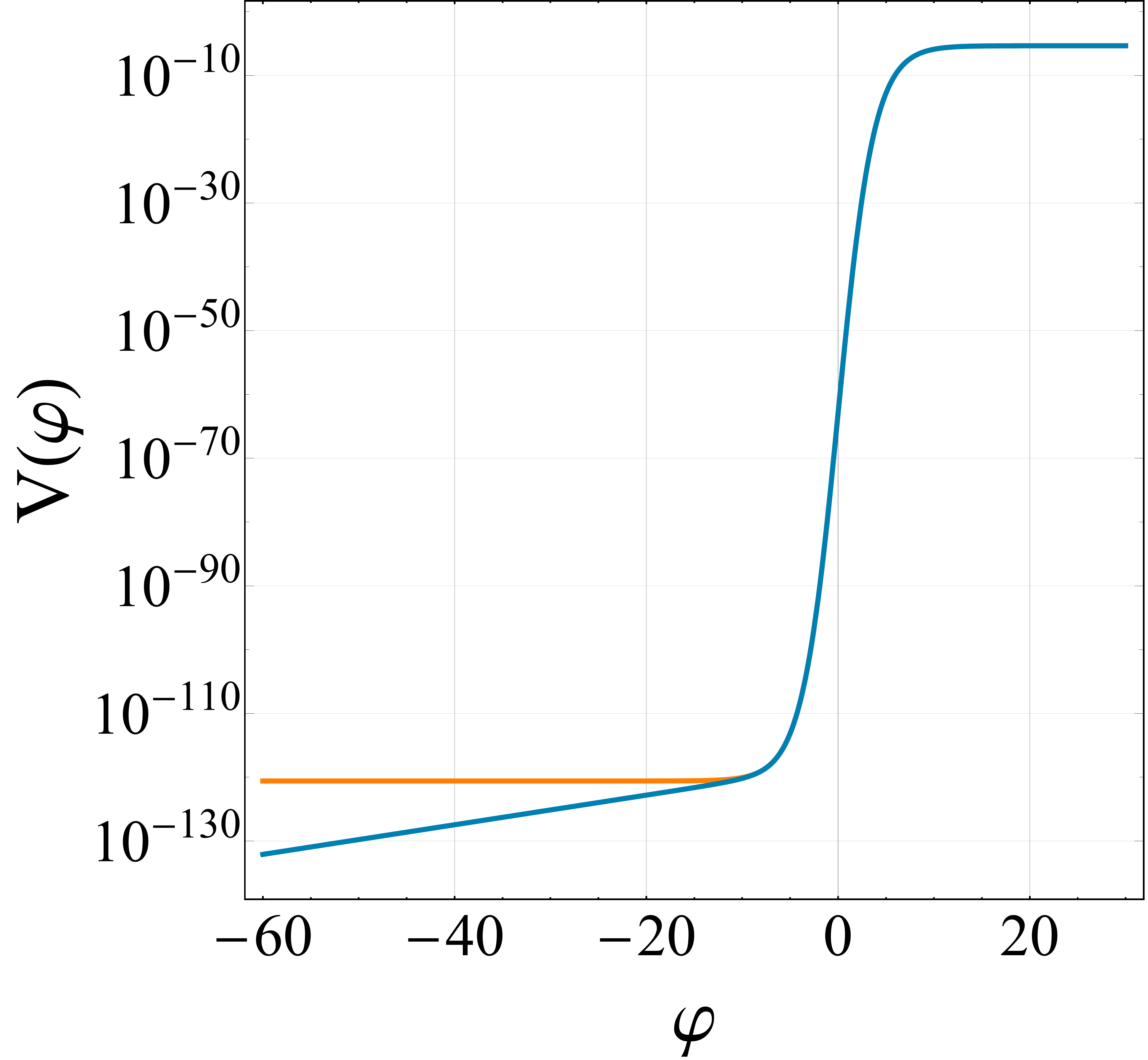}
\caption{\label{fig:model0-model1-shape} \footnotesize The two quintessential inflation models with an exponential potential studied in this work: Exp-model I (orange curve) with the form $M^{2}e^{\gamma(\tanh{\vp\over\sqrt{6\alpha}}-1)}$, and a constant, nonzero asymptotic value for $\varphi\to\--\infty$, and Exp-model II (blue curve) with the form $M^{2} e^{{-2\gamma}} \Big(e^{\gamma\big(\tanh{\vp\over\sqrt{6\alpha}}+1\big)}- 1\Big)$ and a vanishing asymptotic value.}
\end{figure}

Clearly, scenarios with other choices of $V_{0}$ (and the resulting cosmological constant $\Lambda$) are also possible in general, but as we will discuss later, our Exp-models I and II are of particular interest, and capture all the interesting features of the exponential potential. The two potentials for our Exp-models I and II are shown in Fig.~\ref{fig:model0-model1-shape}. Exp-model I (orange curve) has a constant, nonzero asymptotic value for large, negative $\varphi$, while Exp-model II (blue curve) decreases to zero when $\varphi\to\--\infty$.

\begin{figure}
\begin{center}
\includegraphics[scale=0.55]{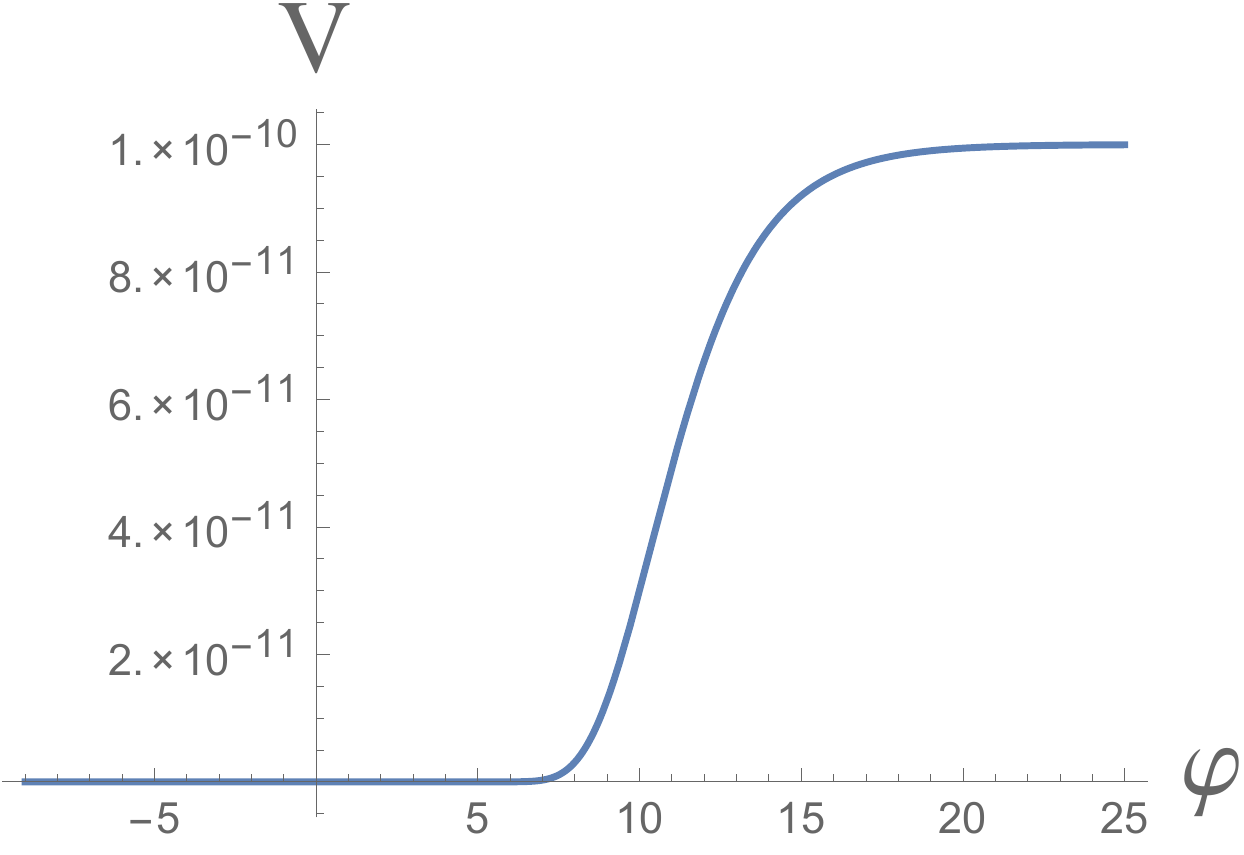}
\end{center}
\caption{\footnotesize The potential $M^{2}e^{\gamma(\tanh{\vp\over\sqrt{6\alpha}}-1)}$ for $\alpha = 7/3$ and $M^{2}= 10^{{-10}}$ in Planckian units.  In this scenario inflation ends at $\vp_{\rm end} \sim 8$, after which the field rapidly falls down and starts the epoch of kinetic energy domination.}
\label{lin1a}
\end{figure}

The figure is shown in logarithmic scale, which is necessary for distinguishing the models of these two types, but this representation hides the steepness of the potential of both of these models at large, positive $\varphi$; see Fig. \ref{lin1a}, where the tiny difference $\sim 10^{{-120}}$ between the two potentials is invisible.

\subsubsection{Inflationary and late-time dynamics}

\begin{figure}[h!]
\center
  \includegraphics[height=5.1cm]{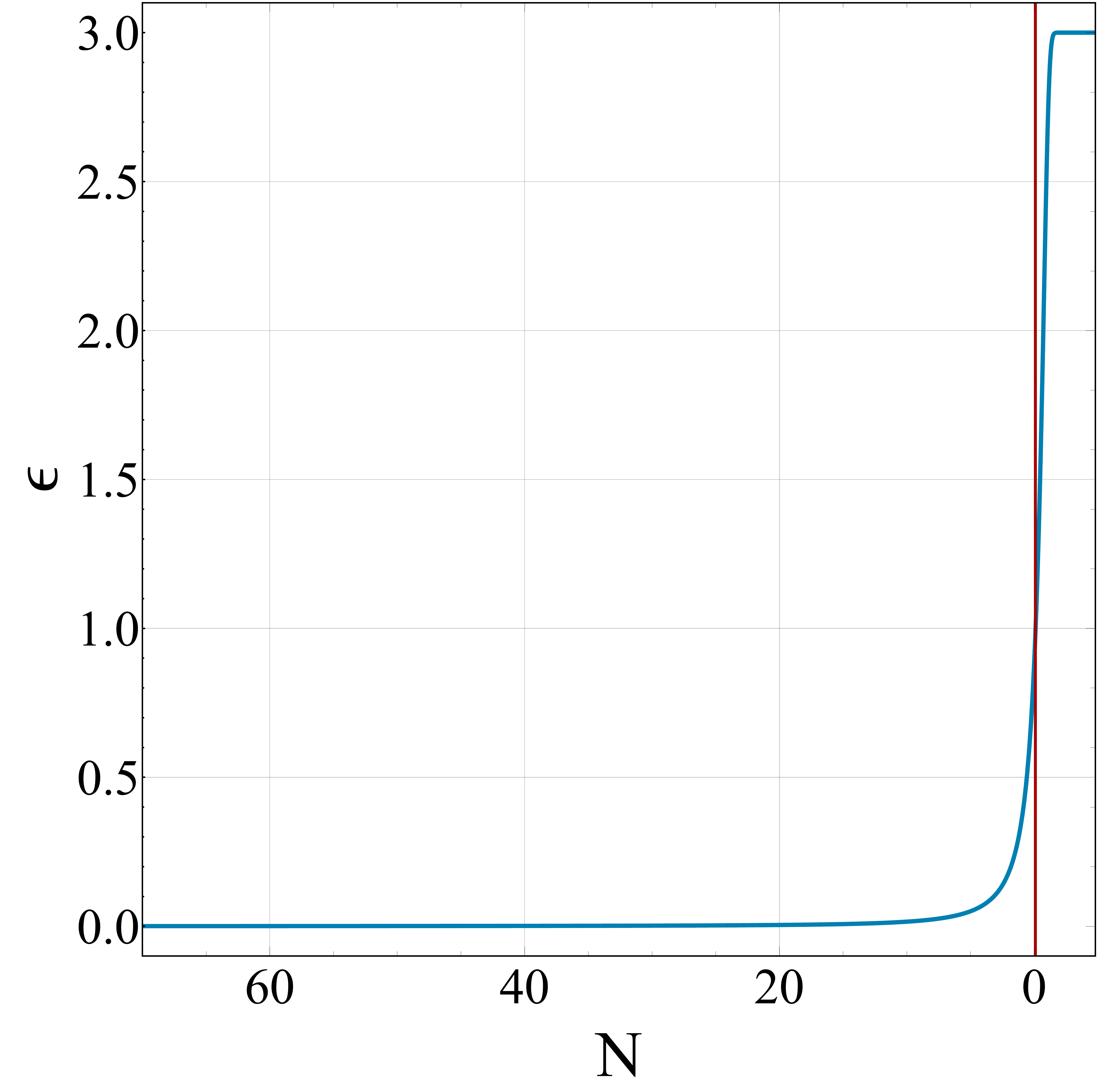}
  \includegraphics[height=5.1cm]{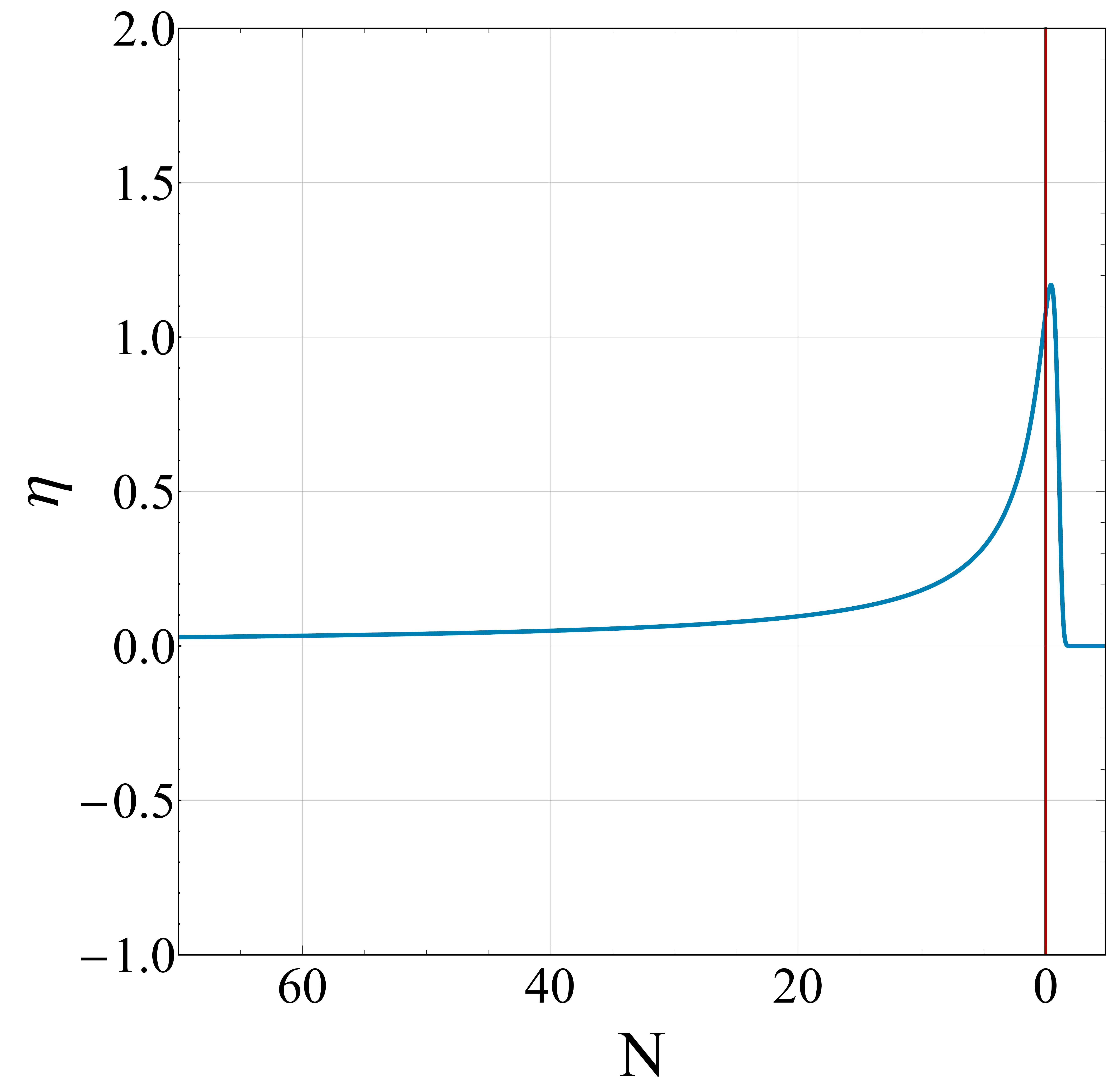}
  \includegraphics[height=5.1cm]{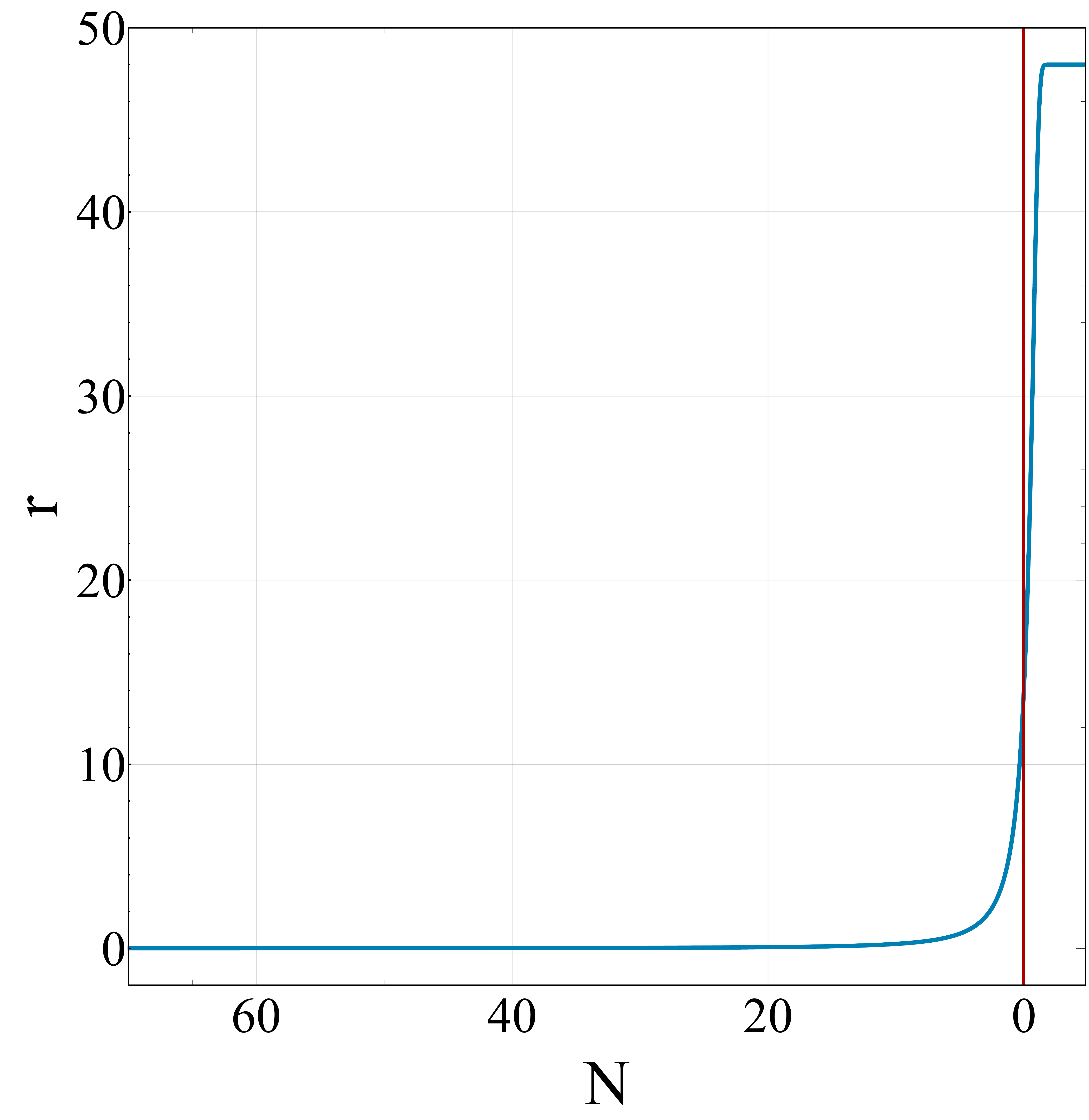}
  \includegraphics[height=5.1cm]{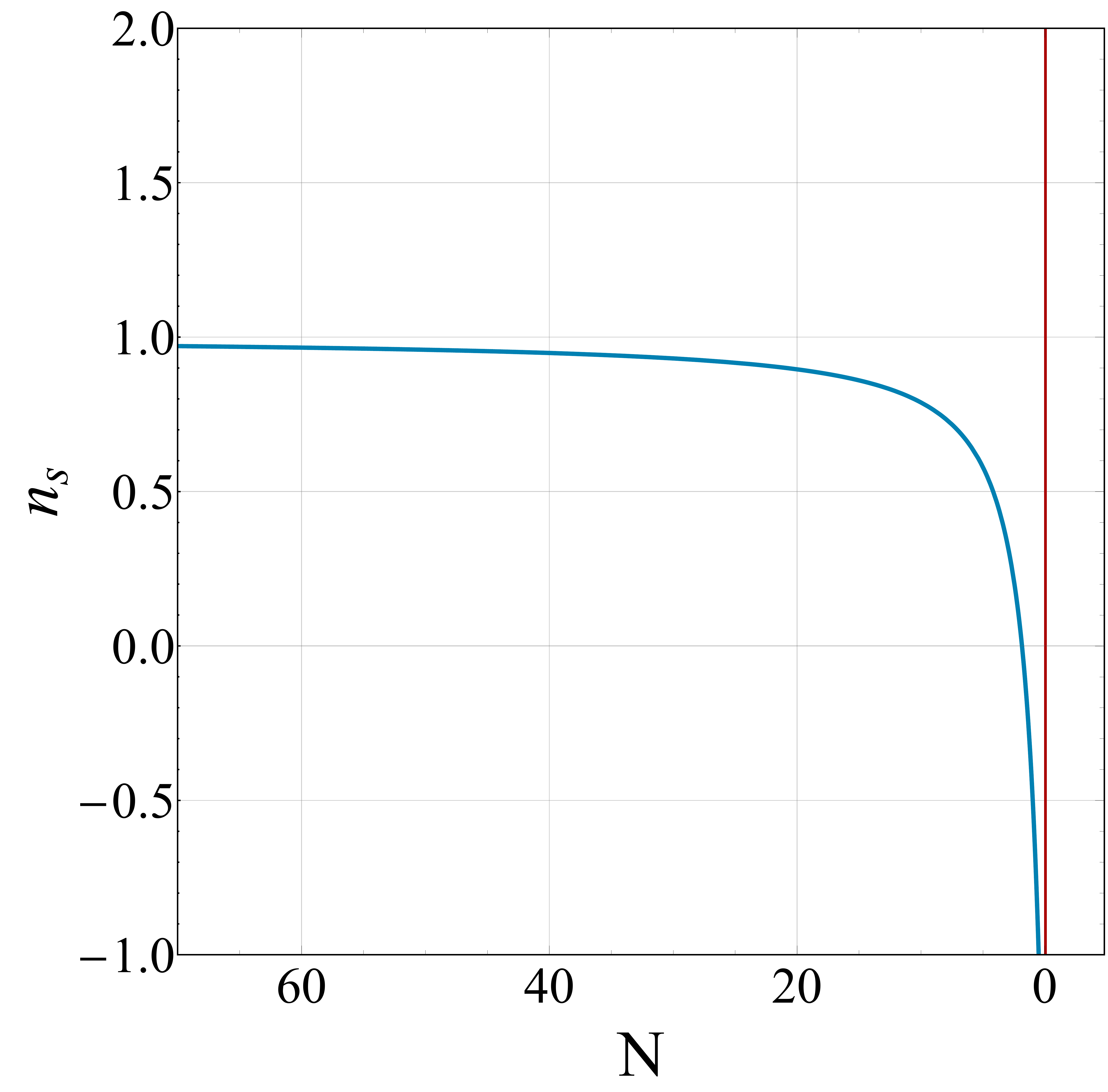}
\caption{\footnotesize\label{fig:model0-inflation} Evolution of the inflationary quantities $\epsilon$, $\eta$, $n_s$, and $r$ as functions of the number of $e$-folds $N$ before the end of inflation for Exp-model II and for a typical set of parameters which give viable late-time cosmological histories. Exp-model I shows similar behavior. In each panel, the red, vertical line depicts the end of inflation (i.e. when $\epsilon$ becomes of $\mathcal{O}(1)$), and $N=63$ corresponds to the moment at which the cosmological scales observed by the CMB experiments had left the horizon. Note that the behavior of $n_s$ and $r$ are correctly shown only during the inflationary period, for which the slow-roll expressions (\ref{eq:nsslowrol}) and (\ref{eq:rslowrol}) hold.}
\end{figure}

Fig.~\ref{fig:model0-inflation} shows an example of the evolution of the inflationary quantities $\epsilon$, $\eta$, $n_s$, and $r$, introduced in section~\ref{sec:evolution}, for Exp-model II and for a typical set of parameters with viable cosmologies. The parameters chosen for the plots are the best-fit ones found through the comparison of the model to the current late-time cosmological observations as described in section~\ref{sec:compdataexp} below. In particular, $\alpha$ has been set to $7/3$. The results for Exp-model I are very similar and we do not present them here. 

In each panel, the red, vertical line shows the end of inflation (i.e. when $\epsilon$ becomes unity), and $N$ is the number of $e$-folds before that, such that the end of inflation is at $N=0$. Both $\epsilon$ and $\eta$ have very small values during the inflationary period. $N\approx 63$ corresponds to the moment at which the cosmological scales observed by the CMB experiments had left the horizon. The duration of the inflationary period depends on the initial conditions for the inflaton field, and must be at least $63$ $e$-folds. In our numerical computations, we have set the initial value of the field such that we obtain much more than $63$ $e$-folds of inflation, but we show only the last $70$ $e$-folds in Fig.~\ref{fig:model0-inflation}. 

Fig.~\ref{fig:model0-inflation} shows that $\epsilon$ at the beginning of the last $63$ $e$-folds has a value very close to zero, and stays almost vanishing for a long period (which is a necessary condition for slow-roll inflation), and then suddenly increases and becomes of $\mathcal{O}(1)$; this ends inflation. The transition of $\epsilon$ from almost zero to $3$ corresponds to a transition from slow roll (where the potential dominates) to a kination period (where the kinetic energy dominates over the potential). This transition is required for inflation to end, and in order to enter a reheating phase.  The second slow-roll parameter, $\eta$, is also small during inflation and becomes of $\mathcal{O}(1)$ at the end of inflation. For both $\epsilon$ and $\eta$ we have computed their exact values over time, i.e. Eqs. (\ref{eq:epsilon1}) and (\ref{eq:eta2}), whereas the {\it slow-roll} values for these two quantities, which can be written in terms of the potential and its derivatives, are valid only during the inflationary period and not in general. The values of $\epsilon$ and $\eta$ measured by the CMB are the ones at $N\sim 63$. We have used the approximate, slow-roll expressions (\ref{eq:nsslowrol}) and (\ref{eq:rslowrol}) for $n_s$ and $r$, which means that their values shown in Fig.~\ref{fig:model0-inflation} are valid for $N\gg \mathcal{O}(1)$. We have checked that the numerical values of $r$ and $n_s$ at  $N= \mathcal{O}(60)$ are in perfect agreement with the approximate, analytical values for $\alpha$-attractor inflation. 

We can also solve the set of Eqs. (\ref{eq:FriedmannLate1})-(\ref{eq:rhoR}) numerically and obtain the cosmic evolution in terms of $H$ for a given set of the free parameters $\Omega_\text{M}$, $\Omega_\text{R}$, $M^{2}$, and $\gamma$. This can then be compared to the cosmological measurements of $H$ and therefore constrain the models. We should however note that one important ingredient in solving the evolution equations is the initial conditions for the field $\varphi$. The initial value of $\vp$ is the freezing value $\vp_\text{F}$ set by the reheating mechanism after inflation, see section~\ref{grpr}.

Let us recap the story. As discussed in section~\ref{grpr}, the field takes positive values during inflation, and rolls down the potential with its value reducing with time and approaching zero. Around this time, and when $\varphi\sim +8$, reheating takes place and matter particles are produced. In case the only reheating at work is gravitational particle production, which is not a very efficient mechanism, the field continues rolling down to values around $-35$ and then freezes. In case other reheating mechanisms, such as instant preheating~\cite{Felder:1998vq,Felder:1999pv,Kofman:2004yc}, are at work in addition to gravitational particle production, reheating will be more efficient and the field will freeze earlier, to values that can be much larger than $-35$; we call this value of the field after reheating $\varphi_\text{F}$, at which $\vp$ is frozen. The field remains frozen at $\varphi_\text{F}$ for sometime after reheating until the Hubble friction becomes so low that the field starts rolling down its potential again. The evolution of the field after reheating and starting from the value $\varphi_\text{F}$ determines the evolution of the universe and cosmic histories at late times, i.e. from radiation domination all the way to the present time.

\begin{figure}[h!]
\center
  \includegraphics[height=5.5cm]{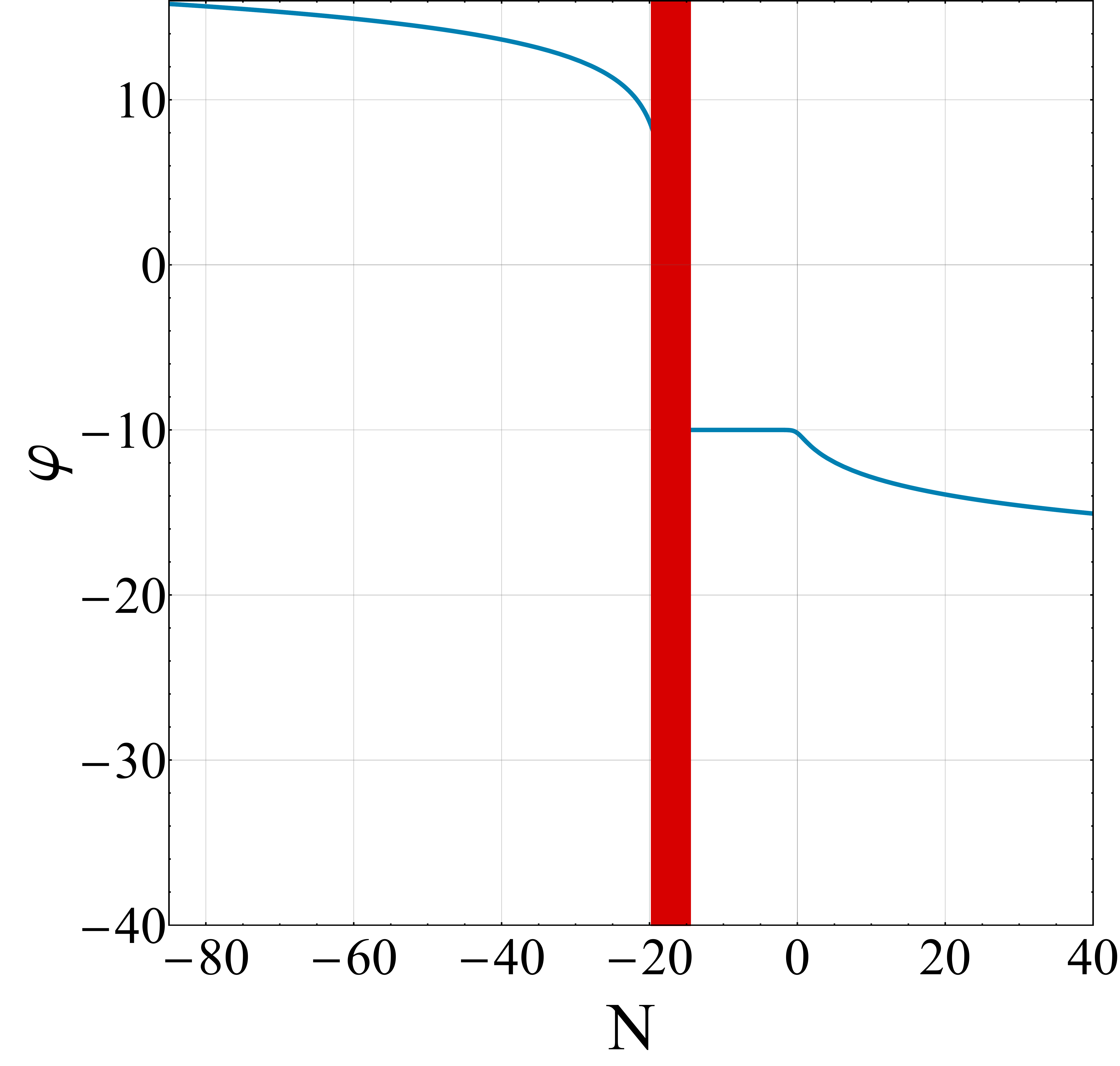}
  \includegraphics[height=5.5cm]{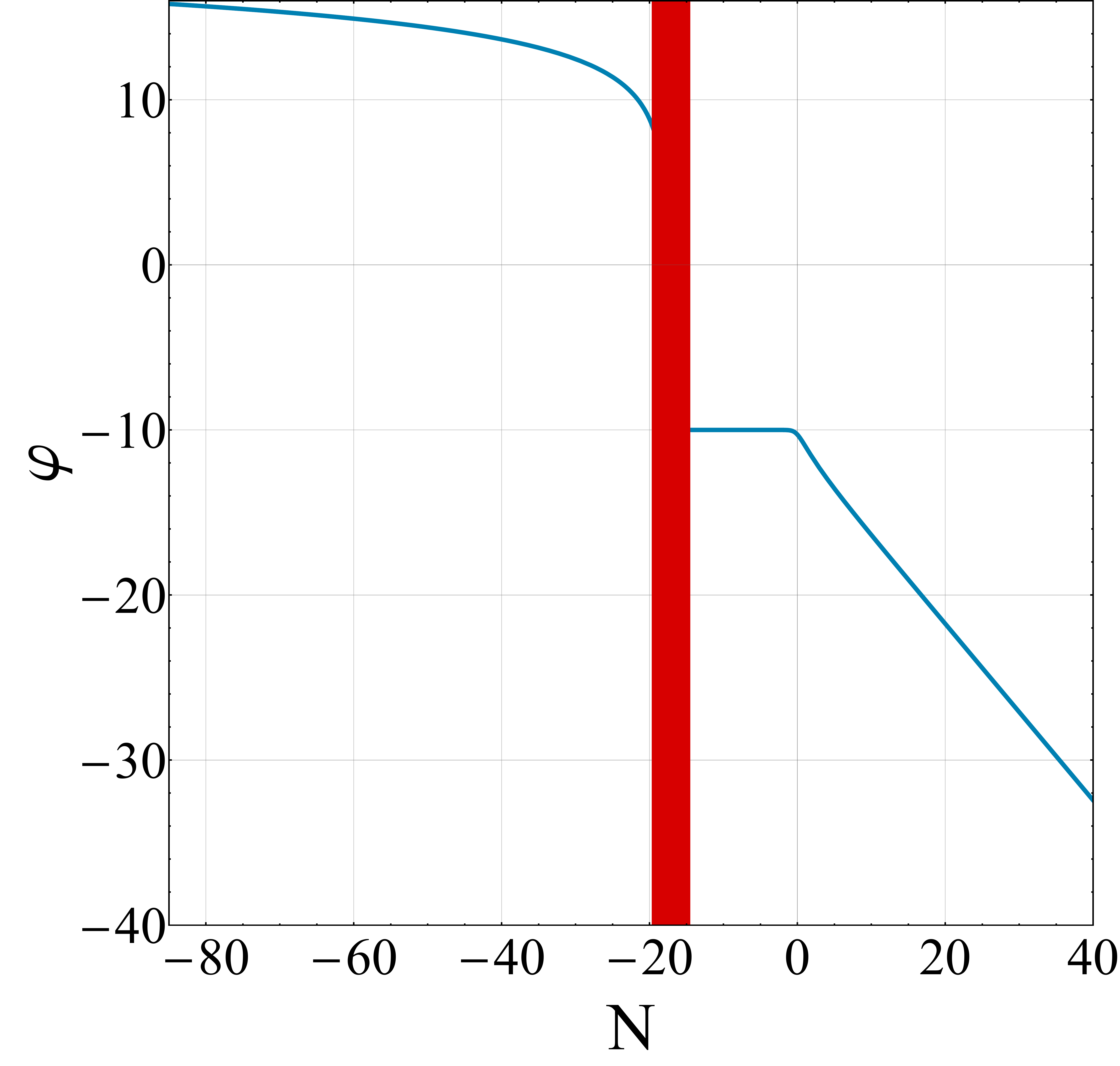}
\caption{\footnotesize\label{fig:model0-model1-phi} {\it Left panel:} Evolution of the scalar field $\varphi$ as a function of the number of $e$-folds $N$ over the entire history of the universe for Exp-model I and for the same set of parameters used for computing the inflationary variables shown in Fig.~\ref{fig:model0-inflation} with a viable late-time cosmological history. The vertical, red bands depict the period after the end of inflation and before the time at which the scalar field freezes, separating the inflationary and late-time periods. This period includes kination and reheating. Note that the field rolls down during inflation and kination (not shown), and then freezes after reheating (to $-10$ in this example), for almost the entire history until very recently when it unfreezes again and starts rolling its potential; this is the onset of dark energy domination. $N=0$ corresponds to the present time. {\it Right panel:} The same as in the left panel, but for Exp-model II. Note the different dynamics for $\varphi$ compared to Exp-model I in the future ($N>0$).}
\end{figure}

Fig.~\ref{fig:model0-model1-phi} depicts an example of the evolution of the scalar field $\varphi$ as a function of the number of $e$-folds $N$ for the entire history of the universe from inflation to late times, for both Exp-models I (left panel) and II (right panel). These have been computed for the same set of parameters as the ones used for computing the inflationary quantities of Fig.~\ref{fig:model0-inflation}, providing viable late-time cosmological histories. We have set the value of $\varphi_\text{F}$ to $-10$ in both cases. 

The vertical, red bands depict the period after the end of inflation and before the time at which the scalar field freezes, separating the inflationary and late-time periods. Note that this period starts with a kination phase, followed by radiation domination, after the occurrence of reheating. Since the exact behavior of the field depends on the details of reheating, we have shown this period of kination plus the start of the radiation domination by a red band. The details of this period are not important for our numerical and statistical analysis later, as long as we have the required information on the initial conditions of the field for our late-time investigation. This boils down to the values of $\varphi_\text{F}$ used in our analysis, which we have ensured to be achievable through our reheating mechanisms. The red bands should therefore be considered only as a sketch for illustrative purposes, while the inflationary evolution and the late-time dynamics shown in Fig.~\ref{fig:model0-model1-phi} are the results of precise numerical computations. The figure shows that the field rolls down its potential during inflation, as well as kination (not shown), and then freezes after reheating, for almost the entire history of the universe, until very recently when it unfreezes again and resumes its rolling down the potential. This unfreezing time is the onset of dark energy domination. Note how the field behaves differently in the future ($N>0$) for the two models.

\begin{figure}[h!]
\center
  \includegraphics[height=4.5cm]{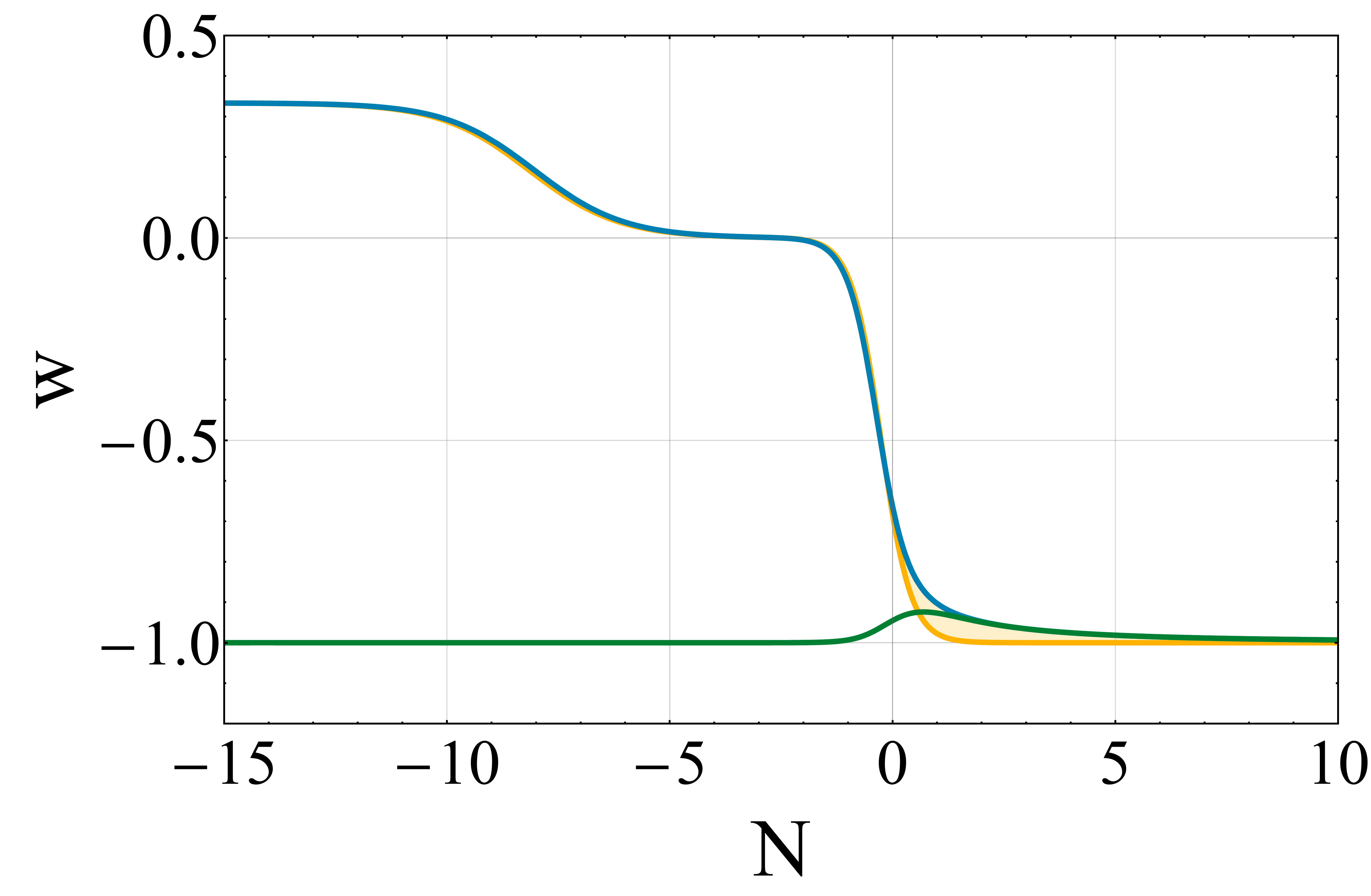}
  \includegraphics[height=4.5cm]{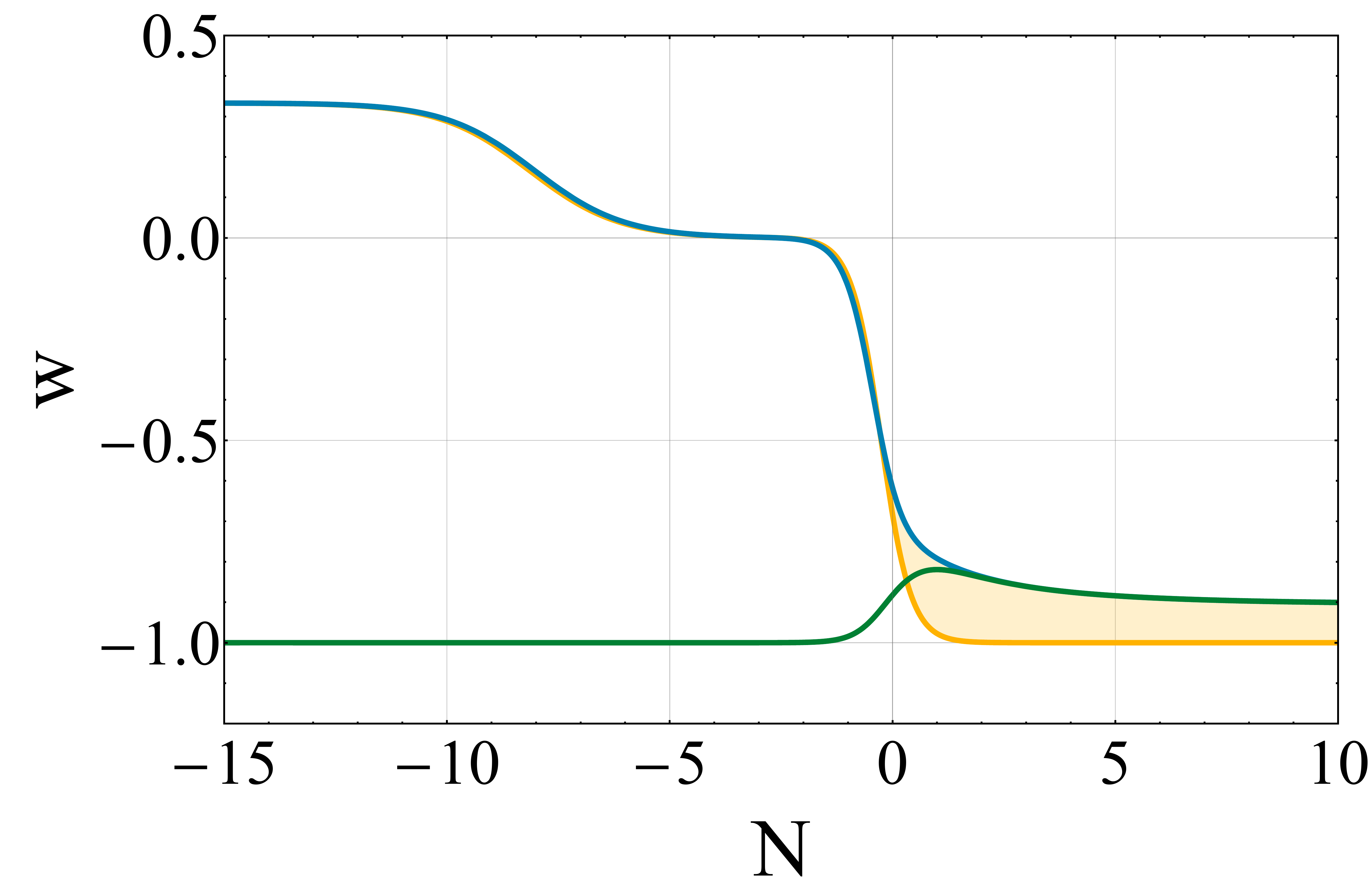}
\caption{\footnotesize\label{fig:model0-model1-w} {\it Left panel:} Evolution of the equation of state as a function of the number of $e$-folds $N$ after reheating for Exp-model I and for the same set of parameters used in Figs.~\ref{fig:model0-inflation} and~\ref{fig:model0-model1-phi} with a viable late-time cosmological history. The blue and green curves show, respectively, the effective equation of state $w_\text{eff}$ and the equation of state of dark energy $w_\text{DE}$. For comparison, the effective equation of state for $\Lambda$CDM is also presented as an orange curve. $N=0$ corresponds to the present time. {\it Right panel:} The same as in the left panel, but for Exp-model II.}
\end{figure}

The evolutions of the effective equation of state $w_\text{eff}$ as well as the equation of state of dark energy $w_\text{DE}$ as functions of the number of $e$-folds $N$ are presented in Fig.~\ref{fig:model0-model1-w} for both Exp-models I (left panel) and II (right panel). The set of parameters used are the same as in Figs.~\ref{fig:model0-inflation} and~\ref{fig:model0-model1-phi} with viable late-time cosmological histories. The blue and green curves depict, respectively, $w_\text{eff}$ and $w_\text{DE}$, and for comparison we have also shown the effective equation of state for the $\Lambda$CDM cosmology (orange curve). $N=0$ corresponds to the present time. For computing these quantities, and for both models, we have again set $\varphi$ to $-10$ and $\varphi^\prime$ to $0$ initially as the initial values of the field and its derivative, respectively. These initial values have been set at $N=-15$, i.e. well inside the radiation domination epoch.

First of all, the figures show that the evolutions of $w_\text{eff}$ for both Exp-models I and II closely follow the one for $\Lambda$CDM in the past, while there are deviations in the future ($N>0$). $w_\text{eff}$ for Exp-model I approaches $-1$ asymptotically (when $N\to+\infty$), just as in $\Lambda$CDM, while its asymptotic value in Exp-model II differs from $-1$. This is expected, as the potential of Exp-model I effectively contains a constant piece $M^{2}e^{-2\gamma}$ which becomes dominant far in the future. This constant piece acts like a cosmological constant, making the dark energy equation of state effectively like that of $\Lambda$, i.e. $w_{\infty} = -1$. 

For Exp-model II, however, the constant piece $M^{2}e^{-2\gamma}$ has been removed by setting $V_{0}$ to the nonzero and negative value $-M^{2}e^{-2\gamma}$, therefore the asymptotic value of $w_\text{DE}$ is no longer $-1$. As we discussed earlier in a related contex, this asymptotic value $w_{\infty}$ for Exp-model II is
\be
w_{\infty}= -1+\frac{2}{3} \, \frac{1}{3\alpha} \,,
\ee 
which is a universal result that does not depend on the values of $M^2$ and $\gamma$; it depends only on the value of $\alpha$. It is this interesting situation discussed in the introduction where one geometric parameter $\alpha$ defines the deviation of $w_{\infty}$ from $-1$, as well as the level of primordial gravity waves from inflation, see Eqs. \rf{cute} and \rf{cute1}.

Another interesting observation in Fig.~\ref{fig:model0-model1-w} is the behavior of the dark energy equation of state $w_\text{DE}$, shown by green curves for both models. Clearly, in both cases, $w_\text{DE}$ today deviates from the equation of state for $\Lambda$, i.e. $-1$, and is also different from its asymptotic value $w_{\infty}$ in the case of Exp-model II.\footnote{These models fall in the class of thawing dark energy models, which have very consistent properties, see e.g. Ref.~\cite{Linder:2015zxa}. Unlike the standard exponential dark energy model with an early-time tracing behavior with $w_\text{DE}=w_\text{M}$ in the high-redshift matter dominated era, here $w_\text{DE}=-1$ rather than $w_\text{DE}=0$ at $z \sim 3-3000$. We thank Eric Linder for pointing this out to us.} We will discuss this in more detail in the next section.

\subsubsection{Comparison to observations, and constraints on parameters}\label{sec:compdataexp}

We perform a statistical analysis of Exp-models I and II in order to understand whether the models are cosmologically viable, how much their parameters are constrained by cosmological observations, and to which extent we expect deviations from the standard model. This will also tell us whether the models can be distinguished from $\Lambda$CDM using the current and upcoming cosmological surveys. For that, as mentioned in section~\ref{sec:intro}, we consider geometrical constraints on the cosmic history at the background level using a combination of the redshift-luminosity relation of supernovae~\cite{Betoule:2014frx}, the observed angular scales of the CMB anisotropies~\cite{Ade:2015xua}, measurements of the baryon acoustic oscillations (BAO)~\cite{Beutler:2011hx,Blake:2011en,Anderson:2012sa,Anderson:2013zyy,Ross:2014qpa}, and the local measurements of the Hubble constant $H_{0}$~\cite{Riess:2016jrr}. 

Our aim in the present work is not an exhaustive and detailed comparison of the models to observations, and the primary goal is to reach a qualitative understanding of the models, their cosmological viability, and their differences in terms of the observational implications. Additionally, contrary to models of modified gravity for cosmic acceleration, minimally coupled quintessence models (including ours) affect observations only through their impacts on the background dynamics, and they do not directly affect clustering and growth of structure as well as other LSS observables such as weak lensing. For these reasons we believe that the geometrical measurements of the cosmic history on their own should provide sufficiently good constraints on our models; we leave an extensive and detailed analysis of the models using all the available cosmological observations, including those involving the constraints from the full CMB temperature and polarization power spectra, as well as galaxy clustering and weak lensing, for future work where a perturbative analysis of the models will be performed and the models will be implemented in a numerical Boltzmann code. Additionally, here we do not perform detailed forecasts for future galaxy surveys using for example a Fisher matrix approach. Our aim here is rather to obtain a relatively good estimate of the predictions of the models, for example through the CPL parameters $w_{0}$ and $w_{a}$, and to check whether the models have the potential of being probed or ruled out by the future surveys; we leave a detailed forecast analysis of the models also for future work. 

Here, therefore, we use only a simple and rough criterion for a model to be testable against $\Lambda$CDM: We assume a point in the parameter space of the model to be distinguishable from $\Lambda$CDM if the corresponding $w_0$ and $w_{a}$ are different from the $\Lambda$CDM values of $-1$ and $0$ more than $\sim 2\%$ and $\sim 4\%$, respectively. These numbers are clearly only rough estimates, and can be different depending on the specific experiments and probes that are being considered. However, we believe that they are good (and perhaps optimistic) estimates of what one will be able to reach using the combination of various probes from the upcoming Stage IV large-scale structure surveys and CMB experiments; see e.g. Ref.~\cite{Laureijs:2011gra} for the values that are targets of one of these experiments. In addition, the situation is more subtle than using only the separate errors on $w_{0}$ and $w_{a}$, for example because of possible correlations between the two parameters --- in fact a more proper way of using these errors is through the 2-dimensional confidence contours for $w_0$ and $w_{a}$. However, since we do not intend to perform a detailed statistical analysis in this paper, and are concerned more with a qualitative analysis of the models, we leave these subtle issues to be addressed in future work.

Before we present and discuss the results of our statistical analysis based on the cosmological data described above, let us use the expression (\ref{eq:COBEgen}) for the COBE/Planck normalization discussed in section~\ref{sec:evolution} and see which constraints we can obtain on the values of the parameters in our potentials solely from early-time (inflationary) physics. We will shortly see that the COBE/Planck normalization indeed provides us with an approximate but a quite strong constraint on the two potentials $M^{2}e^{\gamma(\tanh{\vp\over\sqrt{6\alpha}}-1)}$ and $M^{2} e^{{-2\gamma}} \Big(e^{\gamma\big(\tanh{\vp\over\sqrt{6\alpha}}+1\big)}- 1\Big)$, for Exp-models I and II.

We should first note that on the tails of the potentials for large and positive $\varphi$, where we assume inflation to take place, the form of the effective potential is approximated by the expression
\begin{eqnarray}\label{eq:approxpotinf}
V(\varphi) = M^{2}(1 - 2 \gamma e^{-\frac{2\varphi}{\sqrt{6 \alpha}}}) +V_{0} + \mathcal{O}(e^{-\frac{4\varphi}{\sqrt{6 \alpha}}}) \,,
\end{eqnarray}
where we have left the cosmological constant undetermined --- again setting $V_{0}$ to $0$ and $-M^{2}e^{-2\gamma}$ gives our Exp-models I and II, respectively, as discussed above. Note that even for Exp-model II with a nonvanishing $V_{0}$, its contribution $M^{2}e^{-2\gamma}$ to the potential (\ref{eq:approxpotinf}) is exponentially small compared to the leading term $M^{2}$, by a factor of $e^{-2\gamma}$. We will see later that we need $\gamma$ to be $\sim 125$ in order to obtain viable cosmic histories for both models, and therefore the contribution from $V_{0}$ to the inflationary potential (\ref{eq:approxpotinf}) is negligible and we can ignore it.

Now let us integrate Eq.~(\ref{eq:phievolvSR}) over an arbitrary interval $[N_{1},N_{2}]$ during the inflationary epoch, 
\begin{eqnarray} 
\int_{\varphi _1}^{\varphi _2} \frac{V(\varphi)}{V_{\varphi}(\varphi)} \dd\varphi = - \int_{N_1}^{N_2}\dd N \label{eq:phiint}\,,
\end{eqnarray}
where $\varphi_{1}$ and $\varphi_{2}$ are the values of the field at $N_{1}$ and $N_{2}$, respectively.
Assuming that both $\varphi_1$ and $\varphi_2$ are sufficiently large, we can use the approximate expression~(\ref{eq:approxpotinf}) and arrive at
\begin{eqnarray} 
\frac{\sqrt{6 \alpha}}{4 \gamma}\Big(\frac{\sqrt{6 \alpha}}{2}\big(e^{\frac{2\varphi _2}{\sqrt{6 \alpha}}} - e^{\frac{2\varphi _1}{\sqrt{6 \alpha}}} \bigr)  -2\gamma(\varphi _2 - \varphi _1)  \Bigr) = N_1 - N_2 \label{eq:phiNcomplete}\,. 
\end{eqnarray}
Now, choosing $N_1$ to be the moment of horizon crossing $N_{\text{crossing}}$ for the observable modes and $N_2$ to correspond to the end of inflation $N_{\text{end}}$ we arrive at the approximate expression
\begin{eqnarray}
e^{\frac{2\varphi_*}{\sqrt{6 \alpha}}} = \frac{4}{3\alpha}\gamma N \label{eq:phiN}\,,
\end{eqnarray}
where $\varphi_*$ is the value of the field at the horizon crossing, and $N\equiv N_{\text{end}}-N_{\text{crossing}}$ is the number of $e$-folds corresponding to the duration of inflation since the moment at which the observable perturbations left the horizon until the end of inflation. In order to obtain Eq.~(\ref{eq:phiN}) we have assumed that the field has travelled at least a few Planck units between the horizon crossing and the end of inflation, and therefore the term proportional to $e^{\frac{2\varphi _1}{\sqrt{6 \alpha}}}$ on the left-hand side of Eq.~(\ref{eq:phiNcomplete}) is the dominant one; we ignore all the other terms. For $\gamma\sim 125$, and assuming $N\approx 63$, Eq.~(\ref{eq:phiN}) gives $\varphi_{*}\sim 15.74$ for $\alpha=7/3$, which is in full agreement with our numerical analysis; note that $\varphi_\text{end}\sim +8$.

Let us now plug the asymptotic expression for our potential (\ref{eq:approxpotinf}) into the COBE/Planck normalization equation (\ref{eq:COBEgen}). Using Eq. (\ref{eq:phiN}) we arrive at
\begin{eqnarray}
M^{2} = \frac{144\pi^2\alpha N}{(2N-3\alpha)^3}\,    \mathcal{P}_{\mathcal{R}}(k) \label{eq:COBEexpXX}\,.
\end{eqnarray}
Taking into account that  $V_{+} \approx M^{2}$ and considering the limit $N \gg \alpha$, we see that  this equation reproduces the previously mentioned general $\alpha$-attractor result  \rf{eq:COBEexp2}. 
 
Thus the COBE/Planck normalization constrains the ratio ${M^{2}/\alpha}$ \cite{Kallosh:2015lwa}. Assuming  $N\approx 63$, using \rf{eq:COBEexpXX}, and applying the measured value of $\mathcal{P}_{\mathcal{R}}$, we find that
\begin{eqnarray}
{M^{2}\over \alpha} \sim 10^{-10}  \label{eq:COBEconst}\,.
\end{eqnarray}
This means that this early-universe condition does not constrain $M^2$ and/or $\alpha$ separately, and the two parameters are degenerate as far as the COBE/Planck normalization is concerned. We will see later that this degeneracy will be broken when the late-time cosmological data are used. 
 
\begin{figure}[h!]
\center
  \includegraphics[height=5cm]{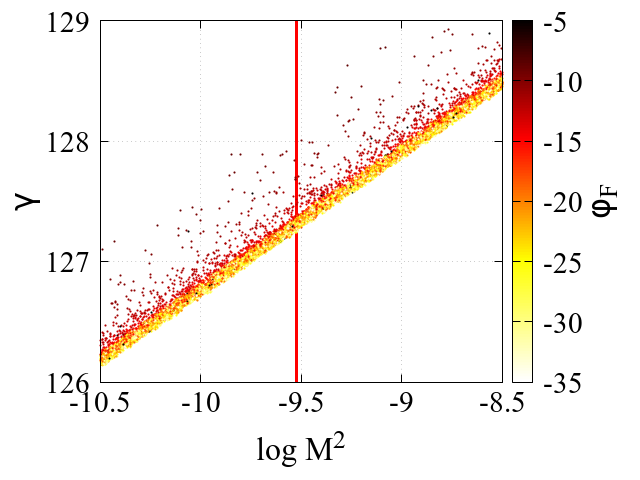}
  \includegraphics[height=5cm]{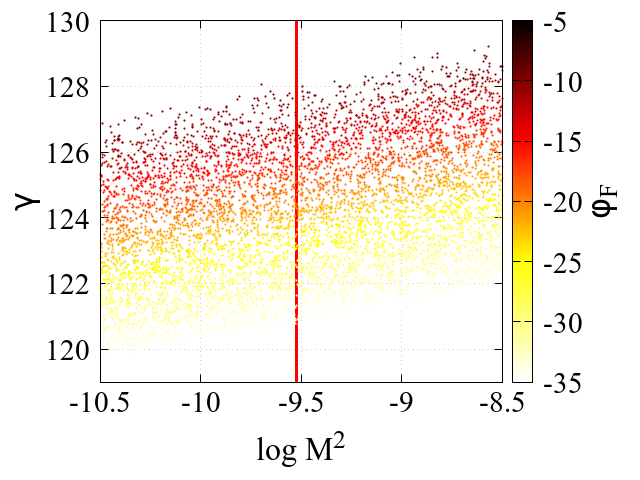}
  \includegraphics[height=5cm]{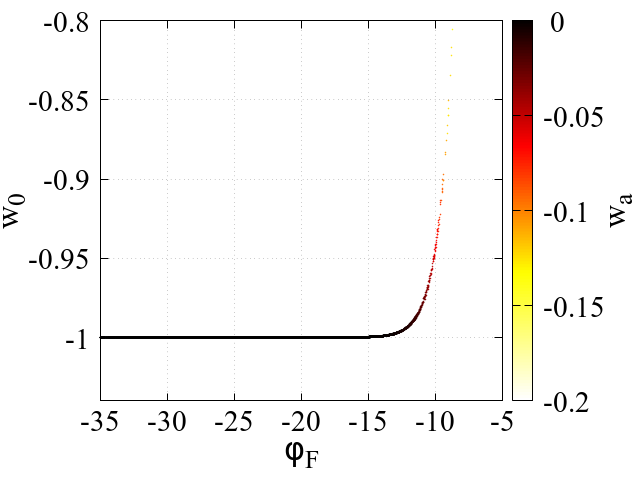}
  \includegraphics[height=5cm]{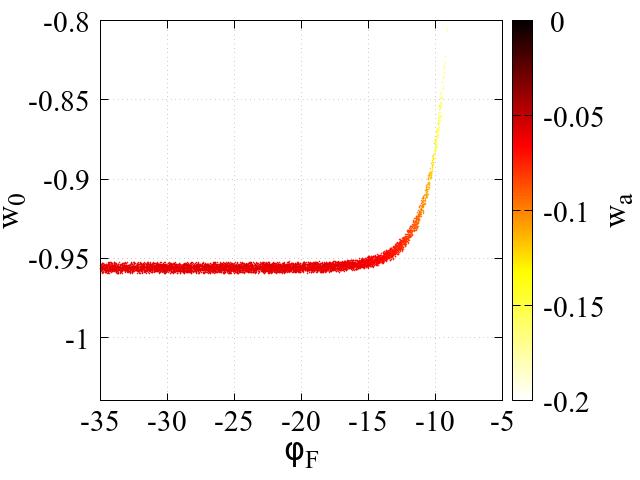}  
\caption{\footnotesize\label{fig:model0-model1-w0-wa-cobe}  {\it Upper panels: } Cosmological constraints on $\log M^{2}$ and $\gamma$ for Exp-model I (left panel) and Exp-model II (right panel) in term of $\varphi_\text{F}$, when it is allowed to vary between $-35$ and $+8$. $\log M^{2}$ has been scanned over only in a range around the COBE/Planck normalization value depicted by the vertical, red lines. {\it Lower panels:} CPL parameters $w_0$ and $w_{a}$ for the dark energy equation of state, for Exp-models I (left panel) and II (right panel) as functions of $\varphi_\text{F}$. The points cluster around $w_0=-1$ (model I) and $w_0\sim-0.96$ (model II) for large, negative values of $\varphi_\text{F}$.}
\end{figure}

Let us first focus on $\alpha=7/3$, which is an interesting case. In that case
$M^{2} \sim 3\times 10^{-10}.$
We will later discuss the dependence of our results on $\alpha$, as well as the constraints on $\alpha$ itself. We first scan over all the free parameters of Exp-models I and II, i.e. $M^{2}$, $\gamma$, $\Omega_\text{M}$, and $\Omega_\text{R}$, as well as the initial value of the field, $\varphi_\text{F}$, comparing the models to the (late-time) cosmological observations described above. Note that although we do not impose the COBE/Planck constraint in our numerical scans, we scan over a range of $\log M^{2}$ around the value given by the COBE/Planck normalization (\ref{eq:COBEconst}). Additionally, as we argued before, we expect $\varphi_\text{F}$ for this potential to be in the range $[-35, +8]$, depending on the reheating mechanism --- this is the range we choose for our numerical analysis. We will see, however, that because of the steepness of the potential for large values of $\varphi_\text{F}$, the effective, viable range for $\varphi_\text{F}$ will be $\sim[-35, -5]$. With all these, we scan over the parameters and compare the cosmic histories with observations. Fig. \ref{fig:model0-model1-w0-wa-cobe} shows the obtained constraints on $\log M^{2}$ and $\gamma$ (upper panels), as well as on the two CPL parameters $w_{0}$ and $w_{a}$ (lower panels) introduced in Eq. (\ref{eq:CPL}). The color assigned to each point corresponds to the value of $\varphi_\text{F}$ and $w_a$ for the upper and lower panels, respectively, and the vertical, red lines depict the value of $\log M^{2}$ given by the COBE/Planck constraint. The figure shows that the constraints on $\gamma$ are quite tight for Exp-model I (left) compared to Exp-model II (right).

We first focus on Exp-model II, which gives a wider region for $\gamma$. The color clearly shows that lower values of $\gamma$ correspond to larger $|\varphi_\text{F}|$. The cut from below comes therefore from the fact that we imposed an upper bound of $35$ on $|\varphi_\text{F}|$ in our scans,  i.e. we did not allow $\varphi_\text{F}$ to become smaller than $-35$ due to gravitational reheating. (This means that in principle there would be no lower bound on $\gamma$ if $|\varphi_\text{F}|$  were allowed   to take arbitrarily large values.) The upper bound on $\gamma$, on the other hand, comes from the fact that if the field does not sufficiently roll down its potential after inflation and before freezing, the model will not provide a viable cosmic history.

Focusing now on the left, upper panel in Fig. \ref{fig:model0-model1-w0-wa-cobe} for Exp-model I, we see that the lower bound on $\gamma$, for a given value of $\log M^{2}$, seems to be highly strict and even increasing $|\varphi_\text{F}|$ will not decrease $\gamma$. This can be understood if we remember again that Exp-model I possesses a cosmological constant limit. Increasing $|\varphi_\text{F}|$ moves the field more and more on the tail of the potential, and the model becomes more and more like $\Lambda$CDM. There is however no possibility of decreasing the total potential energy of the field further, as the scalar field only contributes with a positive energy on top of the cosmological constant. That is why there is a lower bound on $\gamma$ for Exp-model I in Fig. \ref{fig:model0-model1-w0-wa-cobe}. $\varphi_\text{F}$ can however take values as large as $\sim+5$, as in Exp-model II, giving larger deviations from $\Lambda$CDM.

The lower panels of Fig. \ref{fig:model0-model1-w0-wa-cobe} show how the CPL parameters $w_{0}$ and $w_{a}$ vary with $\varphi_\text{F}$ in both models.  First note that the viability regions are quite thin, and already tight as far as the constraints from the cosmological data are concerned. We have checked that by imposing the full COBE/Planck constraint (\ref{eq:COBEconst}) these regions become only slightly thiner, which means that the late-time data are quite constraining on their own, independently of the strong constraint on the model imposed by the COBE/Planck normalization --- see Appendix~\ref{sec:appendix} for a discussion of how these constraints are affected if the inflationary priors on $M^2$ are fully relaxed.   Second, we can clearly see that the models deviate more and more from $\Lambda$CDM by increasing $\vp_\text{F}$ to less and less negative values, as illustrated by the deviations in $w_0$ and $w_{a}$ from $-1$ and $0$, respectively. Note that all the points shown in Fig.~\ref{fig:model0-model1-w0-wa-cobe} are cosmologically viable, and therefore, by having a sufficiently efficient reheating to stop the field from rolling too much after inflation, we can expect a relatively large deviation from $\Lambda$CDM, detectable by future cosmological surveys.  The deviations are already quite large around $\vp_\text{F}=-8$ so that we do not obtain viable cosmologies for larger values of $\vp_\text{F}$. In addition, it is important to note that for Exp-model II, the model does not predict the asymptotic value of $w_{\infty}=-1+\frac{2}{9\alpha}$ ($\sim -0.9$ in this case for $\alpha=7/3$) for the present value of the dark energy equation of state. The closest value to $w_{\infty}$ it can reach is $\sim -0.96$ for large, negative $\vp_\text{F}$, and deviates more and more from it when $\vp_\text{F}$ increases.

Let us now restrict ourselves to specific values of $\varphi_\text{F}$ to see how much deviation from $\Lambda$CDM we can expect for Exp-models I and II by decreasing $|\varphi_\text{F}|$. This is interesting because specific, observed deviations from $w_{0}=-1$ and $w_{a}=0$ may constrain the initial value of the field after reheating, and therefore in turn constrain the reheating mechanism itself within the framework of these models.

\begin{figure}[h!]
\center
  \includegraphics[height=5cm]{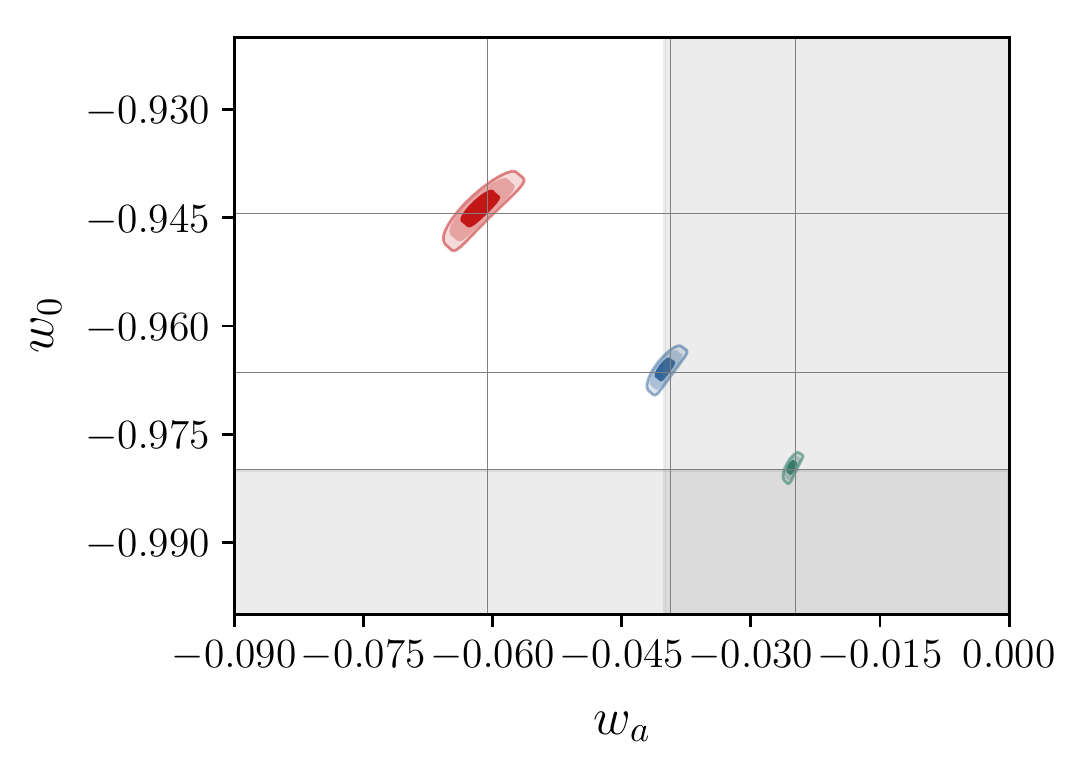}
  \includegraphics[height=5cm]{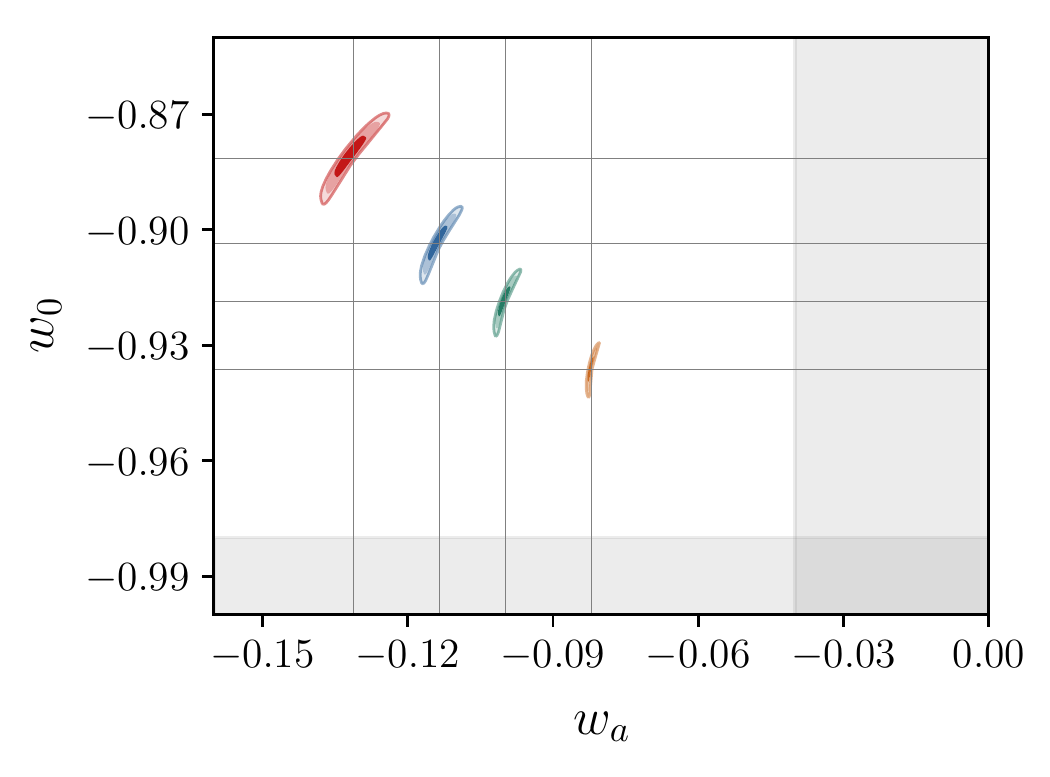}
\caption{\footnotesize\label{fig:model1-w0-wa-cobe}  {\it Left panel:} Constraints on $w_0$ and $w_{a}$ for Exp-model I, and for three cases of $\varphi_\text{F}=-10$ (red contours), $\varphi_\text{F}=-10.5$ (blue contours), and $\varphi_\text{F}=-11$ (green contours). The shaded, grey regions indicate a rough estimate of the target sensitivity for Stage IV large-scale structure surveys in combination with CMB experiments, expected to detect deviations of up to $\sim 2\%$ and $\sim 4\%$ in $w_{0}$ and $w_{a}$, respectively, from the $\Lambda$CDM values. {\it Right panel:} The same as in the left panel, but for Exp-model II. Here, red, blue, green, and orange contours correspond to $\varphi_\text{F}=-10$, $\varphi_\text{F}=-10.5$, $\varphi_\text{F}=-11$, and $\varphi_\text{F}=-12$, respectively. Note that all these cases for Exp-model II show detectable deviations from $\Lambda$CDM.} 
\end{figure}

The left panel of Fig.~\ref{fig:model1-w0-wa-cobe} shows the results of our scans of Exp-model I when $\varphi_\text{F}$ has been fixed to three values $-10$ (red contours), $-10.5$ (blue contours), and $-11$ (green contours). Each set of contours shows $1\sigma$, $2\sigma$, and $3\sigma$ confidence regions. The shaded, grey regions indicate the planned sensitivity of the upcoming Stage IV large-scale structure surveys in combination with the CMB measurements, which are expected to detect deviations of up to $\sim 2\%$ and $\sim 4\%$ in $w_{0}$ and $w_{a}$, respectively, from the $\Lambda$CDM values; see the discussion earlier in this section.  

We first notice that the three sets of contours are extremely tight and $w_0$ and $w_{a}$ are strongly constrained, even though $M^2$ in these numerical scans is not set to the exact COBE/Planck normalization value, and the range is relatively large. The constraints are already quite strong that even though constraining $M^2$ to the COBE/Planck-normalization value makes the contours even smaller, it will not affect the results significantly --- see Appendix~\ref{sec:appendix} for cases where no inflationary constraints are imposed on the mass scale $M$. Our results show that $|\varphi_\text{F}|$ of around $10$ or smaller will be detectable by future LSS experiments. It is also interesting to note that the changes in $w_0$ and $w_{a}$ are highly sensitive to the value of $\varphi_\text{F}$; we do not expect to detect any deviations from $\Lambda$CDM for $|\varphi_\text{F}|$ larger than $\sim10.5$ in Exp-model I using the next generation of the LSS surveys. Our analysis also shows that for values smaller than $\sim10$, on the other hand, it becomes difficult to obtain viable late-time cosmologies. 

The right panel of Fig.~\ref{fig:model1-w0-wa-cobe} shows the same as in the left panel, but for Exp-model II, where red, blue, green, and orange contours correspond to $-10$, $-10.5$, $-11$, and $-12$ for $\varphi_\text{F}$, respectively. The deviations from $\Lambda$CDM in this model are generically larger compared to Exp-model I, and are therefore more easily detectable by upcoming surveys; note how all four contours are located outside the shaded, grey regions.

\begin{figure}
\center
  \includegraphics[height=4cm]{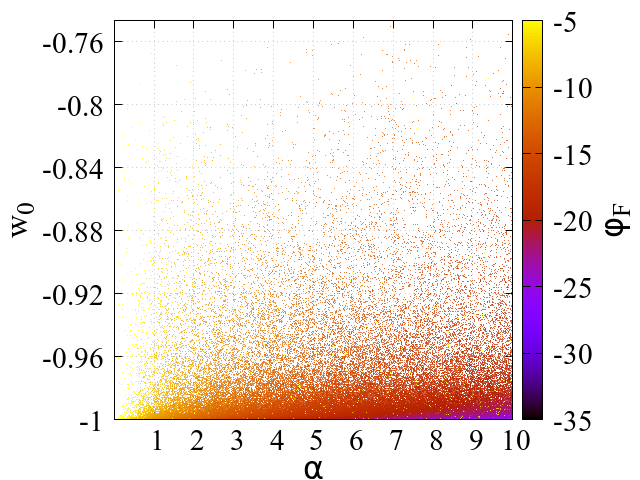}
  \includegraphics[height=4cm]{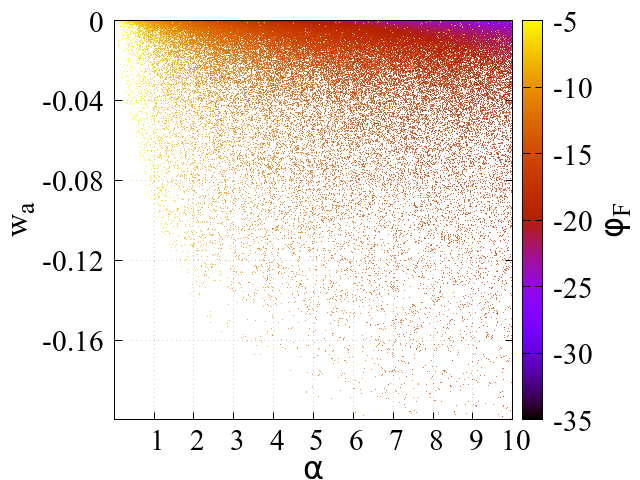}
  \includegraphics[height=4cm]{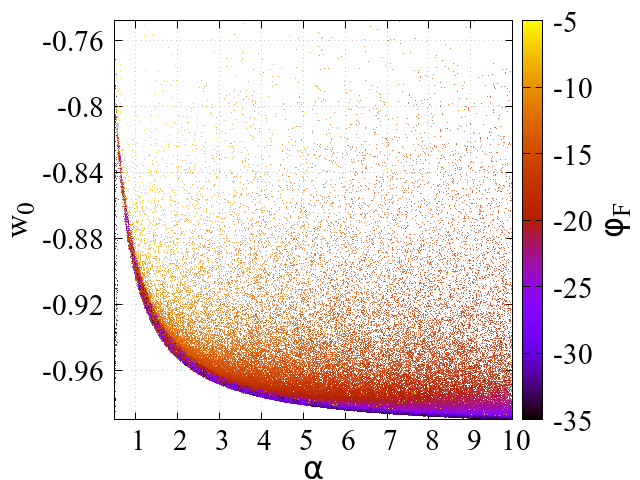}
  \includegraphics[height=4cm]{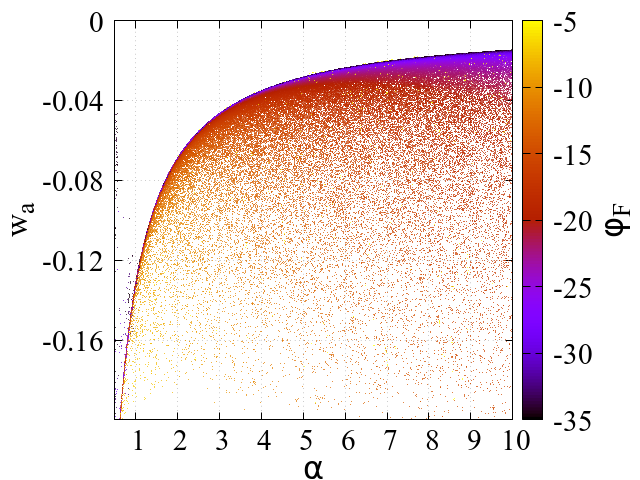}
\caption{\footnotesize\label{fig:model0-model1-w0-wa-alpha-phirehfree} {\it Upper left:} Present value of the dark energy equation of state, $w_{0}$, versus $\alpha$ for Exp-model I. Different colors show different values of $\varphi_\text{F}$.  Here the COBE/Planck normalization has been imposed and the values of $M^2$ are fully fixed in terms of $\alpha$ as $M^{2} \approx 10^{-10} \alpha$.  {\it Upper right:} The same as in the upper left panel but for the CPL parameter $w_{a}$. {\it Lower panels:} The same as in the upper panels but for Exp-model II.}
\end{figure}

Finally, we study the effects of varying $\alpha$ in our two exponential models I and II, by leaving it as a free parameter and scanning over it together with the other parameters of our models, including $\varphi_\text{F}$. In order to include the constraints imposed by the COBE/Planck normalization, here we use (\ref{eq:COBEconst}) and fix the value of $M^2$ in terms of $\alpha$ as $M^{2} \approx 10^{-10} \alpha$. This means that $M^2$ is no longer a free parameter in our numerical scans, and is fully fixed by $\alpha$ --- see Appendix~\ref{sec:appendix} for a discussion of the results when the COBE/Planck constraint is fully relaxed.  Fig.~\ref{fig:model0-model1-w0-wa-alpha-phirehfree} shows the results of the comparison to data for Exp-model I (upper panels) and Exp-model II (lower panels), and for the two CPL parameters $w_{0}$ (left panels) and $w_{a}$ (right panels). As expected, we see that the points cluster around the $\Lambda$CDM values $w_{0}=-1$ and $w_{a}=0$ for Exp-model I, for {\it all} values of $\alpha$. Our detailed numerical results show that the smaller the value of $\alpha$, the closer the cosmology to that of $\Lambda$CDM. For very small values of $\alpha$ it is possible to obtain deviations, but $|\varphi_\text{F}|$ will also need to be quite small. For example, for $\alpha \lesssim 0.02$ we do not see any deviations from the $\Lambda$CDM values for $\varphi_\text{F}\in[-35,-1.4]$, detectable by the future Stage IV LSS experiments. By increasing $\varphi_\text{F}$ we start seeing deviations, and for example $\varphi_\text{F}=-1$ gives $w_{0}\sim-0.971$ and $w_{a}\sim0.806$, which should be detectable in the future. It is interesting to compare this model with the linear model of section~\ref{linsec}. Although in both cases the potential contains an effective cosmological constant piece, the linear model can provide dark energy that is distinguishable from a pure $\Lambda$ for $\alpha$ as small as $0.005$, as we discussed in section~\ref{linsec}, whereas Exp-model I is effectively equivalent to $\Lambda$CDM for such very small $\alpha$.

The story is however different for Exp-model II, as can be seen clearly from the figure. There are {\it forbidden} regions for both $w_{0}$ and $w_{a}$ for a given $\alpha$, which are specified through the curved boundaries set by the maximum value of $35$ for $|\varphi_\text{F}|$. As expected, in this case, the smaller the value of $\alpha$, the more difficult to obtain viable cosmologies, as $w_{0}$ and $w_{a}$ deviate more and more from the $\Lambda$CDM values by decreasing $\alpha$. For the values of $\alpha$ smaller than $\sim 0.5$, the deviations are already too large, and our numerical results show that it becomes almost impossible to obtain viable cosmologies with $\alpha$ smaller than $\sim 0.5$. We therefore conclude that Exp-model II is consistent with current observational constraints for all $\alpha \gtrsim 0.5$. This lower bound on $\alpha$ seems to be in disagreement with the bound presented in Ref.~\cite{Dimopoulos:2017tud} (i.e. $\alpha > 1.5$) for the same model. The reason could be due to the tight observational constraints that the authors have imposed on the equation of state of dark energy. That constraint is however valid only when $w_\text{DE}$ is assumed to be a constant, which is clearly not the case here. The values of $w_{0}$ and $w_{a}$ that we find here for $\alpha \gtrsim 0.5$ are in perfect agreement with current observational constraints. 

\begin{figure}
\center
  \includegraphics[height=4cm]{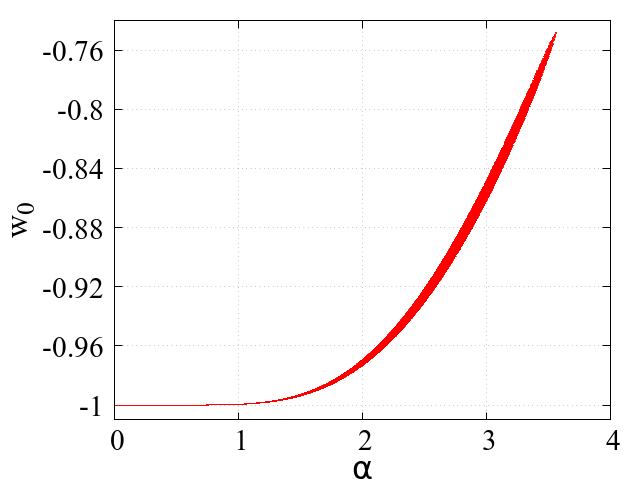}
  \includegraphics[height=4cm]{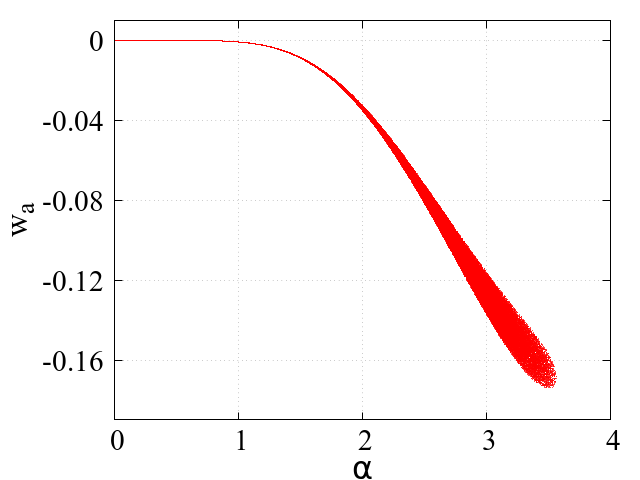}
  \includegraphics[height=4cm]{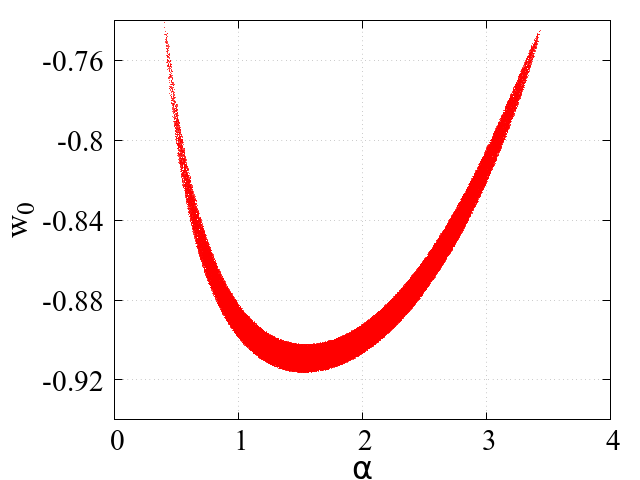}
  \includegraphics[height=4cm]{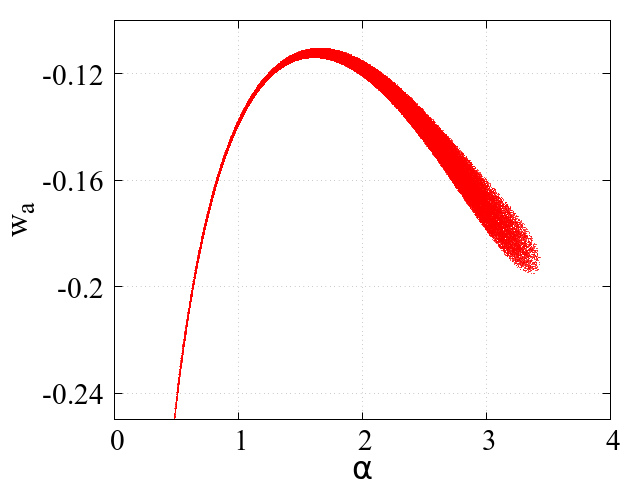}
\caption{\footnotesize\label{fig:model0-model1-w0-wa-alpha-phirehm10} The same as in Fig.~\ref{fig:model0-model1-w0-wa-alpha-phirehfree} but when $\varphi_\text{F}$ has been fixed to $-10$. Upper and lower panels again correspond to Exp-model I and Exp-model II, respectively.}
\end{figure}

Let us now fix $\varphi_\text{F}$ to $-10$ and see how the four panels of Fig.~\ref{fig:model0-model1-w0-wa-alpha-phirehfree} change. The results are presented in Fig.~\ref{fig:model0-model1-w0-wa-alpha-phirehm10}; again the upper panels correspond to Exp-model I, and the lower panels correspond to Exp-model II. For Exp-model I, we now see that there is an upper bound of $\sim 3.5$ on $\alpha$ in order for the model to provide cosmic histories consistent with current data; $\alpha$ can however take any values smaller than this bound. Exp-model II, on the other hand, now allows only values of $\alpha$ in the approximate range of $[0.5,3.3]$ when $\varphi_\text{F}$ is fixed to $-10$. In addition, it is interesting to see that both $w_{0}$ and $w_{a}$ show different behavior in terms of $\alpha$ for the two models. The upper panels of Fig.~\ref{fig:model0-model1-w0-wa-alpha-phirehm10} show that increasing $\alpha$ enhances the deviation from $\Lambda$CDM in Exp-model I, while the lower panels show that for Exp-model II both $w_{0}$ and $w_{a}$ are extremized around some intermediate values of $\alpha\sim 1.5$, below and above which the deviations from $\Lambda$CDM are larger.

 Before we end this section let us point out that our analysis shows that the results presented in Figs.~\ref{fig:model0-model1-w0-wa-alpha-phirehfree} and~\ref{fig:model0-model1-w0-wa-alpha-phirehm10} do not change even if the COBE/Planck normalization is not strictly imposed, i.e. even if we allow $M^2$ to be one or two orders of magnitude larger or smaller than the COBE/Planck-normalized value for each $\alpha$. However, fully relaxing the COBE/Planck normalization does affect the results, as discussed in Appendix~\ref{sec:appendix}.

\section{\Large {2-field quintessential inflation models}}\label{twofield}

\subsection{\boldmath{Dark energy and exponential potentials}}

As we discussed in section~\ref{basic}, the asymptotic expression for the alpha-attractor potential at large negative $\vp$ \rf{plateaumin} after a change of variables and a redefinition  $ \sqrt{2\over 3\alpha}\to \lambda$ can be represented in a more familiar way $V(\vp) = \Lambda + e^{\lambda \varphi }$.  These models with a vanishing cosmological constant $\Lambda = 0$ were among the first candidates for the role of dark energy, see e.g. Refs.~\cite{Wetterich:1987fm,Ratra:1987rm}. However, unlike the dark energy model with the linear potential, which was proposed a year earlier   \cite{Linde:1986dq}, the original models with exponential potentials discussed in Refs.~\cite{Wetterich:1987fm,Ratra:1987rm} did not provide a solution to the cosmological constant problem. Some progress in this direction was achieved only much later, in the models with the potential \rf{plateaumin2} and $\Lambda < 0$  \cite{Kallosh:2003mt}. 
Even though the models considered in Ref.~\cite{Kallosh:2003mt} described single field exponential potentials, the context of this theory was similar to the linear model of Refs.~\cite{Linde:1986dq,Kallosh:2003bq}, which presumed the prior stage of inflation driven by another field. Therefore, before discussing dark energy in the context of two-field $\alpha$-attractors, we describe and generalize the results of Ref.~\cite{Kallosh:2003mt}, in the light of the  string theory landscape developments. 

Let us first consider the simplest case of $\Lambda = 0$.  For $\lambda \ll 1$ ($\alpha \gg 1/3$), the potential is flat, the energy density of  normal matter decreases faster than $V$, and the system eventually enters the asymptotic regime of power-law inflation with 
\be\label{walpha3}
w_{\infty} =  -1 + {\lambda^{2}\over 3} = -1 + {2\over 9\alpha} \ .
\ee 
Meanwhile in the models with a dS plateau, $\Lambda > 0$, the asymptotic value of $w$ is $ -1$, but for large $\alpha$ the transition from $w =-1 + {2\over 9\alpha}$ to $w = -1$ may take a long time. In the models with $\Lambda < 0$, the universe eventually collapses, but if    $\lambda \ll 1$ and $|\Lambda| \ll 10^{{-120}}$, there is a very long interval, longer than the present age of the universe, when life as we know it can exist, and $w$ is very close to $-1$  \cite{Kallosh:2003mt}.
Thus, one could argue that exponential potentials, as well as $\alpha$-attractors, can easily provide us with viable dark energy models with $w$ very close to $-1$, but still noticeably different from it. However, a more detailed investigation shows that the situation is much more nuanced.

First of all, models with exponential potentials cannot simultaneously describe inflation {\it and} quintessence. They support inflation for $\lambda \ll 1$, but then inflation never ends. A way around it is to  assume, along the lines of Ref.~\cite{Linde:1986dq}, that the potential of the dark energy field $\vp$ is given by $V(\vp) \sim  e^{\lambda \varphi } + \Lambda$, but inflation is driven by some other field. %(The situation where inflation was driven by the field $\vp$ was considered in section~\ref{sec:1field-quint-inf}.) 
Then, because of inflationary fluctuations of the ultra-light field $\vp$, after inflation the universe becomes divided into exponentially many exponentially large parts where $\vp$ takes different values, so that its potential energy $V(\vp)$ takes all possible values of $\Lambda$, including values many orders of magnitude higher than $10^{{-120}}$. In each of these parts, the field $\vp$ is locally very homogeneous. Thus, just as in the linear model of Ref.~\cite{Linde:1986dq}, the universe becomes divided into many parts with different values of the effective cosmological constant $\Lambda + e^{\lambda \varphi }$. Therefore all values of the field $\vp$ with $\Lambda + e^{\lambda \varphi } \gg 10^{{-120}}$ are anthropically forbidden.

Indeed, in the parts of the post-inflationary universe models with $\lambda \ll 1$ and $|\Lambda| \ll 10^{{-120}}$, the scalar field starts moving (very slowly,  because $V' \sim \lambda V \ll V$) when the density of cold (and hot) matter of the universe, which rapidly decreases during  its expansion, becomes smaller than $V(\vp)$. If the field was frozen and starts moving at $V(\vp) \gg 10^{{-120}}$, the universe enters the regime of quasi-exponential expansion too early, which disrupts galaxy formation.

If $\Lambda$ is negative, but the initial value of $V(\vp) \sim  e^{\lambda \varphi } + \Lambda$ was positive, the universe in these models may enter the stage of accelerated expansion which may continue for a few billion years after that, until the universe collapses \cite{Kallosh:2003mt}. However, this regime is possible only for $\lambda \lesssim 1$, and only in some  finite ($\lambda$-dependent)  range of $\Lambda<0$ and post-inflationary values of the field $\vp$    \cite{Kallosh:2003mt}. 

On the other hand, if $\Lambda$ is small but positive, $\Lambda  \sim +10^{{-120}}$, the universe may enter the stage consistent with the presently available data for {\it any} value of $\lambda$, and for an infinitely large range of post-inflationary values of the field $\vp$ such that $e^{\lambda \varphi } \lesssim 10^{{-120}}$.  Only in a finite part of this range of $\vp$ does one have $e^{\lambda \varphi } \sim \Lambda$ and $w$ close to -1  but distinctly different from it. Meanwhile in the infinitely large range of $\vp$, all  the way down to $-\infty$, one  has $e^{\lambda \varphi}  \ll \Lambda$. Therefore, for any given  $\lambda$, the anthropically viable ``phase space'' of $\Lambda$ and $\vp$ is dominated by positive $\Lambda \sim +10^{{-120}}$ and by indefinitely large negative $\vp$, where dark energy is indistinguishable from the cosmological constant, and the equation of state is given by $w = -1$ with an exponentially good accuracy. A similar conclusion was reached in Ref.~\cite{Garriga:2002tq} for a broad class of dark energy models, though some exceptions from this rule are possible, see e.g. Refs.~\cite{Kallosh:2002gg,Garriga:2003hj}.

\subsection{Non-interacting $\alpha$-attractors}

A similar conclusion can be reached  in many models of two-field $\alpha$-attractors, if one assumes, as we did before, that the potential of the field $\phi$ responsible for dark energy  is very small, and inflation is driven by some other field $\chi$, not interacting with the field $\phi$. To illustrate this possibility, we consider here a toy model of two non-interacting fields.

\begin{figure}[h!]
\begin{center}
\includegraphics[scale=0.35]{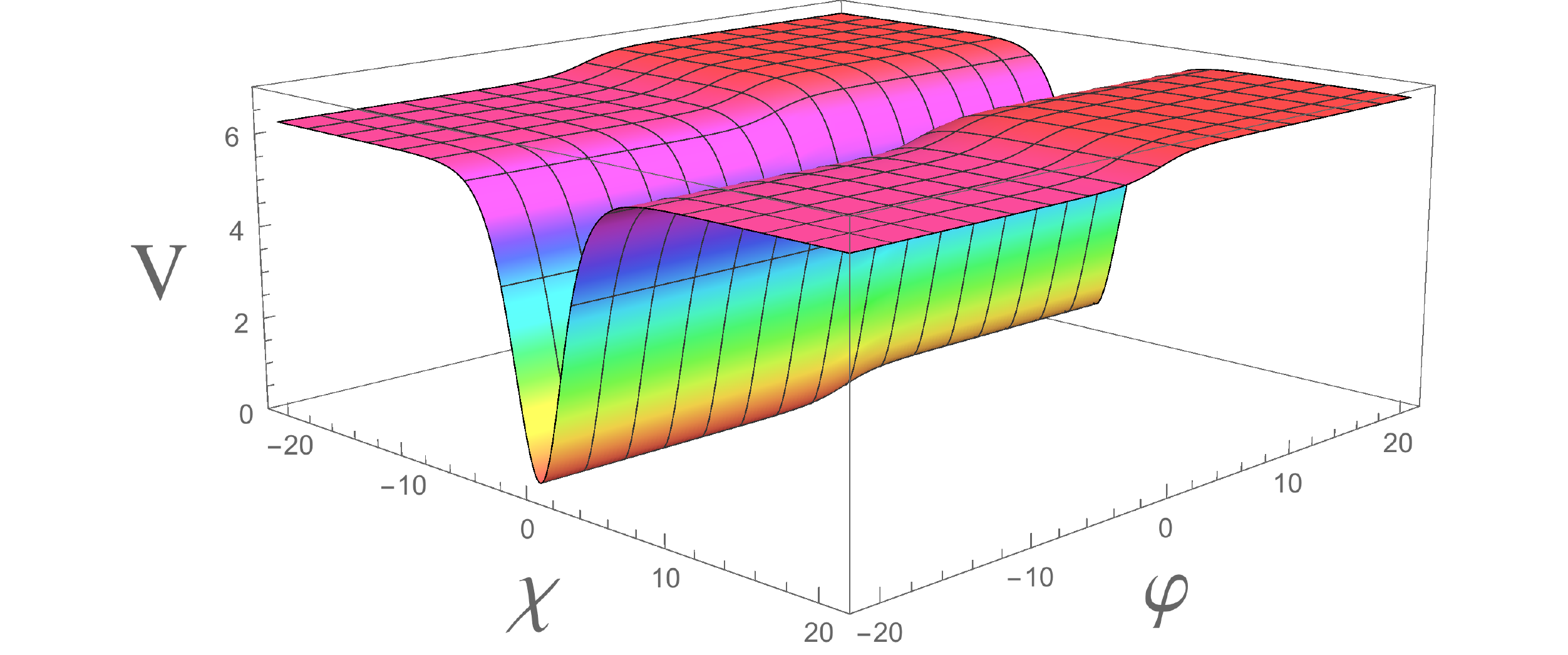}
\end{center}
\caption{\footnotesize The shape of the potential $V(\vp,\chi)$ \rf{cosmo4} for $\alpha =  \beta  = 1$, $3\beta  m^{2 } \gg \gamma \sqrt{6\alpha} $, and $V_{0} \approx \gamma \sqrt{6\alpha}$. }
\label{F4}
\end{figure}

Let us consider an extended version of the $\alpha$-attractor model, adding to it a scalar field $\sigma$ with a non-canonical kinetic term: 
 \be
 {1\over \sqrt{-g}} \mathcal{L} = { R\over 2}   -  {(\partial_{\mu} \phi)^2\over 2(1-{\phi^{2}\over 6\alpha})^{2}} -   {(\partial_{\mu} \sigma)^2\over 2(1-{\phi^{2}\over 6\beta})^{2}}   - {m^{2}\over 2} \sigma^{2}   - \gamma  \phi - V_{0} \,.
\label{cosmo2}\ee
As before, one can represent this theory in terms of two canonically normalized fields,
\be\label{tanh3} 
\phi = \sqrt {6 \alpha}\, \tanh{\varphi\over\sqrt {6 \alpha}}, \quad \sigma = \sqrt {6 \beta}\, \tanh{\chi\over\sqrt {6 \beta}}  \ .
\ee
The inflaton potential in terms of the canonically normalized fields $\vp$ and $\chi$ becomes
 \be
 V(\vp,\chi) =     3\beta  m^{2 }  \, \tanh^{2}{\chi\over\sqrt {6 \beta}} +  \gamma \sqrt{6\alpha}  \tanh {\varphi\over\sqrt {6 \alpha}} + V_{0} \,.
\label{cosmo4}\ee
We illustrate the general structure of this potential for $\alpha =  \beta  = 1$ and some particular (non-realistic) values of parameters such that $3\beta  m^{2 } \gg \gamma \sqrt{6\alpha} $, and $ V_{0} \approx \gamma \sqrt{6\alpha}$, see Fig. \ref{F4}. In that case the  term $ 3\beta  m^{2 }  \, \tanh^{2}{\varphi\over\sqrt {6 \beta}}$ is responsible for inflation in this model, the dark energy potential $ \gamma \sqrt{6\alpha}  \tanh {\varphi\over\sqrt {6 \alpha}} + V_{0}$ is very shallow, and it approaches a small cosmological constant $V_{-} = V_{0} -\gamma \sqrt{6\alpha}$ in the limit $\vp \to -\infty$, and $V_{+} = V_{0} +\gamma \sqrt{6\alpha}$ in the limit $\vp \to \infty$.

Inflation begins at the plateau with $V(\vp,\chi) =     3\beta  m^{2 } \gg V_{+} $. This plateau is almost exactly flat, so inflation may begin with an equally large probability at any point of the plateau with $\chi \gg \sqrt {6 \beta}$ \cite{Linde:2017pwt}. It ends when the field $\chi$ falls down to the dark energy valley with $\chi = 0$. Since the field $\vp$ at the beginning of inflation can take any value with (almost exactly) equal probability because of a (nearly exact) shift symmetry of the potential in the $\vp$ direction, all values of the field $\vp$ after inflation will be equally probable as well. 

In that case, one can use the same argument as the one we used for the theory with exponential potential: After inflation, the fields roll down either to the right plateau, or to the left plateau,  but it is most probable that it will end up extremely far from $\vp =0$. By a proper choice of parameters, including adjustment of the parameter $V_{0}$, one can easily have the regime of acceleration at the time $t \sim 10^{10}$ years. However, with an overwhelmingly large probability the absolute value of the field $\vp$ after inflation will be extremely large, and therefore this stage will be indistinguishable from the pure cosmological constant with $w = -1$.

The same conclusion is valid for most of the dark energy models based on the $\alpha$-attractors with $V(\vp)$ much smaller than the energy density of the inflaton field $\chi$ during inflation.  Indeed, for most of such models the asymptotic behavior of the potential $V(\vp)$ in the limit $|\vp| \to \infty$ is given by one of the two asymptotic expressions \rf{plateau} or \rf{plateaumin}. The asymptotic values of the cosmological constant $\Lambda$ along the two shoulders of the potential is given either by $V_{-}$ or by $V_{+}$. By adding a constant to the potential, one can adjust at least one of these parameters to belong to the anthropic range $|\Lambda| \lesssim 10^{{-120}}$. Then all arguments given above apply.

Thus we see that one can easily obtain a viable dark energy model in any model of $\alpha$-attractors, with a very broad range of  parameters and potentials, as long as the value of dark energy potential $V(\vp)$ is sufficiently small. But the observational consequences of these models for the most general class of initial conditions are practically indistinguishable from the predictions of the simplest cosmological constant models. This is good news from the point of view of generality of the predictions, but perhaps not very good news from the point of view of observers.

\begin{figure}[h!]
\begin{center}
\includegraphics[scale=0.4]{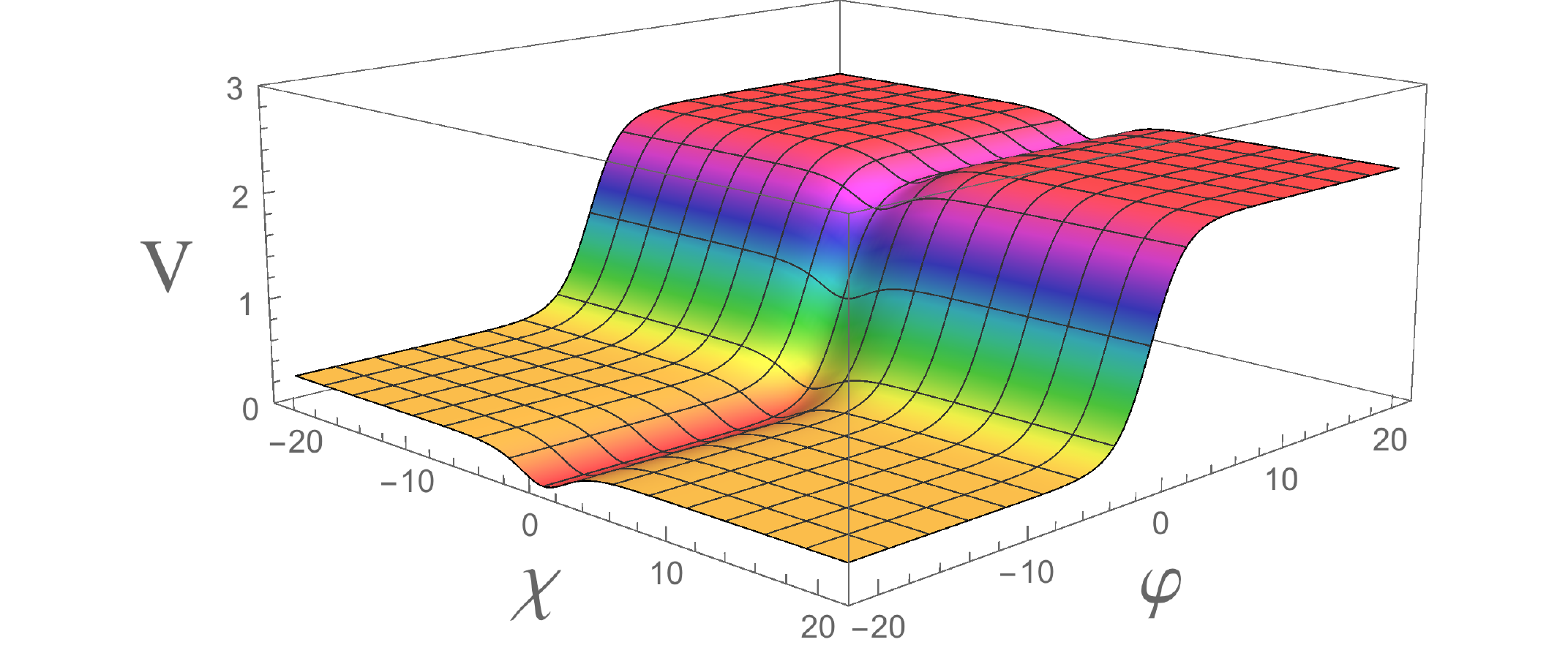}
\end{center}
\caption{\footnotesize The shape of the potential $V(\vp,\chi)$ \rf{cosmo4} for $\alpha =  \beta  = 1$, $3\beta  m^{2 } \ll \gamma \sqrt{6\alpha} $, and $ V_{0} \approx \gamma \sqrt{6\alpha}$. }
\label{F5}
\end{figure}

However, these conclusions were obtained under the conditions some of which can be relaxed. For example, consider the same model as before, but instead of the regime with $3\beta  m^{2 } \gg \gamma \sqrt{6\alpha} $ we may investigate an opposite regime $3\beta  m^{2 } \ll \gamma \sqrt{6\alpha} $. The potential in this case is shown in  Fig. \ref{F5}.

In this model, the potential at the first stage of inflation is dominated by the quintessence potential $V(\vp) =\gamma \sqrt{6\alpha}  \tanh {\varphi\over\sqrt {6 \alpha}} + V_{0}$, falling from the high (red) plateau. Depending on initial conditions, inflationary scenario can be realized in two distinct ways. In the first scenario, the initial value of the field $\chi$ is extremely large, and its potential is very flat. In that case, the  fields will first roll in the $\vp$ direction and fall from the cliff to the yellow plateau determined by the term $3\beta  m^{2 }  \, \tanh^{2}{\chi\over\sqrt {6 \beta}}$. Then there will be a second stage of inflation driven by the field $\chi$, which ends at $\chi = 0$. We call this scenario `cascade inflation' \cite{Kallosh:2017wnt}.  The value of the field $\vp$ at the end of inflation will be determined by the initial conditions, and by the two stages of cascade inflation, including (for some initial conditions) a stage of eternal inflation. 

On the other hand, if the initial value of the field $\chi$ is relatively small, and the field $\vp$ is very large, then in the beginning of inflation, the field $\chi$ rolls down the valley with $\chi = 0$, and the subsequent stage of inflation and quintessential evolution will be determined by the single field evolution of the field $\vp$.

In the next section we will briefly describe a simple model of two interacting attractors; as we will see taking into account interactions may open many other possibilities.

\subsection{Interacting $\alpha$-attractors}

Now we add an interaction term $g^{2}\phi^{2}\sigma^{2}$ to the potential of the model \rf{cosmo2},
 \be
 {1\over \sqrt{-g}} \mathcal{L} = { R\over 2}   -  {(\partial_{\mu} \phi)^2\over 2(1-{\phi^{2}\over 6\alpha})^{2}} -   {(\partial_{\mu} \sigma)^2\over 2(1-{\phi^{2}\over 6\beta})^{2}}   - {m^{2}\over 2} \sigma^{2}  -g^{2}\phi^{2}\sigma^{2} - \gamma  \phi - V_{0} .
\label{cosmo5}\ee
The inflaton potential in terms of the canonically normalized fields $\vp$ and $\chi$ becomes
 \be
 V(\vp,\chi) =   36 \alpha \beta g^{2} \tanh^{2}{\vp\over\sqrt {6 \alpha}}\, \tanh^{2}{\chi\over\sqrt {6 \beta}}+  3\beta  m^{2 }  \, \tanh^{2}{\chi\over\sqrt {6 \beta}} +  \gamma \sqrt{6\alpha}  \tanh {\varphi\over\sqrt {6 \alpha}} + V_{0}.
\label{cosmo6}\ee

\begin{figure}[h!]
\begin{center}
\vskip -5pt
\includegraphics[scale=0.4]{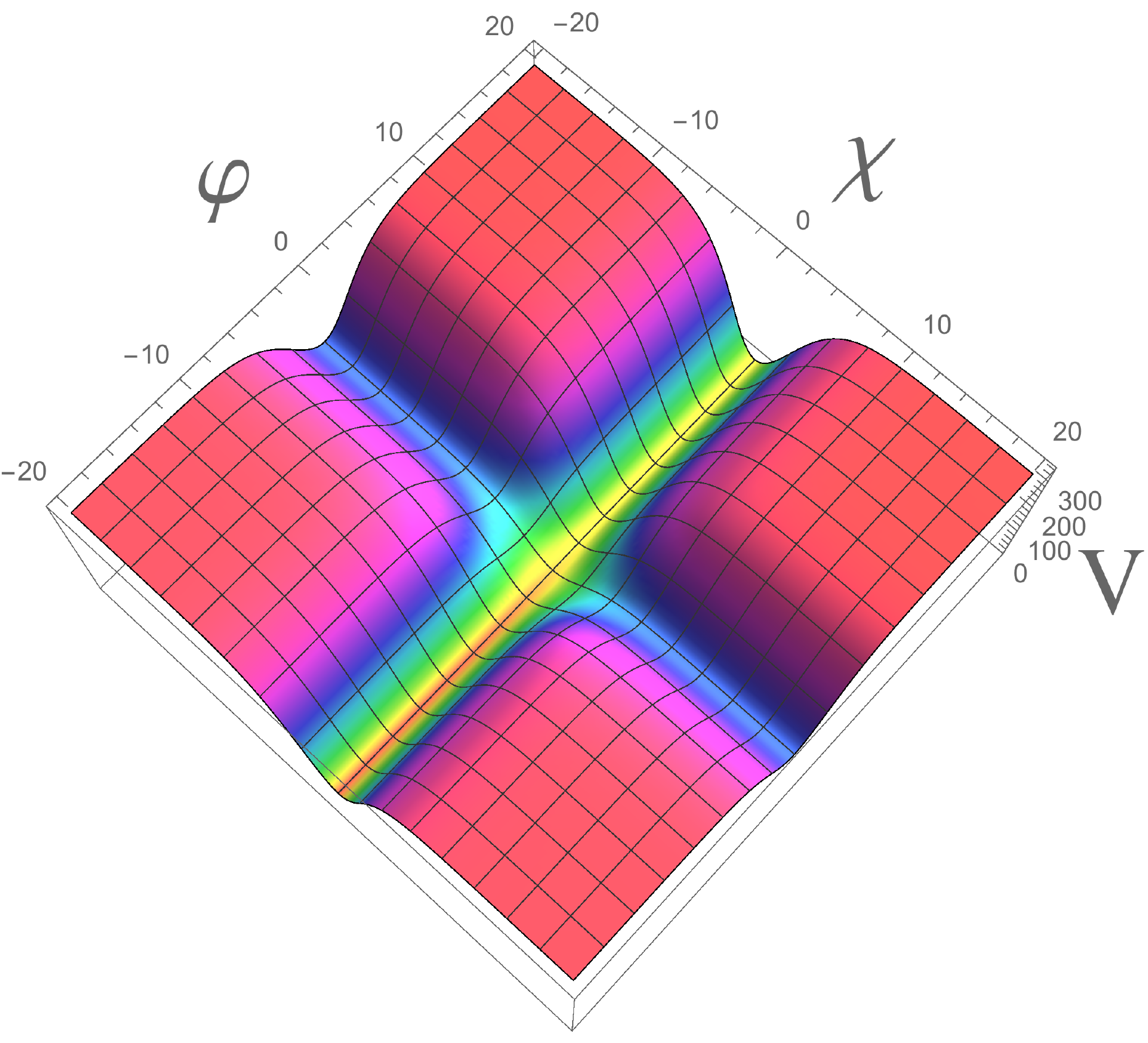}
\end{center}
\vskip -15pt
\caption{\footnotesize The shape of the potential $V(\vp,\chi)$ \rf{cosmo6} for $\alpha =  \beta  = 1$ and $36 \alpha \beta g^{2} \gg 3\beta  m^{2 } \gg \gamma \sqrt{6\alpha}, V_{0}$. The green valley corresponds to quintessence with the linear potential $V =\gamma\phi+V_{0} = \gamma \sqrt{6\alpha}  \tanh {\varphi\over\sqrt {6 \alpha}} + V_{0}$.}
\label{F6}
\end{figure}

We will take the parameters such that $36 \alpha \beta g^{2} \gg 3\beta  m^{2 } \gg \gamma \sqrt{6\alpha}, V_{0}$. In that case, the potential can be illustrated (not to scale) by Fig.~\ref{F6}. Inflation begins at one of the high red plateaus of the height approximately given by $ 36 \alpha \beta g^{2}$. The blue valley describes the $\alpha$-attractor inflationary potential $V(\chi) =3\beta  m^{2 }  \, \tanh^{2}{\chi\over\sqrt {6 \beta}}  + V_{0}$. The green valley corresponds to the dark energy potential $ \gamma \sqrt{6\alpha}  \tanh {\varphi\over\sqrt {6 \alpha}} + V_{0}$.

One can show that about half of all inflationary trajectories starting at the red plateau describe the fields falling directly to the dark energy valley. We assume that $3\beta  m^{2 } \sim 10^{-10}$ and $36 \alpha \beta g^{2}$ is much greater, possibly even as large as $\mathcal{O}(1)$ in Planck units, then the inflationary trajectories falling directly to the dark energy valley produce parts of the universe with too large perturbations of density, which make such parts of the universe anthropically disfavored.

Another half of all inflationary trajectories starting at the red plateau describe the fields falling towards the blue inflationary valley. Then the inflaton field $\chi$ rolls along this valley, which generates perturbations of the proper magnitude in accordance with the $\alpha$-attractor scenario. The process of reheating occurs due to oscillations of the field $\chi$ near the point $\vp = \chi = 0$. At this point, the potential has a tiny slope which pushes the dark energy field $\vp$ towards its large negative values, but this field does not start rolling until the density of particles produced by reheating drops down substantially. When this happens, the field $\vp$ starts moving towards $\vp \to -\infty$. 

Consider the simplest case of $V_{0} = \gamma \sqrt{6\alpha}  \sim 10^{{-120}}$. Then the dark energy potential $\gamma \sqrt{6\alpha}  \tanh {\varphi\over\sqrt {6 \alpha}} + V_{0}$ is given by $V_{0} \sim 10^{{-120}}$ at $\vp = 0$, and vanishes in the limit $\vp \to -\infty$. To give a particular example, one may consider $\alpha = 7/3$. Then, just like in the theory with exponential potential, the asymptotic value of $w$ for dark energy will be about $0.905$, but its initial value at the moment when the field $\vp$ starts moving down will be given (almost) exactly by -1. By taking $V_{0}$ slightly greater than $\gamma \sqrt{6\alpha}$, one can make $w$ much closer to $-1$.
This model represents a simple  $\alpha$-attractor version of the dark energy model with the linear potential proposed in Ref.~\cite{Linde:1986dq}.

%\comLA{\bf Things to check and discuss: Fine-tuning of $\gamma \sqrt{6\alpha}  \sim 10^{{-120}}$ seems to be unavoidable, on top of the fine-tuning of the cosmological constant.   }

%\subsection{An example with $w_{\infty} \neq -1$}

\subsection{\boldmath{Quintessence with a linear potential}}

Inspired by our discussions in the previous section, let us now consider a concrete example of the 2-field, interacting, $\alpha$-attractor scenario where the simplest linear potential for the quintessence field $\phi$ has the form given in Eq. (\ref{lin}), i.e.
\begin{equation}
V(\vp) =  \gamma \sqrt{6\alpha}  (\tanh {\varphi\over\sqrt {6 \alpha}}+1) + \Lambda\ ,\label{eq:interacting_quintpot3}
\end{equation}
in terms of the canonical field $\vp$, with $\Lambda$ being a constant. We additionally assume $36 \alpha \beta g^{2} \gg 3\beta m^{2 } \gg \gamma \sqrt{6\alpha},\Lambda$. As discussed in the previous section, we further assume that the inflationary trajectory starts at the red plateau of Fig.~\ref{F6} at large values of the field $\chi$, and then the fields $\varphi$ and $\chi$ fall towards the blue inflationary valley at $\vp=0$. The inflaton field $\chi$ then rolls along the valley, and reheating occurs through the oscillations of $\chi$ near the point $\phi=\chi=0$. At this point, the tiny slope in the dark energy potential pushes the quintessence field $\varphi$ towards its negative values. As stated before, in this scenario inflation is {\it not} driven by $\varphi$, and it only sets the value of $\varphi$ to something around $0$ as the initial value of the dark energy field for the late-time evolution of the universe, contrary to the quintessential inflation models, studied in section~\ref{sec:1field-quint-inf}, which could accommodate a wide range of initial conditions for the quintessence field $\varphi$ that was also responsible for inflation.

\begin{figure}[h!]
\center
  \includegraphics[height=4cm]{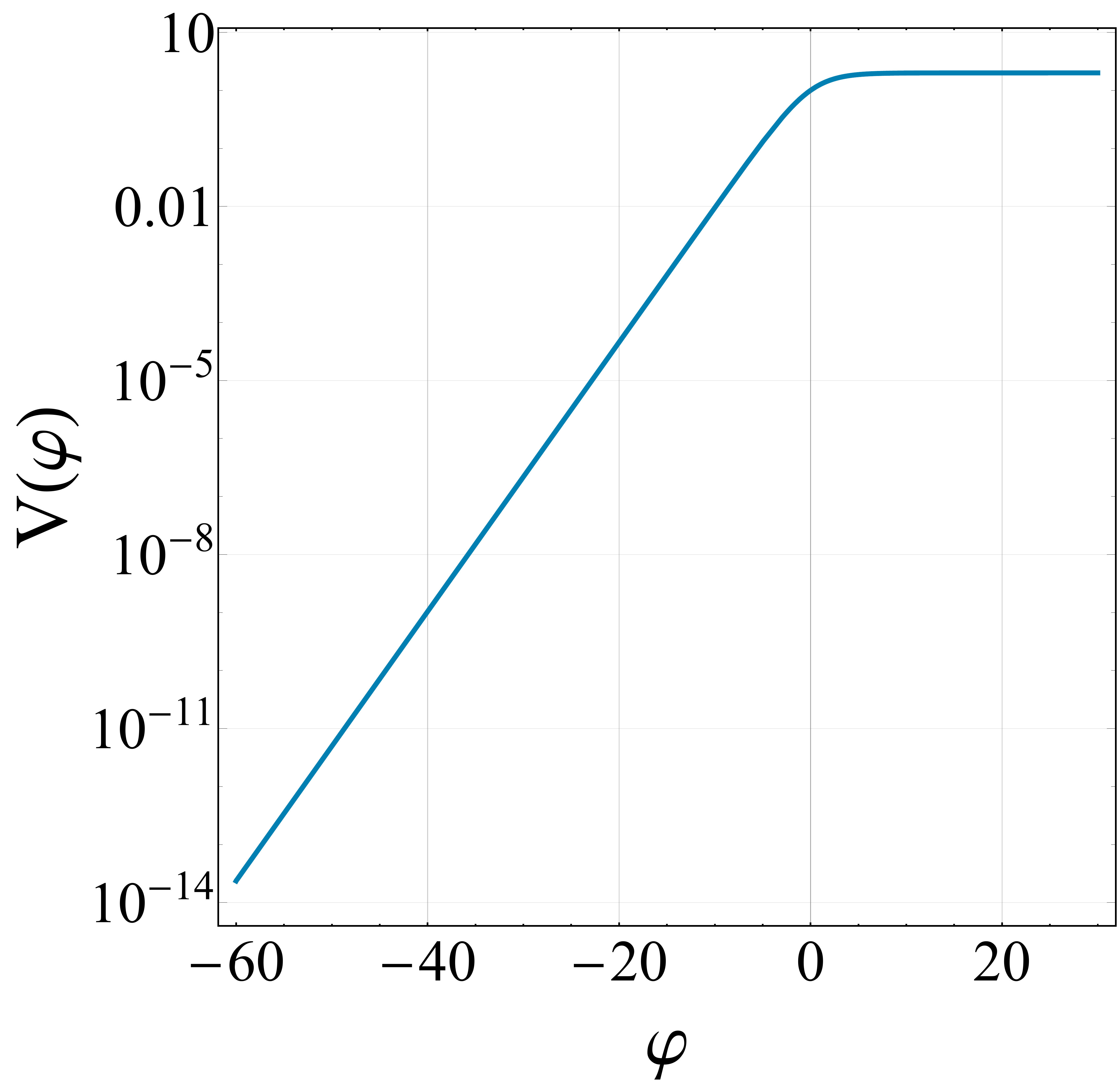}
  \includegraphics[height=4cm]{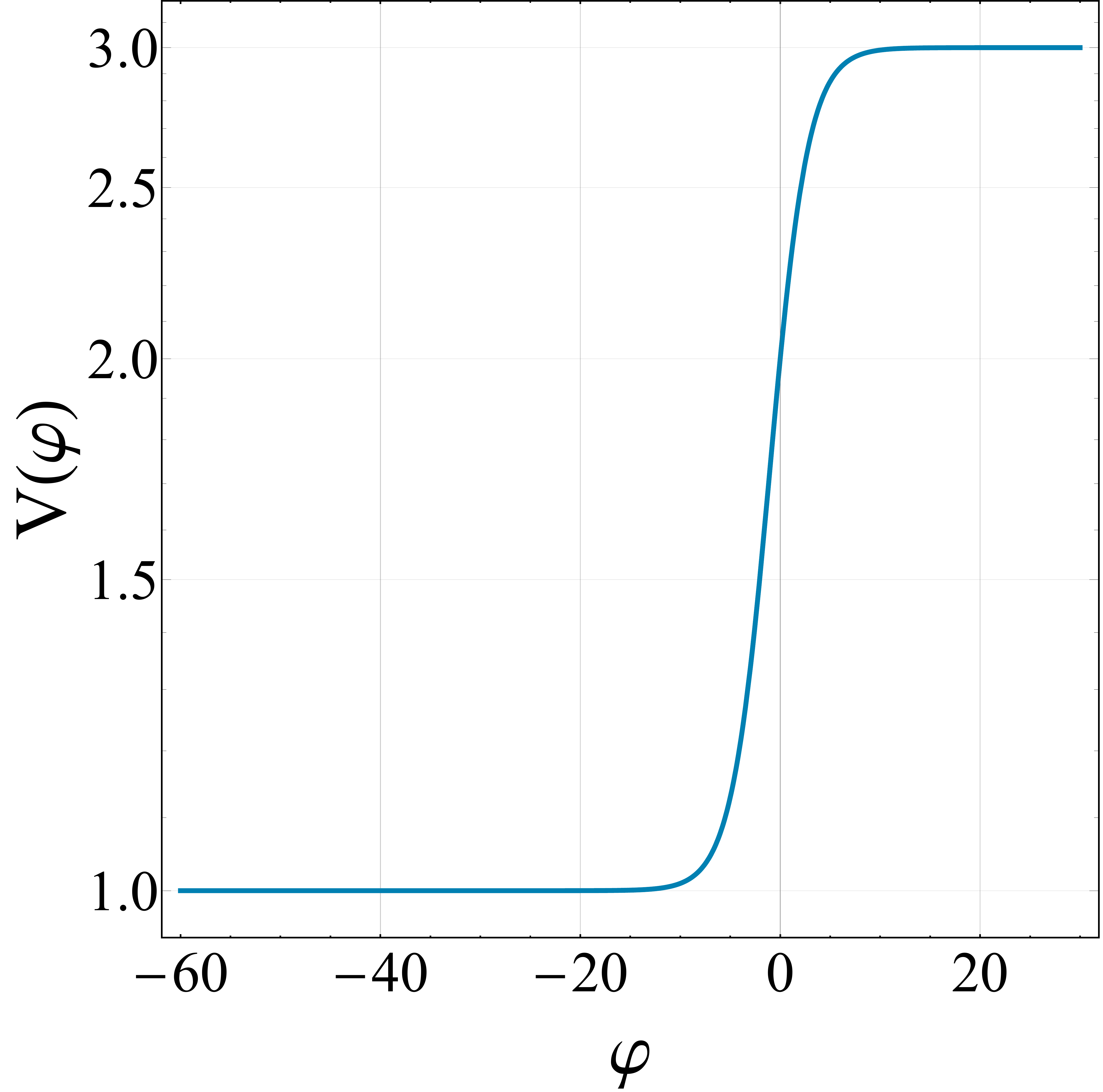}
\caption{\footnotesize\label{fig:modelPL-shape} The shape of the potential $V(\vp) =  \gamma \sqrt{6\alpha}  (\tanh {\varphi\over\sqrt {6 \alpha}}+1) + \Lambda$ for $\Lambda=0$ (left panel) and $\Lambda=\gamma \sqrt{6\alpha}$ (right panel). Here we have set $\gamma \sqrt{6\alpha}$ to $10^{{-120}}$ and $\alpha$ to $7/3$. The values of the potentials on the $y$-axes are normalized to $10^{{-120}}$.}
\end{figure}

Now we consider the case with both $\gamma \sqrt{6\alpha}$ and $\Lambda$ being of $\mathcal{O}(10^{{-120}})$. Note that the potential approaches a cosmological constant $V_{-}=\Lambda$ for large, negative $\varphi$, and therefore $\Lambda=0$ corresponds to a potential with a vanishing asymptotic value in the limit $\vp \to -\infty$. The potential has been shown in Fig.~\ref{fig:modelPL-shape} for $\Lambda=0$ (left panel) and $\Lambda=\gamma \sqrt{6\alpha}$ (right panel); we have set $\alpha=7/3$ for both cases. The figure shows that the potential monotonically decreases for $\Lambda=0$ and takes an asymptotic, constant value for $\Lambda=\gamma \sqrt{6\alpha}$ at large, negative $\varphi$. The value of $\gamma$ has been chosen such that the asymptotic value of the potential gives $10^{{-120}}$.

The asymptotic value for the equation of state of dark energy, $w_\text{DE}$, in this model can be obtained by assuming a slow-roll approximation. As we discussed before, this asymptotic value for $\Lambda=0$ is
\begin{equation}
w_\infty=-1+\frac{2}{9\alpha} \ ,
\label{eq:wasymp_lin}
\end{equation}
which depends only on $\alpha$. The asymptotic value for $\Lambda\ne0$ is $-1$.

\begin{figure}
\center
  \includegraphics[height=4cm]{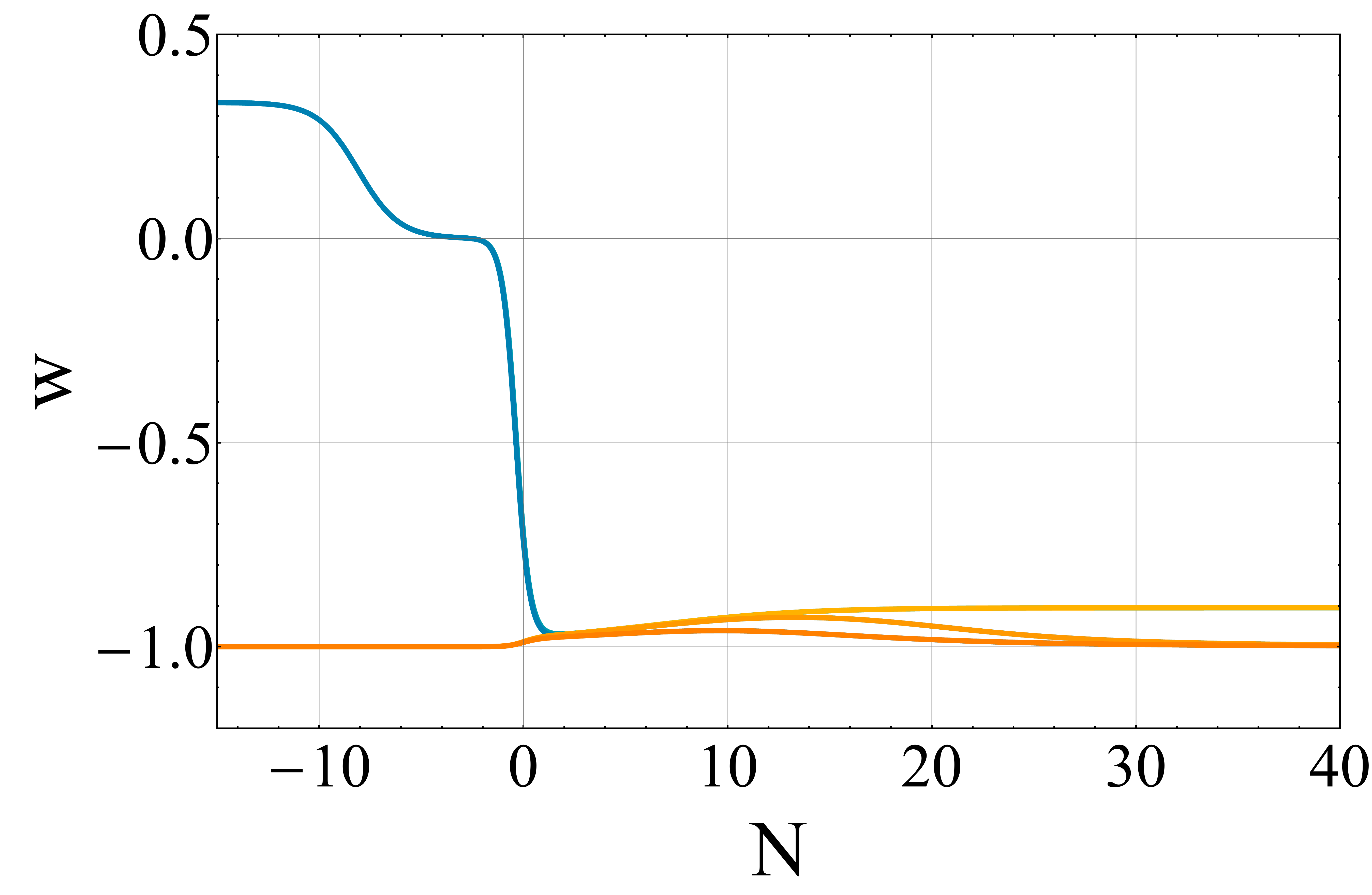}
\caption{\footnotesize\label{fig:modelPL-wN-1} Evolution of the equation of state as a function of the number of $e$-folds $N$ after reheating for the linear potential $V(\vp) =  \gamma \sqrt{6\alpha}  (\tanh {\varphi\over\sqrt {6 \alpha}}+1) + \Lambda$ in the framework of the interacting, 2-field $\alpha$-attractors. The three yellow-to-orange curves show the dark energy equation of state $w_\text{DE}$ for $\Lambda=0$, $10^{-2}\times\gamma \sqrt{6\alpha}$, and $10^{-1}\times\gamma \sqrt{6\alpha}$, respectively. The effective equation of state $w_\text{eff}$ is almost identical for all values of $\Lambda$ in the past (shown collectively by a blue curve), is different from $w_\text{DE}$ in the past, and becomes identical to it in the future when the field $\varphi$ becomes dominant. $N=0$ corresponds to the present time, $\gamma \sqrt{6\alpha}$ has been set to $10^{{-120}}$, and $\alpha$ has been set to $7/3$ for all the cases.}
\end{figure}

Let us now study the time evolution of $w_\text{eff}$ as well as $w_\text{DE}$ for a few values of $\Lambda$ and for $\alpha=7/3$. The results have been presented in Fig.~\ref{fig:modelPL-wN-1} for $\Lambda=0$, $10^{-2}\times\gamma \sqrt{6\alpha}$, and $10^{-1}\times\gamma \sqrt{6\alpha}$. Note that $w_\text{eff}$ is almost identical in the past ($N<0$) for all the cases (blue curve), and shows different behavior for the future ($N>0$). Note also that $w_\text{eff}$ is different from $w_\text{DE}$ in the past, and becomes identical to it in the future, when the field $\varphi$ dominates. In addition, as expected, the figure shows that the deviation from $\Lambda$CDM is maximal when $\Lambda=0$, and decreases when $\Lambda$ increases. For the specific case of $\Lambda=0$, $w$ has an asymptotic value of $\sim -0.905$, in full agreement with our analytical expression (\ref{eq:wasymp_lin}), while for any other values of $\Lambda$ the asymptotic value is $-1$.

\subsection{Comparison to observations, and constraints on parameters}

With the qualitative discussions of the previous section, let us now study our 2-field, interacting, $\alpha$-attractor model in a rigorous way and through the comparison of the late-time predictions of the model to the observations. The potential is of the form given in Eq. (\ref{eq:interacting_quintpot3}). We scan over the parameters of the model, i.e. $\gamma$, $\alpha$, and $\Lambda$, and compare the evolution of the background cosmological observables to the data. We set $\varphi_\text{F}$ to $0$ in all our scans.
\begin{figure}
\center
  \includegraphics[height=3.9cm]{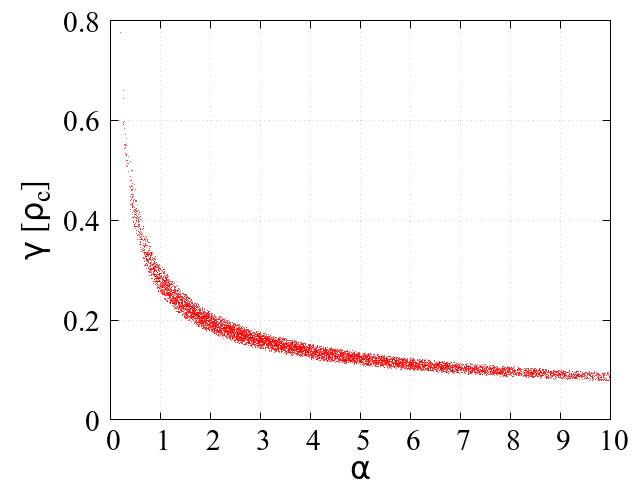}
  \includegraphics[height=3.9cm]{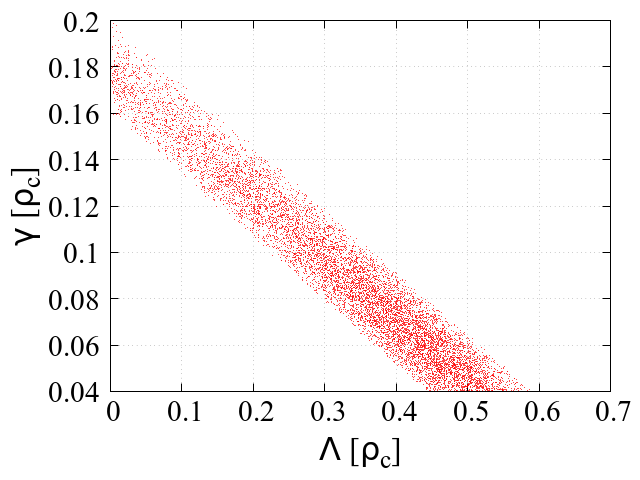}
  \includegraphics[height=3.9cm]{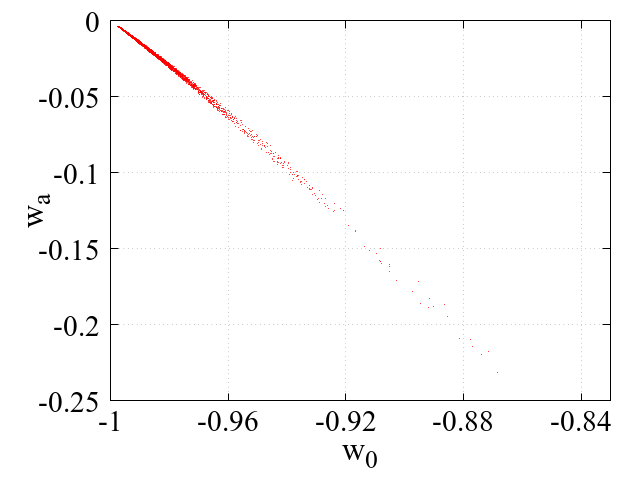}
  \includegraphics[height=3.9cm]{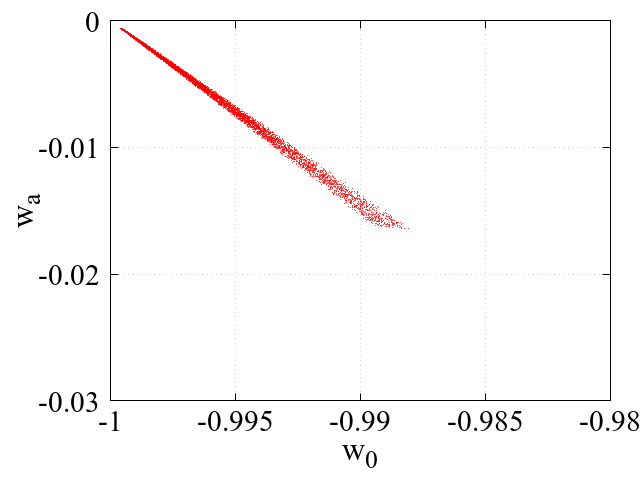}
\caption{\footnotesize\label{fig:modelLin_p123} {\it Upper panels:} Constraints on $\gamma$, $\alpha$ and $\Lambda$ for the linear, interacting, $\alpha$-attractor model with the linear potential $V(\vp) =  \gamma \sqrt{6\alpha}  (\tanh {\varphi\over\sqrt {6 \alpha}}+1) + \Lambda$, when $\Lambda$ is fixed to $0$ (left) and when $\alpha$ is fixed to $7/3$ (right). Note that both $\gamma$ and $\Lambda$ are presented in units of the critical density today, i.e. $\rho_c\equiv 3H_0^2$. {\it Lower panels:} Similar to the upper panels, but for constraints on the CPL parameters $w_{0}$ and $w_{a}$.}
\end{figure}

The upper panels of Fig.~\ref{fig:modelLin_p123} present our results for $\gamma$ versus $\alpha$ (left panel) and $\Lambda$ (right panel). Note that the values of $\gamma$ and $\Lambda$ are given in units of the critical density today, i.e. $\rho_c\equiv 3H_0^2$. For the left panel, where $\alpha$ is kept free, the value of $\Lambda$ has been set to $0$, while for the right panel, with $\Lambda$ being scanned over, $\alpha$ has been fixed to $7/3$. The value of $\gamma$ is correlated with both $\alpha$ and $\Lambda$. In order to see this correlation clearly, let us first focus on the left panel with $\Lambda$ being fixed to $0$, i.e. when the potential is $V(\vp) =  \gamma \sqrt{6\alpha}  (\tanh {\varphi\over\sqrt {6 \alpha}}+1)$. The figure shows that $\gamma$ increases by decreasing $\alpha$. When $\alpha$ becomes very small, we know that the potential rapidly decreases and the $\tanh{\frac{\varphi}{\sqrt{6\alpha}}}$ piece in the potential drops quickly to $\sim -1$. This will be largely cancelled by the constant piece $\gamma\sqrt{6\alpha}$, and one therefore would need an enormous value of $\gamma$ to compensate for that and to obtain the required amount of dark energy given by observations. This may mean that we should in principle be able to obtain good fits to the data for very small $\alpha$ with very large $\gamma$. However, the figure tells us that even though $\gamma$ indeed seems to be increasing at small $\alpha$, very small $\alpha$ ($\lesssim 0.3$) are disfavored by our analysis. This can be understood by looking at the lower, left panel of the figure, where the constraints on the two CPL parameters $w_{0}$ and $w_{a}$ are presented for the same scan. This shows that increasing $\alpha$ forces $w_0$ and $w_a$ to become closer to their $\Lambda$CDM values of $-1$ and $0$, respectively, and reducing $\alpha$ to small values corresponds to larger deviations form $\Lambda$CDM. This illustrates why $\alpha$ cannot be smaller than $\sim 0.3$ for this $\Lambda=0$ case, as the model predicts an equation of state for dark energy with present values that deviate too much from the observed values, and the number of viable points is therefore almost vanishing for very small $\alpha$. Therefore, even though the required {\it amount} of dark energy can be provided by the model for small $\alpha$, it does not produce the correct behavior for the dark energy {\it equation of state}. Clearly, by increasing $\Lambda$ to nonzero values, which is equivalent to adding a cosmological constant to the potential, small $\alpha$ can also provide viable models of dark energy. 

Let us now investigate the effect of changing $\Lambda$ on the predictions of the model, by focusing on the right panels of Fig.~\ref{fig:modelLin_p123}, where $\alpha$ has been fixed to $7/3$ and $\Lambda$ has been allowed to vary. The figure shows that the larger the value of $\Lambda$, the smaller the value of $\gamma$. This behavior is easily understood, as the total dark energy in our model is a combination of the $\varphi$-dependent piece and the cosmological constant $\Lambda$, and by increasing $\Lambda$ the contribution from the $\varphi$-dependent piece should reduce in order for the model to produce the correct, total amount of dark energy consistent with observations, i.e. to provide $\Omega_\text{DE}\approx0.7$. $\Omega_\text{DE}$ in general includes two pieces, one from the dynamics of the scalar field (i.e. the field-dependent part of the potential plus the kinetic energy of the field), and one from the cosmological constant $\Lambda$. Here therefore, by increasing the contribution from the cosmological constant the contribution from the field needs to drop in order to have the total amount of $\Omega_\text{DE}\approx0.7$. Decreasing $\gamma$ to zero in the upper, right panel of Fig.~\ref{fig:modelLin_p123} will make $\Lambda$ take a value of $\approx 0.7$ in units of $\rho_c$, which is what we expect. Note also that, as expected, increasing $\Lambda$ makes $w_0$ and $w_a$ closer to their $\Lambda$CDM values $-1$ and $0$, respectively (as shown in the lower, right panel of Fig.~\ref{fig:modelLin_p123}), which is consistent with our illustration in Fig.~\ref{fig:modelPL-wN-1}. 

Our conclusion, based on these results, is that this class of 2-field, interacting models, can provide interesting cosmological evolutions perfectly consistent with the current data. The deviations from the $\Lambda$CDM model depend however on the value of $\alpha$. For relatively large $\alpha$, such as $7/3$, the deviations are not large enough to be detected by the next generation of the LSS experiments, as $w_{0}$ and $w_{a}$ are not sufficiently different from the $\Lambda$CDM values, but  (depending on the value of $\Lambda$)  decreasing $\alpha$ can make the deviations larger and potentially detectable. This class of models, therefore, has predictions that in some cases can be tested, verified, or ruled out by the future cosmological surveys.

\section{Conclusions}
\label{sec:conclusions}

In this paper we constructed several viable models of dark energy based on the theory of $\alpha$-attractors, using the flexibility of choosing the cosmological constant provided by the string theory landscape. We studied a broad variety of the models, such as the models of quintessential inflation, where a single field $\vp$ plays the double role of the inflaton and the quintessence. The simplest of these models is the $\alpha$-attractor version of the theory with a linear potential described  in section \ref{linsec}. We also performed a detailed investigation of the models with exponential potential in sections \ref{should} and \ref{sec:exp-pot}.

The asymptotic flatness of the plateau potential in $\alpha$-attractors and the possibility to avoid the fifth force problem, see section~\ref{sec:fifth}, make these models particularly suitable candidates  for the role of dark energy. In several different  models  with the asymptotically vanishing height of the potential $V_{-} = \Lambda = 0$, we have a universal $\alpha$-dependent  prediction relating to each other the tensor to scalar ratio $r$ and the asymptotic value of the equation of state $w_{\infty}$:
\be\label{rw}
r = {12\alpha\over N^{2}}, \qquad  w_{\infty} = -1+\frac{2}{9\alpha} \, ;
\ee
see Eqs.~\rf{cute} and \rf{cute1}. This is a rather interesting correlation between $r$ and $w_{\infty}$, which  may seem to be suggesting  a possible way to test these models using a combination of the upcoming Stage IV cosmological experiments aiming at measuring both the B-mode polarization of the CMB and the growth and evolution of large-scale structure in the universe.  One should however note that, as we have shown in this paper for various models of quintessential inflation, $w_\infty$ is only the ultimate value of the dark energy equation of state parameter and not its present value. This means that $w_\infty$ cannot be used directly to test the models, and one needs a detailed analysis in order to compare the predictions of the models to the cosmological observations. 

 Moreover, if one accepts the simplest interpretation of the predictions of the string theory landscape, one is free to add to the potential any constant that  keeps the effective value of  $\Lambda$ within the anthropically allowed range of $|\Lambda| \lesssim 10^{{-120}}$. If, for example, one adds a positive cosmological constant $\Lambda  \lesssim 10^{{-120}}$, the last prediction in \rf{rw} changes to $w_{\infty}=-1$, without altering the prediction for $r$ and the spectral index $n_{s}$. In other words, by combining quintessential inflation with the string theory landscape, we have a possibility to describe a broad range of outcomes for $w$ without altering the inflationary predictions of the models.

We also studied $\alpha$-attractor models where inflaton and quintessence  are described by two different fields. From the point of view of model building, these models described in section \ref{twofield} can be quite simple, but they allow much greater flexibility, which deserves a more detailed investigation.

Thus, we constructed a class of models which provide a good fit to the existing observational data related to inflation and  dark energy. None of these models  solve the cosmological constant problem without the help of the ideas based on anthropic considerations and inflationary multiverse/string theory landscape. This seems to be a general problem of various presently existing alternatives to the simple cosmological constant scenario \cite{Brax:2017idh}. The construction of dark energy  models with flat directions and $w = -1$ is relatively simple, and some of these models do not require much fine-tuning in addition to the required fine-tuning of the present value of dark energy.  The construction of models with $w$ close to $-1$, but distinctly different from it, is more complicated and requires additional fine-tuning. In some models, including the models studied in sections \ref{linsec} and \ref{sec:exp-pot}, this extra fine-tuning can be relatively modest. For example, the main fine-tuning required in the simple linear model studied in section \ref{linsec} is the choice of $\alpha \lesssim 0.02$ and a proper adjustment of the mechanism of reheating.

An interesting byproduct of our investigation of $\alpha$-attractors is the realization that their universal prediction $n_{s} = 1-2/N$ may give distinctly different numerical results for the quintessential $\alpha$-attractors as compared to the usual $\alpha$-attractors with a conventional reheating mechanism. We noticed that for some of the quintessential $\alpha$-attractors with gravitational reheating, the required number of inflationary $e$-folds $N$ can be greater than the required number of $e$-folds in more conventional models by $\Delta N \sim 10$, which increases the value of $n_{s}$ by about $0.006$. This increase coincides with the Planck $1\sigma$ error bar  for $n_{s}$ \cite{Ade:2015lrj}. Therefore with the future improvement in the accuracy of CMB observations we might be able to distinguish  the conventional inflationary models where the field after inflation oscillates and  relaxes at the minimum of its potential, from the simplest models of quintessential inflation, even if these models predict $w = -1$.

\acknowledgments We thank A.~Ach\'ucarro, P.~Bull, P.~Creminelli, R.~Flauger, G.~Hinshaw, E.~Komatsu, E.~Linder, M.~Martinelli, L.~Senatore, E.~Silverstein, A.~Silvestri, D.-G.~Wang, and Y.~Yamada for their interest in this work and for helpful discussions and comments. Y.A. acknowledges support from the Netherlands Organization for Scientific Research (NWO) and the Dutch Ministry of Education, Culture and Science (OCW), and also from the D-ITP consortium, a program of the NWO that is funded by the OCW. The work of R.K. and  A.L. is supported by SITP at Stanford, and by the US National Science Foundation grant PHY-1720397. R.K. and A.L. are grateful to the Lorentz Center in Leiden for the hospitality when this work was initiated.
V.V. is supported by a de Sitter PhD fellowship of NWO. 

\appendix

\section{Constraints on exponential models without relying on COBE normalization}
\label{sec:appendix}

In this appendix we reproduce the results of section~\ref{sec:compdataexp} when the COBE/Planck normalization is fully ignored, and therefore no inflationary constraints are placed on $M^2$. This is interesting for two reasons. First of all, it is instructive to see the constraints on the models when only the late-time cosmological data are used without combining them with the inflationary ones. This shows more clearly how a combination of the two sets of constraints (early- and late-time) helps us to more strongly constrain the models, and what are the effects of the two classes of observations individually. In addition, one may want to consider our scalar-field models as regular quintessence scenarios without connecting them to inflation in the framework of quintessential inflation. In that case, there are no constraints on the scale $M$, and therefore the final predictions of the models might be different. This will show how powerful a description of both early and late times in a unified framework can be.

\begin{figure}[h!]
\center
  \includegraphics[height=4cm]{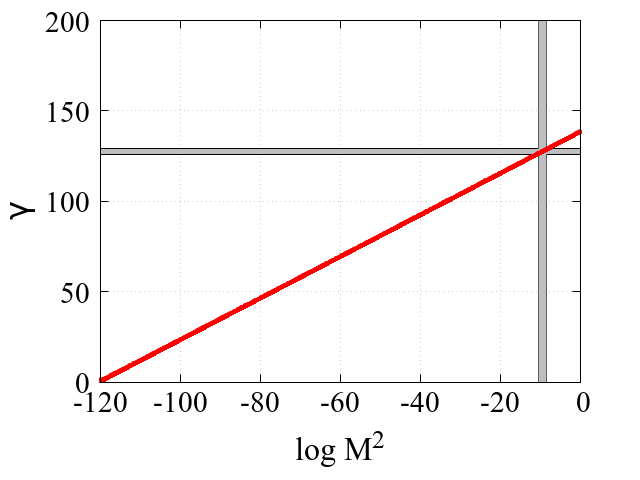}
  \includegraphics[height=4cm]{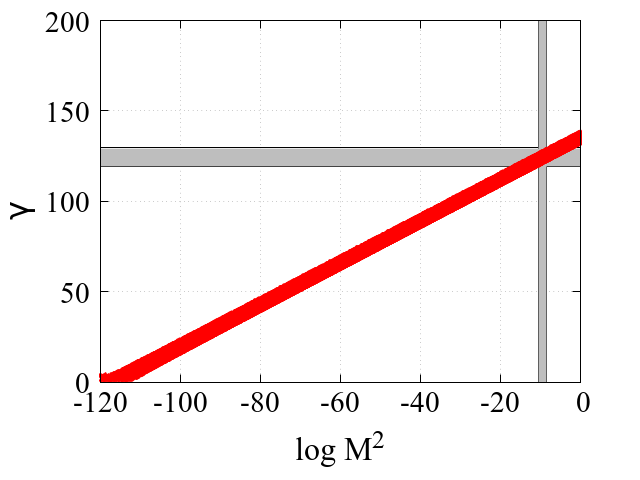}
  \includegraphics[height=4cm]{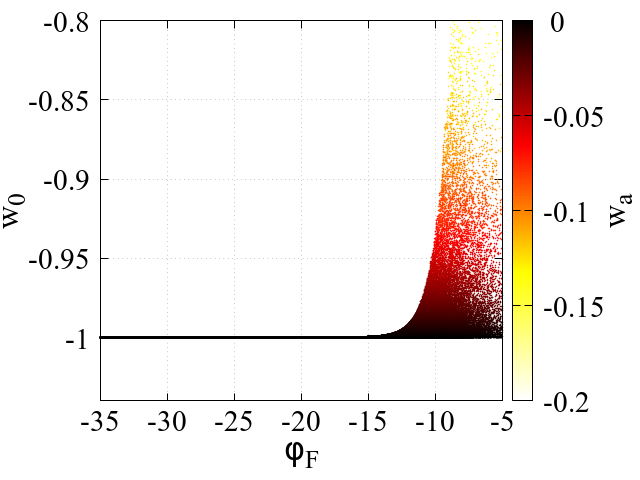}
  \includegraphics[height=4cm]{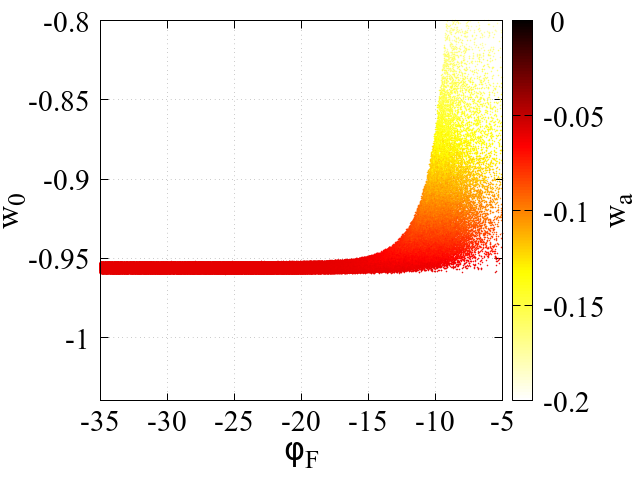}
\caption{\footnotesize\label{fig:model0-V0-n-NOcobe} {\it Upper panels:} Cosmological constraints on $\log M^{2}$ and $\gamma$ for Exp-model I (left panel) and Exp-model II (right panel) when $\varphi_\text{F}$ is allowed to vary between $-35$ and $+8$. The thin, red regions show the values compatible with current constraints on cosmic histories, and the vertical and horizontal, grey bands show, respectively, the ranges of $\log M^{2}$ and $\gamma$ used in the analysis performed in section~\ref{sec:compdataexp}. {\it Lower panels:} CPL parameters $w_0$ and $w_{a}$ for the dark energy equation of state, for Exp-models I (left panel) and II (right panel) as functions of $\varphi_\text{F}$. The points cluster around $w_0=-1$ (model I) and $w_0\sim-0.96$ (model II) for large, negative values of $\varphi_\text{F}$.}
\end{figure}

Let us therefore first fix $\alpha$ to $7/3$ as we did in section~\ref{sec:compdataexp}, and scan over a very wide range of values for $M^{2}$ and $\gamma$. We choose the ranges $[-120,0]$ and $[0,300]$ for $\log M^{2}$ and $\gamma$, respectively. Fig.~\ref{fig:model0-V0-n-NOcobe} shows the constraints we obtain on $\log M^{2}$ and $\gamma$ for Exp-model I (upper, left panel) and Exp-model II (upper, right panel); compare these with the corresponding panels in Fig.~\ref{fig:model0-model1-w0-wa-cobe}. The red region in each case presents the values of the two parameters which are compatible with the cosmological constraints used in our numerical scans, and clearly shows a strong correlation between $M^{2}$ and $\gamma$. For all the values of $\log M^{2}$ and $\gamma$ in this region we find cosmologies that are in perfect agreement with all the background cosmological data. Note that the point $(\gamma,\log M^{2})=(0,-120)$ corresponds to a cosmological constant. The two vertical and horizontal, grey bands show, respectively, the ranges of $\log M^{2}$ and $\gamma$ used in scanning the parameter space for the discussion of section~\ref{sec:compdataexp}. Note how narrow these bands are compared to the ranges considered here, although there we had allowed $M^{2}$ to vary within two orders of magnitude around the COBE/Planck value given by (\ref{eq:COBEconst}). Another observation is that the red region is thiner for Exp-model I compared to Exp-model II.

Let us now try to understand the (red) degeneracy lines in Fig.~\ref{fig:model0-V0-n-NOcobe} (upper panels) by studying analytically the behavior of the potentials for large and negative $\vp$, i.e. on the tails of the potentials corresponding to the late-time evolution of the universe. The potentials in this $\vp\to - \infty$ limit become
\begin{align}
&\text{Exp-model I:}~~~~~~~~~~~V(\vp)=M^{2}e^{-2\gamma}(1+2\gamma e^{2\frac{\vp}{\sqrt{6\alpha}}})\, ,\\
&\text{Exp-model II:}~~~~~~~~~~V(\vp)=M^{2}e^{-2\gamma}\gamma e^{2\frac{\vp}{\sqrt{6\alpha}}}\, .
\end{align}
For Exp-model I, the leading term is $M^{2}e^{-2\gamma}$, which is the quantity that is constrained by the data. The value of this quantity should be close to the observed cosmological constant $\Lambda_\text{obs}$, therefore
\begin{equation}
M^{2}e^{-2\gamma} \approx \Lambda_\text{obs} \Rightarrow \log M^{2} \approx 0.869\gamma +\log \Lambda_\text{obs}\, ,\label{eq:degmodelI}
\end{equation}
which is in very good agreement with the red line in the upper, left panel of Fig.~\ref{fig:model0-V0-n-NOcobe}. The same argument holds for Exp-model II with the entire $M^{2}e^{-2\gamma}\gamma e^{2\frac{\vp}{\sqrt{6\alpha}}}$ being the leading term. There are now two extra contributions to $\log M^{2}$ in (\ref{eq:degmodelI}): $\log\gamma$ and $\log e^{2\frac{\vp}{\sqrt{6\alpha}}}$. The former is a small number, of $\mathcal{O}(2)$, and the latter is also small, as $|\vp|$ is quite large. That is why the red degeneracy line in the upper, right panel of Fig.~\ref{fig:model0-V0-n-NOcobe} for Exp-model II has a slope almost identical to the one in the left panel for Exp-model I. It is however interesting to note the slight dependence of the degeneracy region on $\varphi$ for Exp-model II. This tells us that we should expect slight changes in the position of the line (moving up and down) by changing the value of $\varphi_\text{F}$, while the line for Exp-model I is expected to be quite insensitive to the choice of $\varphi_\text{F}$ --- one can see that this is indeed the case by looking at the {\it zoomed} version of the degeneracy regions around the COBE/Planck constraints presented in Fig.~\ref{fig:model0-model1-w0-wa-cobe} of section~\ref{sec:compdataexp}. In addition, this explains why the line for Exp-model II is thicker compared to Exp-model I.

Let us now take a look at the behavior of the dark energy component by computing the two CPL quantities $w_{0}$ and $w_{a}$. The results are shown in Fig.~\ref{fig:model0-V0-n-NOcobe} (lower panels); compare these with the lower panels of Fig.~\ref{fig:model0-model1-w0-wa-cobe}  of section~\ref{sec:compdataexp}. First of all, both models again show strong levels of clustering of viable models around $w_0=-1$ (for Exp-model I) and $w_0\sim-0.96$ (for Exp-model II) when $|\varphi_\text{F}|$ becomes large. Exp-model I, as discussed before, behaves asymptotically like a cosmological constant, and that is why $w_0$ and $w_{a}$ approach $-1$ and $0$, respectively, corresponding to a $\Lambda$-like dark energy. Exp-model II, on the other hand, does not have a cosmological constant asymptote, and not only $w_\text{DE}$ in that model approaches a non-$\Lambda$ (universal) value when $N\to\infty$ (as discussed earlier in this paper), its present value $w_0$ also behaves like an attractor, independently of the values of $M^{2}$ and $\gamma$. Note that the value of $w_\text{DE}$ today for this model for large, negative $\varphi_\text{F}$ is different from the asymptotic value $w_{\infty}$ (which is $\sim -0.905$ for $\alpha=7/3$ considered here), as discussed before. The main difference between the results here and the ones presented in Fig.~\ref{fig:model0-model1-w0-wa-cobe} is for small $|\varphi_\text{F}|$. Here, contrary to Fig.~\ref{fig:model0-model1-w0-wa-cobe}, the scatter plots do not show thin regions for $w_0$ at small $|\varphi_\text{F}|$, and $w_0$ can take a variety of values equal or larger than $-1$. Reducing the allowed range of $M^2$ cuts out the scatter plots from the lower, right corners, and eventually, by restricting $M^2$ to only take the values constrained by the COBE/Planck normalization, the scatter plots turn into thin regions, as shown in Fig.~\ref{fig:model0-model1-w0-wa-cobe}.

\begin{figure}[h!]
\center
  \includegraphics[height=4.5cm]{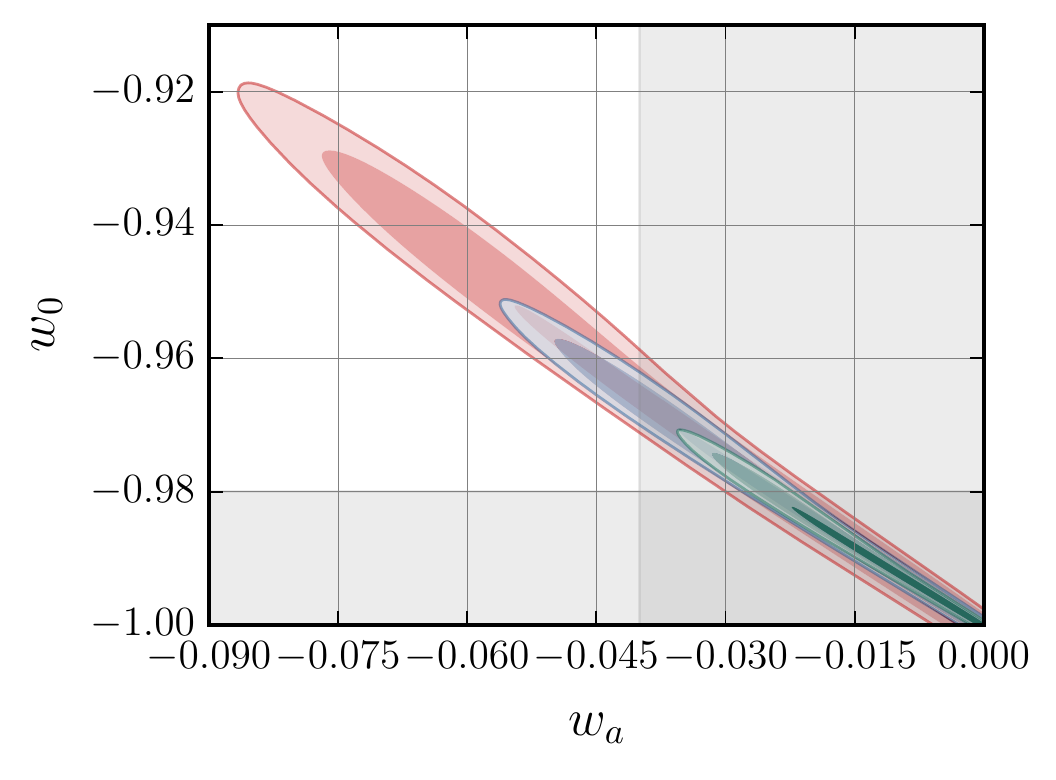}
\caption{\footnotesize\label{fig:model0-w0-wa-NOcobe} The same as in the left panel of Fig.~\ref{fig:model1-w0-wa-cobe}, but when no inflationary constraints have been imposed on $M^2$.}
\end{figure}

We can now fix the value of $\varphi_\text{F}$, and see how the constraints on $w_0$ and $w_a$ are affected compared to Fig.~\ref{fig:model1-w0-wa-cobe} of section~\ref{sec:compdataexp}, when the inflationary constraints on $M^2$ are relaxed. Fig.~\ref{fig:model0-w0-wa-NOcobe} shows the results of our scans for Exp-model I when $\varphi_\text{F}$ is fixed to the same three values of $-10$ (red contours), $-10.5$ (blue contours), and $-11$ (green contours), as in section~\ref{sec:compdataexp}. First of all, the figure shows that the deviations can be as large as about $10\%$ for both $w_{0}$ and $w_{a}$ if $|\varphi_\text{F}|$ is allowed to take values as low as about $10$. More importantly, since here we have not imposed any inflationary constraints on $M^2$, the contours are continuously connected to the $\Lambda$CDM values $w_{0}=-1$ and $w_{a}=0$, in contrast to the results of Fig.~\ref{fig:model1-w0-wa-cobe}. We find similar results for Exp-model II, with the only main difference that in that case the contours are no longer connected to the $\Lambda$CDM point, as expected; we do not show them here for brevity.

\begin{figure}[h!]
\center
  \includegraphics[height=4cm]{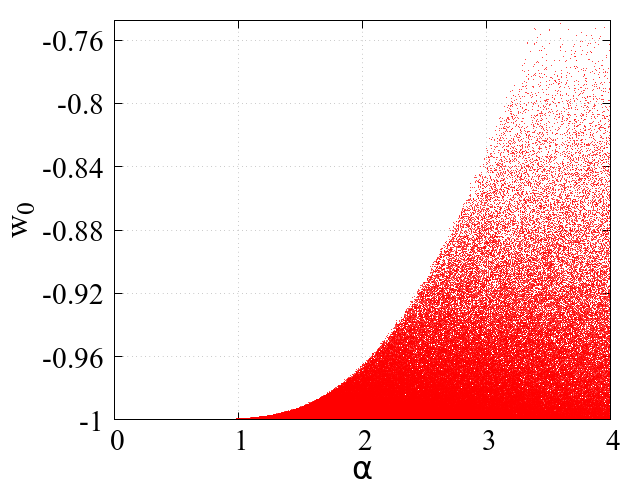}
  \includegraphics[height=4cm]{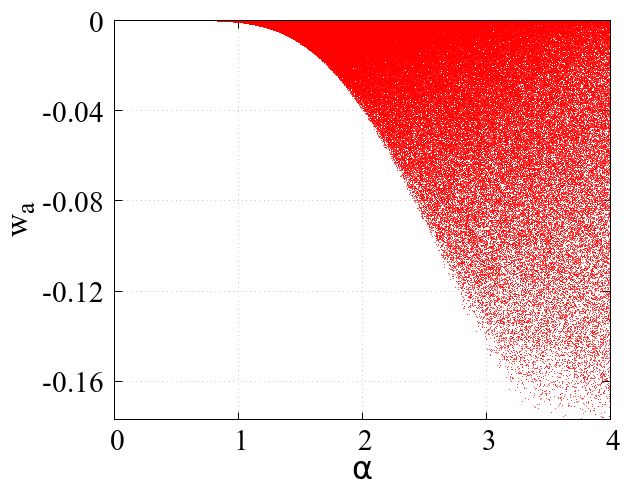}
  \includegraphics[height=4cm]{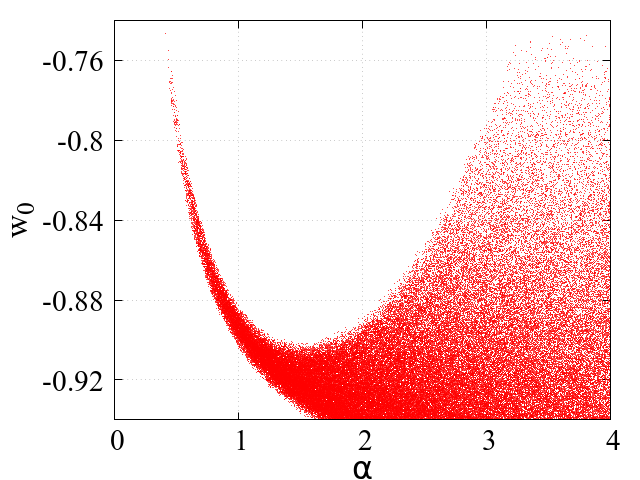}
  \includegraphics[height=4cm]{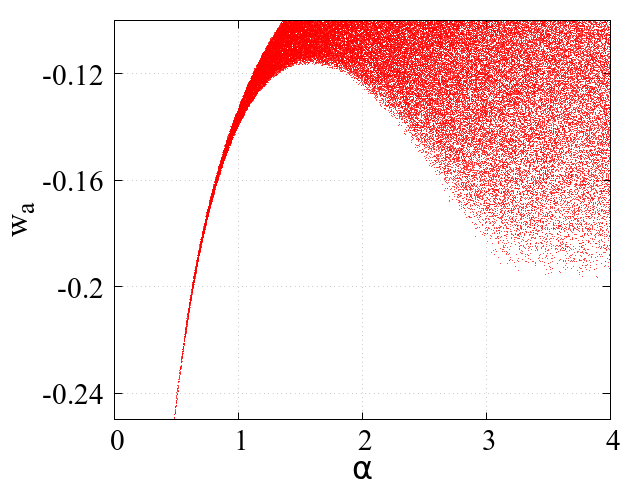}
\caption{\footnotesize\label{fig:model0-model1-w0-wa-alpha-phirehm10-noCOBE} The same as in Fig.~\ref{fig:model0-model1-w0-wa-alpha-phirehm10}, but when the COBE/Planck normalization has not been imposed and $M^2$ has been allowed to freely vary. Upper and lower panels again correspond to Exp-model I and Exp-model II, respectively.}
\end{figure}

Finally, let us present the results of our analysis for cases where $\alpha$ is not fixed, i.e. it is allowed to vary. Here, in contrast to the analysis of section~\ref{sec:compdataexp}, we do not impose the COBE/Planck condition $M^{2} \approx 10^{-10} \alpha$, leave both $M^2$ and $\alpha$ completely free, and scan over a wide range of $M^2$. Fig.~\ref{fig:model0-model1-w0-wa-alpha-phirehm10-noCOBE} shows the results of our numerical scans for both Exp-model I (upper panels) and Exp-model II (lower panels) when $\varphi_\text{F}$ has been fixed to $-10$, to be compared with Fig.~\ref{fig:model0-model1-w0-wa-alpha-phirehm10} of section~\ref{sec:compdataexp}. The figure shows that now there are larger uncertainties in the values of $w_0$ and $w_a$ for large values of $\alpha$.

\bibliographystyle{JHEP}

\bibliography{lindekalloshrefs,akramirefs}

\end{document}